\documentclass[pra, reprint]{revtex4-1} 
\usepackage[T1]{fontenc}    
\usepackage[utf8]{inputenc} 
\usepackage{graphicx}       
\usepackage{amssymb}        
\usepackage{amsmath}        
\usepackage{siunitx}        
\usepackage{hyperref}       
\usepackage[final]{changes}        
\newcommand{\stkout}[1]{\ifmmode\text{\sout{\ensuremath{#1}}}\else\sout{#1}\fi}
\setdeletedmarkup{\stkout{#1}} 

\begin{document}

\makeatletter                   
\renewcommand\@biblabel[1]{#1.} 
\makeatother                    

\title{Self-induced ultrafast electron-hole plasma temperature oscillations in nanowire lasers}
\author{Andreas Thurn}\email{andreas.thurn@wsi.tum.de}
\author{Jochen Bissinger}%
\affiliation{Walter Schottky Institut, Technische Universität München, Am Coulombwall 4, 85748 Garching, Germany.}
\author{Stefan Meinecke}
\affiliation{Institut für Theoretische Physik, Technische Universität Berlin, Hardenbergstraße 36, 10623 Berlin, Germany.}
\author{Paul Schmiedeke}
\affiliation{Walter Schottky Institut, Technische Universität München, Am Coulombwall 4, 85748 Garching, Germany.}
\author{Sang Soon Oh}
\affiliation{School of Physics and Astronomy, Cardiff University, Cardiff CF24 3AA, UK.}
\author{Weng W. Chow}
\affiliation{Sandia National Laboratories, Albuquerque, New Mexico 87185-1086, USA.}
\author{Kathy Lüdge}
\affiliation{Institut für Physik, Technische Universität Ilmenau, Weimarer Straße 25, 98693 Ilmenau, Germany.}
\author{Gregor Koblmüller}
\author{Jonathan J. Finley}\email{finley@wsi.tum.de}
\affiliation{Walter Schottky Institut, Technische Universität München, Am Coulombwall 4, 85748 Garching, Germany.}


\begin{abstract}
Nanowire lasers can be monolithically and site-selectively integrated onto silicon photonic circuits. To assess their full potential for ultrafast opto-electronic devices, a detailed understanding of their lasing dynamics is crucial. However, the roles played by their resonator geometry and the microscopic processes that mediate energy exchange between the photonic, electronic, and phononic \added{sub}systems are largely unexplored. \replaced{Here, we study the dynamics of GaAs-AlGaAs core-shell nanowire lasers at cryogenic temperatures using a combined experimental and theoretical approach. Our results indicate that these NW lasers exhibit sustained intensity oscillations with frequencies ranging from $\SI{160}{GHz}$ to $\SI{260}{GHz}$. As the underlying physical mechanism, we identified self-induced electron-hole plasma temperature oscillations resulting from a dynamic competition between photoinduced carrier heating and cooling via phonon scattering. These dynamics are intimately linked to the strong interaction between the lasing mode and the gain material, which arises from the wavelength-scale dimensions of these lasers. We anticipate that our results could lead to new approaches for ultrafast intensity and phase modulation of chip-integrated nanoscale semiconductor lasers.}{Here, we apply femtosecond pump-probe spectroscopy to show that GaAs-AlGaAs core-shell nanowire lasers exhibit sustained intensity oscillations with frequencies ranging from $\SI{160}{GHz}$ to $\SI{260}{GHz}$. These dynamics are intricately linked to the strong interaction between the lasing mode and the gain material arising from their wavelength-scale dimensions. Combined with dynamic competition between photoinduced carrier heating and cooling via phonon scattering, this enables self-induced electron-hole plasma temperature oscillations, which modulate the laser output. We anticipate that our results will lead to new approaches for ultrafast intensity and phase modulation of chip-integrated nanoscale semiconductor lasers.}
\end{abstract}

\maketitle



\section{Introduction}
Nanowires (NW) provide a unique approach to site-selectively and epitaxially integrate direct gap semiconductors onto silicon~\cite{Chen2011_NanopillarsOnSilicon,Sun2014_RTLaserOnSilicon,Mayer2016_GaAsLaserOnSilicon,Stettner2017_WaveguidePaper}. Optically pumped lasing has been demonstrated at room temperature using III-V and II-VI compound semiconductors~\cite{Mayer2013_RTLaserGaAs,Saxena2013_RTLaserGaAs,Zimmler2008_FirstZnORTSingleNWLasing}, and group-III nitrides~\cite{Johnson2002_RTLaserGaN}, with emission wavelengths that can be tuned from the ultraviolet to the near infrared spectral range~\cite{Huang2001_RTLaserZnO,Johnson2002_RTLaserGaN,Duan2003_RTLaserCdSElectricallyDriven,chin_near-infrared_2006,Zimmler2008_FirstZnORTSingleNWLasing,Geburt2012_CdSRTOpticallyPumped,Mayer2013_RTLaserGaAs,Saxena2013_RTLaserGaAs,Stettner2016_RTGaAs/AlGaAsMQW,Saxena2016_RTLaserGaAs/AlGaAsMQW,Stettner2018_RTGaAs/InGaAsMQW,Schmiedeke2021_TelecomLasingNW}. Thus, they are promising candidates for on-chip lasers in photonic integrated circuits. To further their development, a detailed understanding of their lasing dynamics is essential. Previous studies laid the groundwork by investigating several aspects of carrier relaxation, gain dynamics, plasmonic \added{and coherent} phenomena, and the role of lattice defects in a number of different material systems~\cite{Sidiropoulos2014_PumpProbePlasmonicLaser,Roder2015_PumpProbeNWLaserCdSZnOGaN,Wille2016_SingleLasingZnOStreakCamera,Blake2016_SingleZnOOKGMicroscope,Hollinger2017_EnsembleZnOSFG,Blake2020_SingleZnOOKGRelaxationOscillation,Mayer2017_PumpProbeGaAsNWLaser}. \deleted{Moreover, evidence for ultrafast intensity oscillations was observed in NW lasers, for which coherent effects have been suggested as a possible explanation~\cite{Mayer2017_PumpProbeGaAsNWLaser}.} Despite these advances, however, the microscopic mechanisms and dynamic processes that ultimately determine, and possibly limit, their potential for ultrafast opto-electronic devices remained largely unexplored.
\setlength{\parskip}{0em}
\par
Here, we investigate the microscopic lasing dynamics of GaAs-AlGaAs core-shell NW lasers~\cite{Mayer2013_RTLaserGaAs,Mayer2016_GaAsLaserOnSilicon,Mayer2017_PumpProbeGaAsNWLaser} and \replaced{present evidence for}{demonstrate that the} ultrafast \added{intensity and phase} oscillations\added{, which} do not stem from coherent effects \added{as previously suggested~\cite{Mayer2017_PumpProbeGaAsNWLaser}}, but \replaced{instead}{in fact} correspond to an exceptionally strong non-equilibrium analog of relaxation oscillations~\cite{Schneider1997_UltrafastDynamicsOfVCSELs}, with frequencies ranging from $\SI{160}{GHz}$ to $\SI{260}{GHz}$. We show that these unique dynamics are enabled by the miniaturized dimensions of the\added{se} laser\added{s} and the resulting competition between carrier heating and cooling \deleted{in the electron-hole plasma} during lasing operation. Our results are supported by complementary microscopic simulations based on a quantum statistical~\cite{Jahnke1995_QuantumStatisticalTheoryPaper} and a semiconductor Bloch model~\cite{Chow1994_SemiconductorLaserPhysicsBook} (Methods).\\

\section{Results}
\subsection{Pump-probe measurement and simulation.}
Femtosecond pump-probe spectroscopy was performed on single NWs with a carefully timed pair of pump and probe pulses, separated by a time delay $\mathit{\Delta}t$. Details on the optical characterization and growth of the NWs are summarized in the Methods section.

\begin{figure*}
    \includegraphics[keepaspectratio]{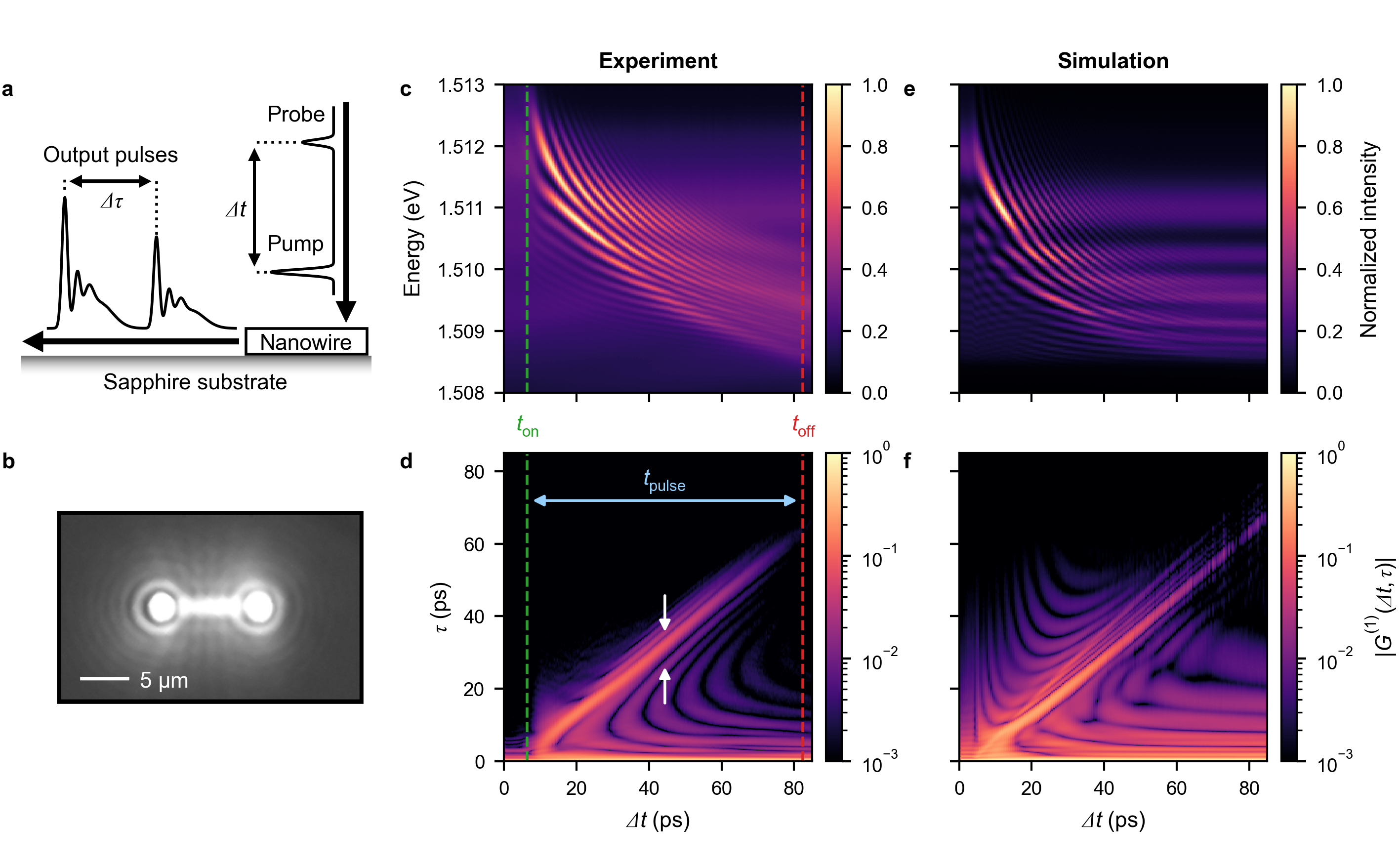}
    \caption{\textbf{Pump-probe measurement and simulation.} \textbf{a}, Schematic of a NW laser subject to pump-probe excitation with delay $\mathit{\Delta} t$, emitting two temporally asymmetric and modulated pulses, separated by $\mathit{\Delta\tau}$. \textbf{b}, Optical microscope image of the NW laser studied in this work, with the excitation laser filtered out. \textbf{c}, Time-integrated spectra, showing typical two-pulse interference fringes with additional beating patterns along the energy axis. \textbf{d}, Normalized magnitude of the electric field autocorrelation ($|G^{(1)}(\mathit{\Delta} t,\tau)|$), exhibiting pronounced oscillations along $\tau$ above and below the main sideband (indicated by white arrows), resulting from the beating patterns in \textbf{c}. In \textbf{c},\textbf{d} we indicated the laser turn-on time ($t_{\mathrm{on}}$) and the disappearance of the sideband ($t_{\mathrm{off}}$). Their difference, $t_{\mathrm{pulse}}\sim t_{\mathrm{off}}-t_{\mathrm{on}}$, gives a measure of the total output pulse length. \textbf{e},\textbf{f}, Corresponding semiconductor Bloch simulation of the respective experimental data in \textbf{c},\textbf{d}.}
    \label{Fig:1}
\end{figure*}

Figure~\ref{Fig:1}a illustrates our pump-probe excitation scheme. The pump pulse power ($P_{\mathrm{pump}}$) and probe pulse power ($P_{\mathrm{probe}}$) are above and below the threshold ($P_{\mathrm{th}}$) of the NW laser, respectively. After excitation, the NW laser emits two modulated output pulses, separated by a time delay $\mathit{\Delta}\tau$. Figure~\ref{Fig:1}b shows an optical microscope image of the NW laser studied in this work (length $L\sim\SI{10}{\micro m}$) during lasing operation. Its emission is dominated by a single mode at an energy of $\sim\SI{1.51}{eV}$ (Supplementary Note I) and is clearly visible from both end-facets. All measurements and simulations were performed at a lattice temperature of $T_{\mathrm{L}}=\SI{10}{K}$ and with an electron excess energy of $\sim\SI{60}{meV}$, unless specified otherwise.

Figure~\ref{Fig:1}c presents typical results of a pump-probe measurement as a function of $\mathit{\Delta} t$. Here, we normalized the spectra to their maximum value and set the excitation power to $P_{\mathrm{pump}}/P_{\mathrm{th}}\sim2.7$ and $P_{\mathrm{probe}}/P_{\mathrm{th}}\sim0.7$, such that the probe pulse alone cannot induce lasing. We observe a delayed onset of two-pulse interference fringes, reflecting the turn-on time ($t_{\mathrm{on}}$) of the laser, which depends on both the excitation conditions \textit{and} the initial relaxation of the photoexcited carriers~\cite{Sidiropoulos2014_PumpProbePlasmonicLaser,Roder2015_PumpProbeNWLaserCdSZnOGaN,Mayer2017_PumpProbeGaAsNWLaser,Roder2018b_ReviewUltrafastNWDynamics}. Experimentally, $t_{\mathrm{on}}=\SI[separate-uncertainty = true]{6.3(8)}{ps}$ was determined from the $\mathit{\Delta} t$ dependence of the spectrally integrated mode intensity (Supplementary Note II), using the transient depletion of the probe absorption~\cite{Sidiropoulos2014_PumpProbePlasmonicLaser}. For increasing $\mathit{\Delta} t$, the interference fringes increase linearly in frequency. Their existence further shows that the weak probe pulse restarts lasing, whereby the probe-induced output pulse partially adopts the phase of the residual electric field in the cavity defined by the previous output pulse~\cite{Mayer2017_PumpProbeGaAsNWLaser}. Hence, the strong pump pulse leaves considerable excitation for an extended period of time, which decays with increasing $\mathit{\Delta} t$. This, in turn, leads to a change in refractive index and, thus, to a pronounced redshift of the interference pattern. Meanwhile, the fringes become weaker and finally disappear at a delay $\mathit{\Delta}t=t_{\mathrm{off}}$. 

For ease of interpretation, we move to the time domain by Fourier transforming the energy axis in Fig.~\ref{Fig:1}c. This yields the electric field autocorrelation $G^{(1)}(\mathit{\Delta} t,\tau)$ as a function of $\mathit{\Delta} t$ and time shift $\tau$~\cite{Sidiropoulos2014_PumpProbePlasmonicLaser,Loudon2000_TheQuantumTheoryOfLight}. Its normalized magnitude $|G^{(1)}(\mathit{\Delta} t,\tau)|$ is presented in Fig.~\ref{Fig:1}d on a logarithmic scale. For normalization, we used the respective maximum value at $\tau=0$ for each $\mathit{\Delta}t$ to facilitate comparison with theory. The fringes in Fig.~\ref{Fig:1}c lead to a linear sideband, with a full width at half maximum (FWHM) of $\sim\SI{3}{ps}$, that sets on at a delay $\mathit{\Delta}t=t_{\mathrm{on}}$ and disappears at $\mathit{\Delta}t=t_{\mathrm{off}}$. We quantified $t_{\mathrm{off}}=\SI[separate-uncertainty = true]{82.4(4)}{ps}$ as the delay time where the main sideband amplitude decreased to $1\%$ of its maximum value. Since the photon lifetime of the resonator is $<\SI{1}{ps}$~\cite{Mayer2013_RTLaserGaAs}, the disappearance of the main sideband marks the termination of the first NW laser output pulse. Thus, $t_{\mathrm{pulse}}\sim t_{\mathrm{off}}-t_{\mathrm{on}}=\SI[separate-uncertainty = true]{76.1(9)}{ps}$ is a measure for the overall output pulse duration. However, in addition to the main sideband, we also observe pronounced oscillations above and below, as indicated by white arrows in Fig.~\ref{Fig:1}d. These reflect the beating patterns in the spectral interference fringes in Fig.~\ref{Fig:1}c. Weak indications of these oscillations can be found in previous work~\cite{Sidiropoulos2014_PumpProbePlasmonicLaser,Mayer2017_PumpProbeGaAsNWLaser}, but their significance has not been scrutinized until now. Together, the long $t_{\mathrm{pulse}}$, the short FWHM of the main sideband, and the oscillations indicate that the NW output pulses are strongly asymmetric and modulated in time.

To understand the origin of the oscillating features in Fig.~\ref{Fig:1}d, we first used the semiconductor Bloch model to simulate the experimental data. For these and all following simulations, we used an end-facet reflectivity of $R=0.5$ and a spontaneous emission coupling factor of $\beta=0.1$\deleted{ (Supplementary Note III)}. Both values were obtained using a quantum statistical simulation of the measured continuous-wave lasing characteristics of the NW under investigation (Supplementary Note III) and are in full accord with literature~\cite{Mayer2013_RTLaserGaAs,Saxena2013_RTLaserGaAs,Saxena2015ModeProfiling,Mayer2016_GaAsLaserOnSilicon}. The result of this semiconductor Bloch approach is presented in Fig.~\ref{Fig:1}e,f, whereby we used the same excitation powers as in the experiment. The simulation exhibits excellent qualitative agreement with all observed features of the experimental data in Fig.~\ref{Fig:1}c,d. It reproduces the delayed onset of interference fringes, the temporal asymmetry of the output pulses and the oscillations both above and below the main sideband. Moreover, the model reveals that these originate from carrier temperature oscillations (Supplementary Note IV).\\

\subsection{Quantum statistical simulation.} 
However, to uncover the full significance of these oscillations and to enable a theoretical description of measurement series within which scattering rates vary, it is necessary to go beyond the relaxation rate approximation used in the semiconductor Bloch model. For this purpose, we used a quantum statistical model that self-consistently calculates the rates for carrier-carrier and carrier-phonon scattering.

\begin{figure*}
    \includegraphics[keepaspectratio]{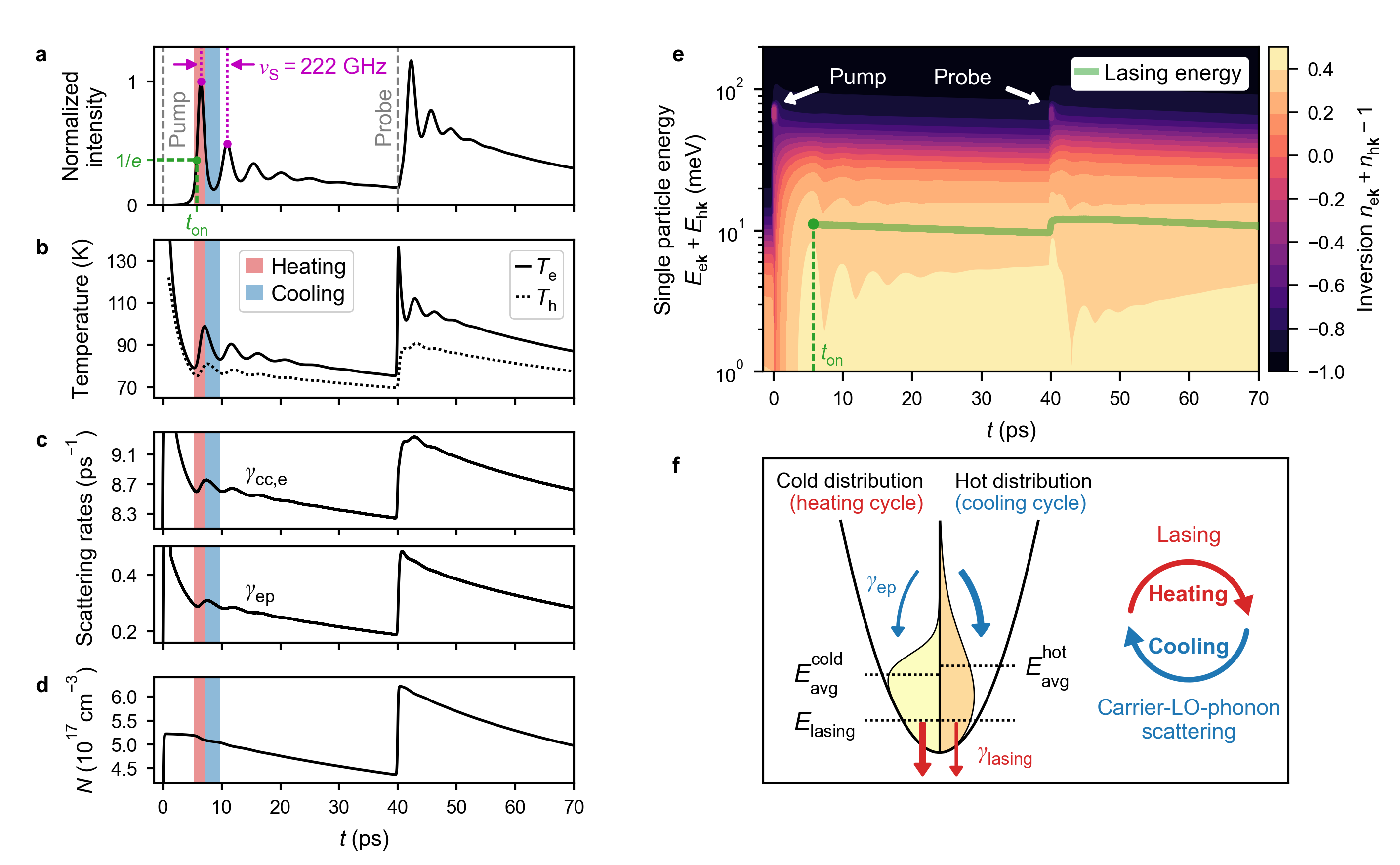}
    \caption{\textbf{Quantum statistical simulation of the time-resolved pump-probe response.} \textbf{a}, Time-dependent output intensity for a pump-probe delay of $\mathit{\Delta}t=\SI{40}{ps}$. From the first output pulse we determine $t_{\mathrm{on}}$ and the oscillation frequency $\nu_{\mathrm{S}}$. \textbf{b-d}, Corresponding time-dependent electron ($T_\mathrm{e}$) and hole ($T_\mathrm{h}$) temperature, carrier-carrier scattering rate of electrons ($\gamma_{\mathrm{cc{,}e}}$), electron-LO-phonon scattering rate ($\gamma_{\mathrm{ep}}$) and carrier density ($N$). In \textbf{a-d}, the shaded areas highlight the first heating and cooling cycle of $T_{\mathrm{e}}$. \textbf{e}, Energy-resolved inversion, illustrating the time-dependent carrier dynamics. Here, $n_{\mathrm{c}\mathbf{k}}$ and $E_{\mathrm{c}\mathbf{k}}$, with $\mathrm{c}\in\{\mathrm{e},\mathrm{h}\}$, are respectively the occupation probability and energy of electrons and holes with wave vector $\mathbf{k}$.  \textbf{f}, Sketch of the oscillation mechanism, enabled by the exceptionally large modal confinement factor $\mathit{\Gamma}$ of NW lasers. Here, $\gamma_{\mathrm{lasing}}$ is the stimulated emission rate of the lasing mode with energy $E_{\mathrm{lasing}}$, while $E_{\mathrm{avg}}^{\mathrm{cold}}$ and $E_{\mathrm{avg}}^{\mathrm{hot}}$ are the average kinetic energies of the respective distributions.}
    \label{Fig:2}
\end{figure*}

Figure~\ref{Fig:2} displays the results of this approach for a pump-probe excitation with $\mathit{\Delta} t=\SI{40}{ps}$. The excitation powers used were the same as in Fig.~\ref{Fig:1}. In Fig. \ref{Fig:2}a, we present the laser intensity as a function of time and determine $t_{\mathrm{on}}\sim\SI{5.7}{ps}$ as the time it takes to reach $1/e$ of the first output pulse maximum. After turn-on, the NW pulses are observed to be strongly asymmetric in time with a pronounced initial peak ($\mathrm{FWHM}\sim\SI{1.3}{ps}$), following oscillations with a frequency of $\nu_{\mathrm{S}}=\SI{222}{GHz}$ and a long tail. This is in qualitative agreement with both the experiment and the simulation in Fig.~\ref{Fig:1}. Following up on the results of the semiconductor Bloch model, we investigated the temporal evolution of the carrier distributions by determining the instantaneous electron ($T_{\mathrm{e}}$) and hole ($T_{\mathrm{h}}$) temperatures for each time step (Supplementary Note V). Figure~\ref{Fig:2}b shows that the carrier temperatures cool to $T_{\mathrm{e}}\sim\SI{79}{K}$ and $T_{\mathrm{h}}\sim\SI{76}{K}$ after $t\sim\SI{5.3}{ps}$, shortly before the laser turns on, which is in full accord with literature~\cite{Leheny1979c_CoolingCurves,Leo1987a_CoolingCurvesBulkGaAs}. This cooling primarily takes place via scattering of carriers with longitudinal optical (LO) phonons~\cite{Shah1969a_LOPhononScatteringDominant,Shah1978a_CarrierRelaxationReview,Shah1999_UltrafastSpectroscopyBook}. After the first temperature minimum at $t\sim\SI{5.3}{ps}$ and during turn-on, we observe a pronounced initial increase of the carrier temperatures by $\mathit{\Delta} T_{\mathrm{e}}\sim\SI{19.8}{K}$ and $\mathit{\Delta} T_{\mathrm{h}}\sim\SI{4.2}{K}$ for electrons and holes, respectively. Subsequently, the carrier temperatures show clear oscillations with a frequency of $\SI{222}{GHz}$, mirrored by the laser intensity in Fig.~\ref{Fig:2}a. Since $T_{\mathrm{e}}$ changes the most, we highlight in Fig.~\mbox{\ref{Fig:2}a-d} its first heating and cooling cycle in red and blue, respectively. As is observed in Fig.~\ref{Fig:2}c, the carrier temperature dynamics are also mirrored in the time dependence of the electron scattering rates near the lasing energy. While these time variations in the scattering rates are not necessary for the existence of the oscillations in Fig.~\ref{Fig:2}a,b, they likely do increase their strength (Supplementary Note IV). Near $t_{\mathrm{on}}$, the carrier-carrier scattering rate of electrons is $\gamma_{\mathrm{cc{,}e}}\sim\SI{8.6}{\per\ps}$, while the electron-LO-phonon scattering rate is $\gamma_{\mathrm{ep}}\sim\SI{0.3}{\per\ps}$. In contrast to the oscillations in Fig.~\mbox{\ref{Fig:2}a-c}, the carrier density in Fig.~\ref{Fig:2}d shows a monotonic, step-like decrease starting from an initial value of $N\sim\SI{5.2e17}{cm^{-3}}$, besides the injection of additional carriers at $t=\SI{40}{ps}$.

To visualize the carrier dynamics in the parabolic bands, Fig.~\ref{Fig:2}e shows the energy-dependent carrier inversion as a function of time in a semi-logarithmic contour plot. The optical pump-probe excitation is indicated by white arrows. Within a few picoseconds, relaxation of the injected carriers (Supplementary Note VI) leads to build up of inversion at the lasing energy (solid green line). After the onset of lasing ($t_{\mathrm{on}}$, green point), clear oscillations of the inversion are visible, which are directly connected to the carrier temperature oscillations in Fig. \ref{Fig:2}b. From Fig.~\ref{Fig:2}e, we further observe no signs of spectral hole burning. This results from the fast carrier-carrier scattering, that continuously thermalizes the electron-hole plasma, and the low end-facet reflectivity $R$, which reduces the stimulated emission rate ($\gamma_{\mathrm{lasing}}$).

To explain the origin of these oscillations, Fig.~\ref{Fig:2}f shows a schematic representation of two electron distributions having similar densities in the conduction band; one characterized by a colder $T_{\mathrm{e}}$ (yellow) than the other (orange). The lasing energy ($E_{\mathrm{lasing}}$) is below the respective average kinetic energy ($E_{\mathrm{avg}}^{\mathrm{cold}}$, $E_{\mathrm{avg}}^{\mathrm{hot}}$). Thus, stimulated emission at a rate $\gamma_{\mathrm{lasing}}$ heats the remaining electron distribution~\cite{Kesler1987_UltrafastTemperatureDynamicsLaser,Stix1986_GainStrongFunctionOfTemperature,Gomatam1990_ExtensionOfBimbergPaperToStimEm,Jahnke1993a_CWSwitchOnVCSEL_Theoryof,Pompe1995_TransientResponseOfVCSELs,Jahnke1996_UltrafastDynamicsOfQWVCSELs}. In case this heating via $\gamma_{\mathrm{lasing}}$ is sufficiently strong to overcome the cooling via $\gamma_{\mathrm{ep}}$, $T_{\mathrm{e}}$ rises. Consequently, $\gamma_{\mathrm{ep}}$ increases, whereas $\gamma_{\mathrm{lasing}}$ decreases due to the reduction in material gain ($G_{\mathrm{mat}}$)~\cite{Stix1986_GainStrongFunctionOfTemperature,Kesler1987_UltrafastTemperatureDynamicsLaser}. Partial recovery of $G_{\mathrm{mat}}$ via $\gamma_{\mathrm{ep}}$ then restarts the cycle. A similar discussion applies to holes.

However, in previous investigations of microcavity and nanoscale lasers this intuitive effect of ultrafast self-induced temperature oscillations was strongly damped, preventing the observation of sustained oscillations~\cite{Michler1995_TransientPulseResponse,Michler1996_EmissionDynamicsOfVCSELs,Jahnke1996_UltrafastDynamicsOfQWVCSELs,Schneider1997_UltrafastDynamicsOfVCSELs,Pompe1995_TransientResponseOfVCSELs}. We explain the observations in Fig.~\mbox{\ref{Fig:2}a-e} for NW lasers by noting that the modal gain ($G_{\mathrm{mod}}$) and $G_{\mathrm{mat}}$ are linked by $G_{\mathrm{mod}}\propto\mathit{\Gamma}\cdot G_{\mathrm{mat}}$, where $\mathit{\Gamma}$ is the modal confinement factor~\cite{Coldren2012}. The corresponding differential gains with respect to carrier temperature ($T_{\mathrm{c}}$, with $\mathrm{c}\in\{\mathrm{e},\mathrm{h}\}$) are, therefore, to a good approximation related by $\partial G_{\mathrm{mod}} / \partial\hspace{.09em}T_{\mathrm{c}}\propto\mathit{\Gamma}\cdot\partial G_{\mathrm{mat}} / \partial\hspace{.09em}T_{\mathrm{c}}$. Thus, large $\mathit{\Gamma}$ leads to enhanced interactions between the lasing mode and the gain material. For NW lasers, this effect is especially pronounced due to their wavelength-scale dimensions and unique resonator geometry, that allows exceptionally strong mode confinement with $\mathit{\Gamma}>1$~\cite{maslov_modal_2004,Saxena2013_RTLaserGaAs,Ning2010a_ConfinementFactor}. We further note that the large $\mathit{\Gamma}$ of NW lasers is, in part, due to the strong lateral optical confinement and the resulting strongly non-paraxial mode propagation~\cite{maslov_modal_2004,Ning2010a_ConfinementFactor}. Moreover, here, there exists a balanced interplay between the rates of stimulated emission, carrier-carrier scattering, and carrier-LO-phonon scattering. As mentioned above, carrier-carrier scattering is sufficiently fast to maintain Fermi-Dirac distributions, and heating via stimulated emission strong enough to compete with cooling via carrier-LO-phonon scattering, such that light-matter coupling is manifested as a strongly oscillating carrier temperature. In our NW lasers, it is precisely this combination of large $\mathit{\Gamma}$ with these carrier dynamics that leads to the observed sustained oscillations. The important role of $\mathit{\Gamma}$ was confirmed by a comparison of the simulated laser dynamics with $\mathit{\Gamma}=1.2$ (used here) and $\mathit{\Gamma}=0.12$, that clearly demonstrates that the oscillations disappear for weaker mode confinement (Supplementary Note VII).\\

\subsection{Excitation power and lattice temperature dependence.} 
To test the above hypothesis for the origin of the observed ultrafast oscillations, we experimentally studied the excitation power and $T_{\mathrm{L}}$ dependence of the NW laser dynamics and compared with the predictions of the quantum statistical model.

\begin{figure*}
    \includegraphics[keepaspectratio]{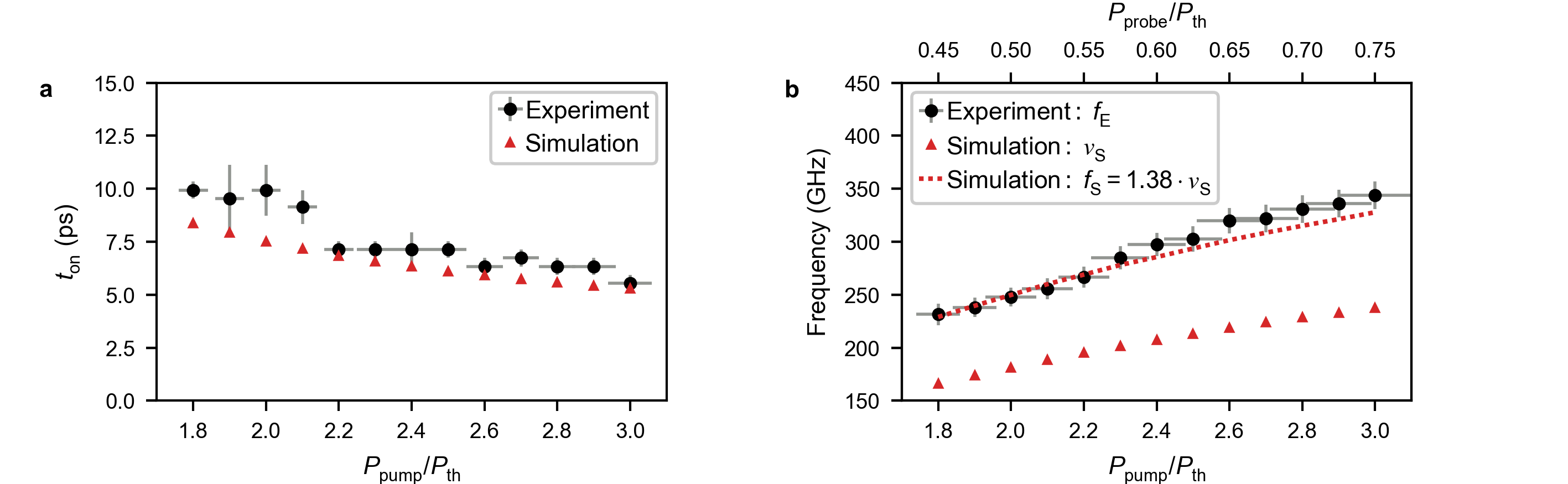}
    \caption{\textbf{Excitation power dependence of the NW laser dynamics.} \textbf{a}, A stronger pump pulse power ($P_{\mathrm{pump}}/P_{\mathrm{th}}$) produces a greater initial carrier density $N$, allowing the laser to turn on faster. Here, $P_{\mathrm{probe}}/P_{\mathrm{th}}\sim0.5$ was kept constant. \textbf{b}, The increase in oscillation frequencies with excitation power depends on a complex interplay of several effects, but is mainly driven by the increased stimulated emission rate $\gamma_{\mathrm{lasing}}$. Here, we kept the ratio $P_{\mathrm{pump}}/P_{\mathrm{probe}}=4:1$ constant to ensure that the intensities of both output pulses remained comparable. All error bars represent $\SI{95}{\%}$ confidence intervals (CIs) of the mean and result from the methods used to determine $P_{\mathrm{th}}$, $t_{\mathrm{on}}$ and $f_{\mathrm{E}}$, as described in Supplementary Note I, II and VIII, respectively. In \textbf{b}, the uncertainty in excitation power refers to the bottom axis.}
    \label{Fig:3}
\end{figure*}

Figure~\ref{Fig:3} presents the results of the excitation power dependent measurement, with which we investigated in detail how the lasing dynamics depend on $N$ and $\gamma_{\mathrm{lasing}}$. As shown in Fig.~\ref{Fig:3}a, $t_{\mathrm{on}}$ decreases from $\SI[separate-uncertainty = true]{9.9(4)}{ps}$ to $\SI[separate-uncertainty = true]{5.5(4)}{ps}$ as the pumping level increases from $P_{\mathrm{pump}}/P_{\mathrm{th}}\sim1.8$ to $\sim3.0$. This decrease is in excellent quantitative agreement with theory. It results from the increase in $N$ with increasing $P_{\mathrm{pump}}/P_{\mathrm{th}}$, allowing the laser to turn on at higher carrier temperatures.

To estimate the frequency ($f_{\mathrm{E}}$) of the oscillations in $|G^{(1)}(\Delta t,\tau)|$ from our measurements, we determined the time interval between the main sideband and the first oscillation above (Supplementary Note VIII). As shown in Fig.~\ref{Fig:3}b, $f_{\mathrm{E}}$ increases with stronger excitation power from $\SI[separate-uncertainty = true]{231(14)}{GHz}$ to $\SI[separate-uncertainty = true]{344(18)}{GHz}$. Since $|G^{(1)}(\Delta t,\tau)|$ and, thus, the simulated oscillation frequency ($f_{\mathrm{S}}$) cannot be directly obtained from the quantum statistical model, we computed the oscillation frequency $\nu_{\mathrm{S}}$. Here, we observe an increase from $\sim\SI{166}{GHz}$ to $\sim\SI{238}{GHz}$, over the same range of excitation powers. Nonetheless, an approximate relationship between $\nu_{\mathrm{S}}$ and $f_{\mathrm{S}}$ can be established using the semiconductor Bloch model. In the present case, $f_{\mathrm{S}}$ is related to $\nu_{\mathrm{S}}$ by $f_{\mathrm{S}}=1.38\cdot\nu_{\mathrm{S}}$ (Supplementary Note IX). Applying this relationship to the frequencies $\nu_{\mathrm{S}}$ in Fig.~\ref{Fig:3}b leads to remarkable quantitative agreement between $f_{\mathrm{S}}$ and $f_{\mathrm{E}}$ over the entire range of excitation powers. The increase of the oscillation frequencies with stronger excitation power results from the complex interplay of various effects. As the main reason, we identified the increased $\gamma_{\mathrm{lasing}}$, leading to enhanced carrier temperature oscillations and, hence, shorter heating and cooling cycles.

\begin{figure*}
    \includegraphics[keepaspectratio]{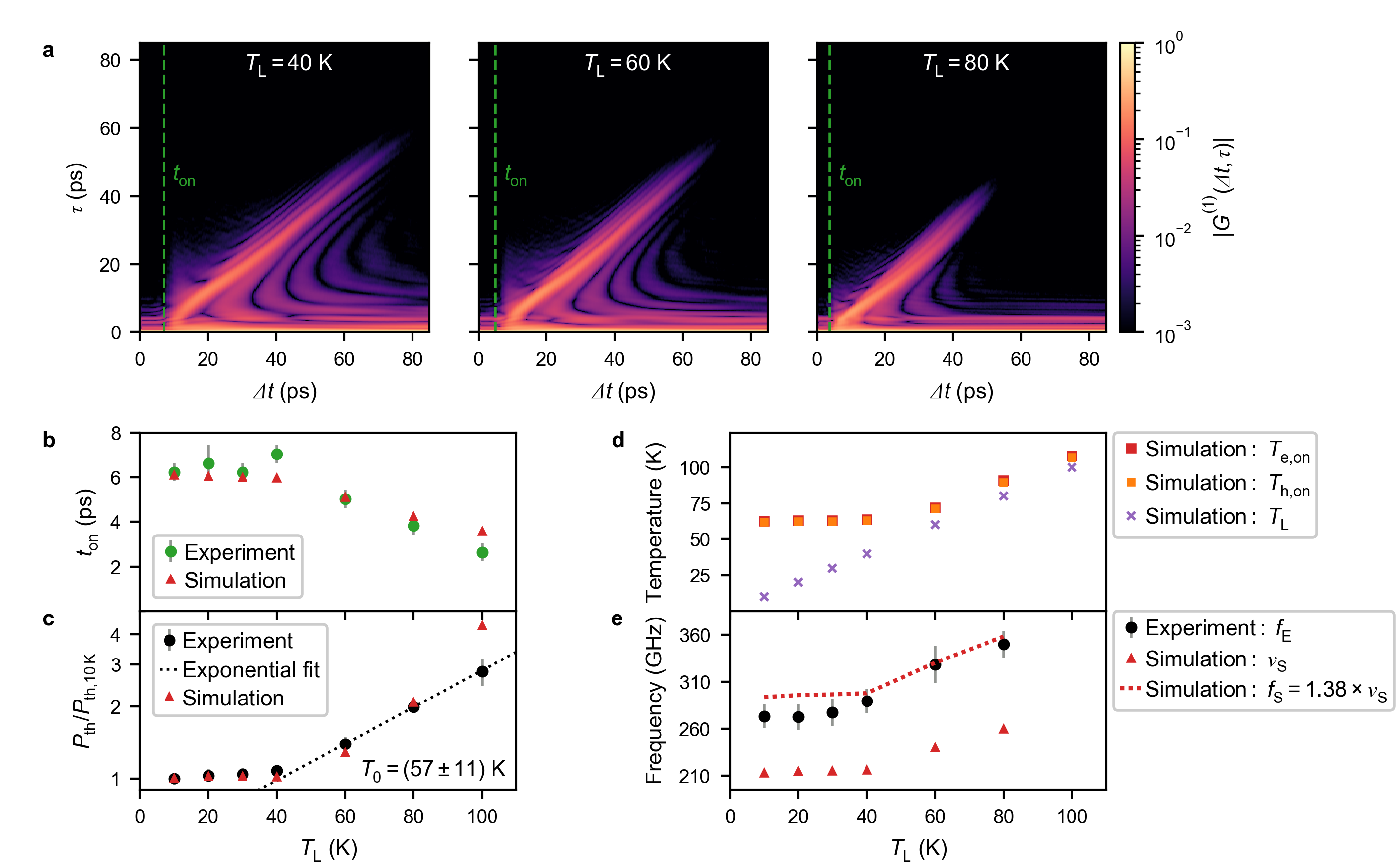}
    \caption{\textbf{Lattice temperature dependence of the NW laser dynamics.} \textbf{a}, Three exemplary pump-probe measurements, illustrating the changes that occur with increasing $T_{\mathrm{L}}$. \textbf{b}, As $T_{\mathrm{L}}$ increases, $t_{\mathrm{on}}$ is at first constant and then decreases. \textbf{c}, Above $\SI{40}{K}$, $P_{\mathrm{th}}$ increases with a characteristic temperature $T_{0}$, as obtained from an exponential fit. This increase leads to the decrease of $t_{\mathrm{on}}$ in \textbf{b}. The data were normalized to the smallest threshold ($P_{\mathrm{th,\SI{10}{K}}}$). \textbf{d}, Electron ($T_{\mathrm{e,on}}$) and hole ($T_{\mathrm{h,on}}$) temperatures at times $t=t_{\mathrm{on}}$ as a function of $T_{\mathrm{L}}$, whereby the excitation power was set to $P_{\mathrm{pump}}/P_{\mathrm{th}}\sim1$. The rise of $T_{\mathrm{e,on}}$ and $T_{\mathrm{h,on}}$ causes the increase of $P_{\mathrm{th}}$ in \textbf{c}. \textbf{e}, The oscillation frequencies increase as the lasing mode shifts towards the high-energy side of the gain spectrum, where $\partial G_{\mathrm{mat}}/\partial\hspace{.09em}T_{\mathrm{c}}$ is larger. All error bars represent $\SI{95}{\%}$ CIs of the mean and result from the methods used to determine $P_{\mathrm{th}}$, $t_{\mathrm{on}}$ and $f_{\mathrm{E}}$, as described in Supplementary Note I, II and VIII, respectively. In \textbf{c}, some error bars are smaller than the symbol size.}
    \label{Fig:4}
\end{figure*}

Complementary to the excitation power series, varying $T_{\mathrm{L}}$ allowed to tune $\partial G_{\mathrm{mat}} / \partial\hspace{.09em}T_{\mathrm{c}}$, and to shift the operating point of the laser towards higher carrier temperatures. All experimental data in Fig.~\ref{Fig:4} were measured with $P_{\mathrm{pump}}/P_{\mathrm{th}}\sim2.5$ and $P_{\mathrm{probe}}/P_{\mathrm{th}}\sim0.5$. For illustration, Fig.~\ref{Fig:4}a presents the measured $|G^{(1)}(\Delta t,\tau)|$ for $T_{\mathrm{L}}=\SI{40}{K},\ \SI{60}{K}$ and $\SI{80}{K}$. As $T_{\mathrm{L}}$ increases, the data show that $t_{\mathrm{on}}$ decreases, $f_{\mathrm{E}}$ increases and $t_{\mathrm{pulse}}$ decreases (Supplementary Note X).

As shown by the data in Fig.~\ref{Fig:4}b, $t_{\mathrm{on}}$ remains approximately constant at $\sim\SI{6}{ps}$ up to $\SI{40}{K}$ and then decreases to $\SI[separate-uncertainty = true]{2.6(4)}{ps}$ as $T_{\mathrm{L}}$ further increases to $\SI{100}{K}$. This behaviour can be understood on the basis of the $T_{\mathrm{L}}$ dependence of $P_{\mathrm{th}}$, shown in Fig.~\ref{Fig:4}c. For $T_{\mathrm{L}}\leq\SI{40}{K}$, $P_{\mathrm{th}}$ remains almost constant, whereas at higher $T_{\mathrm{L}}$ it increases exponentially with a characteristic temperature $T_{0}=\SI[separate-uncertainty=true]{57(11)}{K}$. Since the pumping level is fixed relative to $P_{\mathrm{th}}$, this increase leads to a larger initial $N$. Consequently, $t_{\mathrm{on}}$ decreases, as observed in Fig.~\ref{Fig:4}b. The simulations presented in Fig.~\ref{Fig:4}b,c quantitatively reproduce the experimental data and, thus, confirm our interpretation.

The $T_{\mathrm{L}}$ dependence of $P_{\mathrm{th}}$ is, in turn, accounted for by the variation of the electron ($T_{\mathrm{e,on}}$) and hole ($T_{\mathrm{h,on}}$) temperatures at laser turn-on, as shown in Fig.~\ref{Fig:4}d. For increasing $T_{\mathrm{L}}$, they are initially constant at $\sim\SI{62}{K}$, reflecting the fact that the cooling rate due to carrier-LO-phonon scattering decreases rapidly for low carrier temperatures ($\lesssim\SI{50}{K}$)~\cite{Shah1969a_LOPhononScatteringDominant,Leheny1979c_CoolingCurves}. Both $T_{\mathrm{e,on}}$ and $T_{\mathrm{h,on}}$ only rise as $T_{\mathrm{L}}$ becomes comparable, shifting the operating point of the laser towards higher carrier temperatures. The carrier distributions therefore spread out towards higher energies, which reduces $G_{\mathrm{mat}}$ at the lasing energy for a given $N$. Hence, $P_{\mathrm{th}}$ increases, explaining the observed trend in Fig.~\ref{Fig:4}c.

We are now in a position to explain the $T_{\mathrm{L}}$ dependence of the oscillation frequencies in Fig.~\ref{Fig:4}e. Up to $\SI{40}{K}$, $f_{\mathrm{E}}$ remains approximately constant at $\sim\SI{270}{GHz}$ and then increases to $\SI[separate-uncertainty = true]{350(20)}{GHz}$ at $T_{\mathrm{L}}=\SI{80}{K}$. Similarly, $\nu_{\mathrm{S}}$ increases from $\SI{213}{GHz}$ to $\SI{259}{GHz}$. We again calculated $f_{\mathrm{S}}$ from $\nu_{\mathrm{S}}$, using the proportionality factor 1.38 stated above, and obtained remarkable agreement with $f_{\mathrm{E}}$. The trend of the oscillation frequencies is strongly influenced by the $T_{\mathrm{L}}$ dependence of $P_{\mathrm{th}}$ in Fig.~\ref{Fig:4}c. As $P_{\mathrm{th}}$ increases, the correspondingly larger initial $N$ blueshifts the lasing mode due to the associated change in refractive index, and broadens the gain spectrum as a result of band gap renormalization and band filling. Simultaneously, the band gap and the lasing mode redshift with increasing $T_{\mathrm{L}}$, whereby the redshift of the lasing mode is smaller than the other effects. In combination, these processes effectively shift the lasing mode towards the high-energy side of the gain spectrum, where $\partial G_{\mathrm{mat}} / \partial\hspace{.09em}T_{\mathrm{c}}$ is larger. This leads to the observed increase in oscillation frequencies. No oscillations were observed for $T_{\mathrm{L}}=\SI{100}{K}$, which coincided with the laser becoming multimodal. In addition, as shown in Fig.~\ref{Fig:4}d, the carrier temperatures increase for higher $T_{\mathrm{L}}$, which reduces $\partial G_{\mathrm{mat}} / \partial\hspace{.09em}T_{\mathrm{c}}$~\cite{Li2000_UltrafastThzModulationLaser}. This counteracts the increases due to the shift of the lasing mode towards the high-energy side of the gain spectrum and at some point will compensate them. The oscillations are therefore expected to disappear at sufficiently high $T_{\mathrm{L}}$, here $>\SI{80}{K}$.\\

\section{Discussion}
In summary, we investigated the microscopic lasing dynamics of GaAs-AlGaAs core-shell NW lasers. Thereby, we demonstrated how a large $\mathit{\Gamma}$, and the consequently large $\partial G_{\mathrm{mod}} / \partial\hspace{.09em}T_{\mathrm{c}}$, can lead to exceptionally strong non-equilibrium laser dynamics. These manifest themselves as carrier temperature oscillations through a dynamic competition between carrier heating via stimulated emission and carrier cooling via carrier-LO-phonon scattering. The results of our combined experimental and theoretical approach are consistent with oscillation frequencies ranging from $\SI{160}{GHz}$ to $\SI{260}{GHz}$. 

\added{Following the promising results of this work, one possible next step would be to directly time-resolve the emission of a single GaAs-AlGaAs core-shell NW laser. Here, the main challenge is to obtain a sufficient signal-to-noise ratio, since the volume and thus the output emission intensity of NW lasers is relatively small~\cite{Roder2018b_ReviewUltrafastNWDynamics}. Such a measurement would allow a direct experimental validation of the simulated data shown in Fig.~\ref{Fig:2}a.}

Since \replaced{the described}{these} dynamics involve oscillations of the carrier temperature and not the carrier density, they circumvent the speed limitations inherent in conventional relaxation oscillations~\cite{Li2000_UltrafastThzModulationLaser}. For direct laser modulation based on changing the carrier density, the relaxation oscillation frequency determines the maximum modulation speed, which is currently limited to $<\SI{60}{GHz}$ without photonic feedback effects~\cite{Yamaoka2021DirectModulation108GHz}. Similarly, we believe that the oscillation frequencies described here determine the ultimate speed limit for laser intensity or phase modulation based on changing the carrier temperature. We therefore anticipate that NW lasers could substantially improve previously proposed laser modulation schemes based on terahertz heating fields~\cite{Li2000_UltrafastThzModulationLaser}. This is because their large $\partial G_{\mathrm{mod}} / \partial\hspace{.09em}T_{\mathrm{c}}$ makes them highly susceptible to applied \added{electric} fields. In such a scheme, an externally applied terahertz heating field modulates the carrier temperature and thus the laser output, avoiding the speed bottleneck of interband modulation~\cite{Li2000_UltrafastThzModulationLaser}. This also creates new opportunities for ultrafast pulse shaping of nanoscale semiconductor lasers and, provided the modulation depth is sufficient, would allow the generation of picosecond pulse trains with extremely high repetition rates $>\SI{100}{GHz}$. In perspective, further research could achieve ultrafast modulation of NW lasers integrated into silicon photonic circuits~\cite{Stettner2017_WaveguidePaper}.

\replaced{The}{Our results conclusively show that the} dynamics described above are substantially different from the expected class B dynamics that are typical for semiconductor lasers and usually described by a standard rate equation approach~\cite{Coldren2012}. In our case, however, such an approach is insufficient. Firstly, the predicted overall output pulse length and shape well above threshold would not correspond well to our experimental observations~\cite{Saxena2013_RTLaserGaAs,Mayer2016_GaAsLaserOnSilicon}. Secondly, relaxation oscillations would be too slow to explain the observed dynamics~\cite{Yamaoka2021DirectModulation108GHz} and, thirdly, not expected following excitation with femtosecond pulses~\cite{Saxena2013_RTLaserGaAs,Mayer2016_GaAsLaserOnSilicon}. This made it necessary to go beyond such a rate equation approach and the corresponding small-signal response theory to be able to describe our experimentally observed dynamics. It is, nevertheless, important to note that the rate equation approach can be augmented to include temperature effects. This can either be done, for example, by introducing a gain saturation parameter~\cite{Coldren2012}, or by including the carrier temperature directly as a dynamical variable~\cite{ning_self-consistent_1995}. However, this requires the introduction of additional phenomenological parameters and would still need a self-consistent inclusion of many-body effects in order to yield predictions that can be quantitatively compared with experiments. Microscopic models, such as those used in the present work~\cite{Jahnke1993c_CWSwitchOnVCSEL_Transient,Chow1994_SemiconductorLaserPhysicsBook}, solve these issues and allow a detailed description of nanoscale lasers on picosecond timescales while minimizing the number of free phenomenological parameters.

We note that our theoretical models~\cite{Jahnke1993c_CWSwitchOnVCSEL_Transient,Chow1994_SemiconductorLaserPhysicsBook} are general and thus allow our findings to be applied to other types of semiconductor lasers as well. There are several points to consider in the development of semiconductor lasers, that are intended to be highly susceptible to externally applied terahertz fields. According to the key insight of this work, such a laser should have a large $\Gamma$. This can be achieved by an optimized spatial overlap of the lasing mode and the gain material\replaced{, which simultaneously also decreases the threshold}{and / or by designing laser cavities with a low group velocity}~\cite{maslov_modal_2004,Ning2010a_ConfinementFactor}. The choice of the gain material is crucial as well, as it directly affects the cooling rate of the carriers via carrier-LO-phonon scattering which is mediated by the Fröhlich interaction. Moreover, the operating point of the carrier density and temperature should be chosen such that $\partial G_{\mathrm{mod}} / \partial\hspace{.09em}T_{\mathrm{c}}$ is maximized. According to a previous theoretical investigation, this is the case when both the carrier density and temperature are as low as possible~\cite{Li2000_UltrafastThzModulationLaser}. The differential modal gain can further be optimized by tuning the cavity length to place the lasing mode on the high energy side of the gain spectrum~\cite{Michler1995_TransientPulseResponse,Michler1996_EmissionDynamicsOfVCSELs,Grabmaier1991_DiffGainIncreasesWithIncreasingEnergy}. Lastly, the laser must have a large enough resonator bandwidth and should preferentially be single mode. Both points can be achieved by making the laser as short as possible, which of course, must be balanced with the simultaneously increasing threshold. Taken together, our work thus \replaced{opens up new approaches to}{shows a new approach of} how the miniaturization of semiconductor lasers can be used to design their ultrafast dynamical properties.

This is complementary to previous research that investigated accelerated laser dynamics based on Purcell enhancement~\cite{Altug2006_UltrafastPhCLaser,romeira_purcell_2018} and plasmonic effects~\cite{Sidiropoulos2014_PumpProbePlasmonicLaser,khurgin_comparative_2014}. We note that no Purcell enhancement is expected for our NW lasers since their effective mode volume is $\sim140$ times larger than that of the photonic crystal laser studied in Ref.~\cite{Altug2006_UltrafastPhCLaser}. Other approaches to ultrafast laser modulation, that typically require larger device structures, include the use of photonic feedback effects~\cite{Yamaoka2021DirectModulation108GHz} and mode field switching~\cite{pellegrino_mode-field_2020}. It is noteworthy that with both photonic feedback on a silicon carbide substrate~\cite{Yamaoka2021DirectModulation108GHz} and Purcell enhancement in a GaAs material system~\cite{Altug2006_UltrafastPhCLaser} modulation frequencies of up to $\sim\SI{100}{GHz}$ have been demonstrated, although the latter could not be confirmed theoretically~\cite{suhr_modulation_2010,gregersen_quantum-dot_2012}.

\replaced{Practical applications of course}{Of course, practical applications} would ultimately require ultrafast modulation of nanolasers at room temperature. The current drawback of the dynamics described in our work is the limitation to $T_{\mathrm{L}}\leq\SI{80}{K}$. However, this could be overcome by using a low-dimensional gain medium, such as multi-quantum wells~\cite{Stettner2016_RTGaAs/AlGaAsMQW}, simultaneously allowing emission wavelength tuning towards the technologically relevant telecom bands~\cite{Stettner2018_RTGaAs/InGaAsMQW,Schmiedeke2021_TelecomLasingNW}.\\

\section{Materials and methods}
\subsection{Growth} The investigated NW lasers were grown in a two-step axial and radial growth process on silicon using molecular beam epitaxy~\cite{Mayer2013_RTLaserGaAs,Mayer2016_GaAsLaserOnSilicon}. These NWs have a length in the range of $\SI{10}{\micro\metre}-\SI{16}{\micro\metre}$ and a diameter of $\sim\SI{340}{nm}$~\cite{Mayer2013_RTLaserGaAs}.\\ 

\subsection{Optical characterization.} For optical investigations the NWs were transferred onto a sapphire substrate, providing excellent heat conductivity at cryogenic temperatures and sufficient refractive index contrast to achieve lasing. The samples were mounted inside a liquid helium flow cryostat and all experiments reported in this work were performed within a temperature range of $\SI{10}{\K}-\SI{100}{\K}$. Single NWs were excited using $\sim\SI{200}{\fs}$ optical pulses (repetition frequency $\sim\SI{82}{\MHz}$), delivered by a mode-locked laser, which were focused to a spot diameter of $\sim\SI{17}{\micro\metre}$, covering the entire NW. The detection spot was centered on one of the endfacets of the NW and had a diameter of $\sim\SI{11}{\micro\metre}$. We determined the threshold ($P_{\mathrm{th}}$) of the NWs using single pulse excitation (Supplementary Note I). To study their ultrafast dynamics we performed non-resonant degenerate pump-probe spectroscopy and measured the spectrally resolved and time-integrated NW laser emission. The spectra were recorded as a function of pump-probe delay $\mathit{\Delta} t$ using a double spectrometer and a CCD, whereby the delay was adjusted using an optical delay line with a bi-directional precision $<\SI{10}{\fs}$. With a spectral resolution of $\mathit{\Delta} E<\SI{40}{\micro\eV}$, interference fringes resulting from pulse separations $>\SI{100}{\ps}$ could be resolved.\\

\subsection{Numerical Modelling.}
With our modelling approach, we reproduced the characteristic features of the experiments and, thereby, made deductions about the underlying carrier and carrier-field interaction dynamics. In a complete semi-classical description, the light-matter interaction is governed by the semiconductor Maxwell-Bloch equations, whereby the carrier-carrier and carrier-phonon interactions are described by the proper collision integrals in the Boltzmann equation framework. However, the direct simulation of the experiment at hand turns out to be too computationally expensive, when using this complete approach. We therefore conducted two different model simplifications, depending on the physical feature that was to be investigated. To study the dynamics of the emitted electric field, we used a relaxation rate approximation for the scattering dynamics, using precalculated values (Supplementary Note IV), while keeping the full semiconductor Maxwell-Bloch equation~\cite{Chow1994_SemiconductorLaserPhysicsBook}. We refer to this as the semiconductor Bloch model (Supplementary Note XI). In contrast, to study the carrier dynamics in detail, we eliminated the polarization dynamics and resorted to the field intensity for the light-matter interaction. While this approach cannot describe the dynamics of the electric field or potential coherent transients, it did allow to keep the full Boltzmann-equation framework for the carrier dynamics with reasonable computational effort~\cite{Henneberger1992_SpectralHoleBurning,Henneberger1992_ManyBodyEffects,Jahnke1993a_CWSwitchOnVCSEL_Theoryof,Jahnke1993b_CWSwitchOnVCSEL_DynamicResponse,Jahnke1993c_CWSwitchOnVCSEL_Transient,Jahnke1995_QuantumStatisticalTheoryPaper,Schneider1997_UltrafastDynamicsOfVCSELs,Chow1994_SemiconductorLaserPhysicsBook}. We refer to this as the quantum statistical model (Supplementary Note XII). Both models use the same set of simulation parameters, tabulated in Supplementary Note III.\\

\section*{Acknowledgements}
We \added{would like to} thank B. Lingnau for contributions towards code used in the semiconductor Bloch model and D. Rudolph for growing the NWs. We gratefully acknowledge the DFG for funding via the project FI 947/4-1, and via the clusters of excellence Munich Center for Quantum Science and Technology (MCQST, EXC 2111) and e-conversion (EXC 2089). Further financial support was provided by the FETOPEN project SiLAS (H2020-FETOPEN 735008), and the ERC project QUANtIC (ID:771747) funded by the European Research Council, and by the European Regional Development Fund (ERDF) via the Welsh Government (80762-CU145 (East)). In addition, we gratefully acknowledge funding via the U.S. Department of Energy’s National Nuclear Security Administration contract DE-NA0003525. This work was performed, in part, at the Center for Integrated Nanotechnologies, an Office of Science User Facility operated for the U.S. Department of Energy (DOE) Office of Science.\\

\bibliography{ms}

\begin{thebibliography}{10}
\expandafter\ifx\csname url\endcsname\relax
  \def\url#1{\texttt{#1}}\fi
\expandafter\ifx\csname urlprefix\endcsname\relax\def\urlprefix{URL }\fi
\providecommand{\bibinfo}[2]{#2}
\providecommand{\eprint}[2][]{\url{#2}}

\bibitem{Chen2011_NanopillarsOnSilicon}
\bibinfo{author}{Chen, R.} \emph{et~al.}
\newblock \bibinfo{title}{{Nanolasers grown on silicon}}.
\newblock \emph{\bibinfo{journal}{Nat. Photonics}}
  \textbf{\bibinfo{volume}{5}}, \bibinfo{pages}{170--175}
  (\bibinfo{year}{2011}).

\bibitem{Sun2014_RTLaserOnSilicon}
\bibinfo{author}{Sun, H.} \emph{et~al.}
\newblock \bibinfo{title}{{Nanopillar Lasers Directly Grown on Silicon with
  Heterostructure Surface Passivation}}.
\newblock \emph{\bibinfo{journal}{ACS Nano}} \textbf{\bibinfo{volume}{8}},
  \bibinfo{pages}{6833--6839} (\bibinfo{year}{2014}).

\bibitem{Mayer2016_GaAsLaserOnSilicon}
\bibinfo{author}{Mayer, B.} \emph{et~al.}
\newblock \bibinfo{title}{{Monolithically Integrated High-$\beta$ Nanowire
  Lasers on Silicon}}.
\newblock \emph{\bibinfo{journal}{Nano Lett.}} \textbf{\bibinfo{volume}{16}},
  \bibinfo{pages}{152--156} (\bibinfo{year}{2016}).

\bibitem{Stettner2017_WaveguidePaper}
\bibinfo{author}{Stettner, T.} \emph{et~al.}
\newblock \bibinfo{title}{{Direct Coupling of Coherent Emission from
  Site-Selectively Grown III–V Nanowire Lasers into Proximal Silicon
  Waveguides}}.
\newblock \emph{\bibinfo{journal}{ACS Photonics}} \textbf{\bibinfo{volume}{4}},
  \bibinfo{pages}{2537--2543} (\bibinfo{year}{2017}).

\bibitem{Mayer2013_RTLaserGaAs}
\bibinfo{author}{Mayer, B.} \emph{et~al.}
\newblock \bibinfo{title}{{Lasing from individual GaAs-AlGaAs core-shell
  nanowires up to room temperature}}.
\newblock \emph{\bibinfo{journal}{Nat. Commun.}} \textbf{\bibinfo{volume}{4}}
  (\bibinfo{year}{2013}).

\bibitem{Saxena2013_RTLaserGaAs}
\bibinfo{author}{Saxena, D.} \emph{et~al.}
\newblock \bibinfo{title}{{Optically pumped room-temperature GaAs nanowire
  lasers}}.
\newblock \emph{\bibinfo{journal}{Nat. Photonics}}
  \textbf{\bibinfo{volume}{7}}, \bibinfo{pages}{963--968}
  (\bibinfo{year}{2013}).

\bibitem{Zimmler2008_FirstZnORTSingleNWLasing}
\bibinfo{author}{Zimmler, M.~A.}, \bibinfo{author}{Bao, J.},
  \bibinfo{author}{Capasso, F.}, \bibinfo{author}{M{\"{u}}ller, S.} \&
  \bibinfo{author}{Ronning, C.}
\newblock \bibinfo{title}{{Laser action in nanowires: Observation of the
  transition from amplified spontaneous emission to laser oscillation}}.
\newblock \emph{\bibinfo{journal}{Appl. Phys. Lett.}}
  \textbf{\bibinfo{volume}{93}}, \bibinfo{pages}{051101}
  (\bibinfo{year}{2008}).

\bibitem{Johnson2002_RTLaserGaN}
\bibinfo{author}{Johnson, J.~C.} \emph{et~al.}
\newblock \bibinfo{title}{{Single gallium nitride nanowire lasers}}.
\newblock \emph{\bibinfo{journal}{Nat. Mater.}} \textbf{\bibinfo{volume}{1}},
  \bibinfo{pages}{106--110} (\bibinfo{year}{2002}).

\bibitem{Huang2001_RTLaserZnO}
\bibinfo{author}{Huang, M.~H.} \emph{et~al.}
\newblock \bibinfo{title}{{Room-Temperature Ultraviolet Nanowire Nanolasers}}.
\newblock \emph{\bibinfo{journal}{Science}} \textbf{\bibinfo{volume}{292}},
  \bibinfo{pages}{1897--1899} (\bibinfo{year}{2001}).

\bibitem{Duan2003_RTLaserCdSElectricallyDriven}
\bibinfo{author}{Duan, X.}, \bibinfo{author}{Huang, Y.},
  \bibinfo{author}{Agarwal, R.} \& \bibinfo{author}{Lieber, C.~M.}
\newblock \bibinfo{title}{{Single-nanowire electrically driven lasers}}.
\newblock \emph{\bibinfo{journal}{Nature}} \textbf{\bibinfo{volume}{421}},
  \bibinfo{pages}{241--245} (\bibinfo{year}{2003}).

\bibitem{chin_near-infrared_2006}
\bibinfo{author}{Chin, A.~H.} \emph{et~al.}
\newblock \bibinfo{title}{Near-infrared semiconductor subwavelength-wire
  lasers}.
\newblock \emph{\bibinfo{journal}{Appl. Phys. Lett.}}
  \textbf{\bibinfo{volume}{88}}, \bibinfo{pages}{163115}
  (\bibinfo{year}{2006}).

\bibitem{Geburt2012_CdSRTOpticallyPumped}
\bibinfo{author}{Geburt, S.} \emph{et~al.}
\newblock \bibinfo{title}{{Low threshold room-temperature lasing of CdS
  nanowires}}.
\newblock \emph{\bibinfo{journal}{Nanotechnology}}
  \textbf{\bibinfo{volume}{23}}, \bibinfo{pages}{365204}
  (\bibinfo{year}{2012}).

\bibitem{Stettner2016_RTGaAs/AlGaAsMQW}
\bibinfo{author}{Stettner, T.} \emph{et~al.}
\newblock \bibinfo{title}{{Coaxial GaAs-AlGaAs core-multishell nanowire lasers
  with epitaxial gain control}}.
\newblock \emph{\bibinfo{journal}{Appl. Phys. Lett.}}
  \textbf{\bibinfo{volume}{108}}, \bibinfo{pages}{011108}
  (\bibinfo{year}{2016}).

\bibitem{Saxena2016_RTLaserGaAs/AlGaAsMQW}
\bibinfo{author}{Saxena, D.} \emph{et~al.}
\newblock \bibinfo{title}{{Design and Room-Temperature Operation of GaAs/AlGaAs
  Multiple Quantum Well Nanowire Lasers}}.
\newblock \emph{\bibinfo{journal}{Nano Lett.}} \textbf{\bibinfo{volume}{16}},
  \bibinfo{pages}{5080--5086} (\bibinfo{year}{2016}).

\bibitem{Stettner2018_RTGaAs/InGaAsMQW}
\bibinfo{author}{Stettner, T.} \emph{et~al.}
\newblock \bibinfo{title}{{Tuning Lasing Emission toward Long Wavelengths in
  GaAs-(In,Al)GaAs Core–Multishell Nanowires}}.
\newblock \emph{\bibinfo{journal}{Nano Lett.}} \textbf{\bibinfo{volume}{18}},
  \bibinfo{pages}{6292--6300} (\bibinfo{year}{2018}).

\bibitem{Schmiedeke2021_TelecomLasingNW}
\bibinfo{author}{Schmiedeke, P.} \emph{et~al.}
\newblock \bibinfo{title}{{Low-threshold strain-compensated InGaAs/(In,Al)GaAs
  multi-quantum well nanowire lasers emitting near 1.3 $\si{\micro m}$ at room
  temperature}}.
\newblock \emph{\bibinfo{journal}{Appl. Phys. Lett.}}
  \textbf{\bibinfo{volume}{118}}, \bibinfo{pages}{221103}
  (\bibinfo{year}{2021}).

\bibitem{Sidiropoulos2014_PumpProbePlasmonicLaser}
\bibinfo{author}{Sidiropoulos, T. P.~H.} \emph{et~al.}
\newblock \bibinfo{title}{{Ultrafast plasmonic nanowire lasers near the surface
  plasmon frequency}}.
\newblock \emph{\bibinfo{journal}{Nat. Phys.}} \textbf{\bibinfo{volume}{10}},
  \bibinfo{pages}{870--876} (\bibinfo{year}{2014}).

\bibitem{Roder2015_PumpProbeNWLaserCdSZnOGaN}
\bibinfo{author}{R{\"{o}}der, R.} \emph{et~al.}
\newblock \bibinfo{title}{{Ultrafast Dynamics of Lasing Semiconductor
  Nanowires}}.
\newblock \emph{\bibinfo{journal}{Nano Lett.}} \textbf{\bibinfo{volume}{15}},
  \bibinfo{pages}{4637--4643} (\bibinfo{year}{2015}).

\bibitem{Wille2016_SingleLasingZnOStreakCamera}
\bibinfo{author}{Wille, M.} \emph{et~al.}
\newblock \bibinfo{title}{{Carrier density driven lasing dynamics in ZnO
  nanowires}}.
\newblock \emph{\bibinfo{journal}{Nanotechnology}}
  \textbf{\bibinfo{volume}{27}}, \bibinfo{pages}{225702}
  (\bibinfo{year}{2016}).

\bibitem{Blake2016_SingleZnOOKGMicroscope}
\bibinfo{author}{Blake, J.~C.}, \bibinfo{author}{Nieto-Pescador, J.},
  \bibinfo{author}{Li, Z.} \& \bibinfo{author}{Gundlach, L.}
\newblock \bibinfo{title}{{Ultraviolet femtosecond Kerr-gated wide-field
  fluorescence microscopy}}.
\newblock \emph{\bibinfo{journal}{Opt. Lett.}} \textbf{\bibinfo{volume}{41}},
  \bibinfo{pages}{2462} (\bibinfo{year}{2016}).

\bibitem{Hollinger2017_EnsembleZnOSFG}
\bibinfo{author}{Hollinger, R.} \emph{et~al.}
\newblock \bibinfo{title}{{Enhanced absorption and cavity effects of
  three-photon pumped ZnO nanowires}}.
\newblock \emph{\bibinfo{journal}{Appl. Phys. Lett.}}
  \textbf{\bibinfo{volume}{111}}, \bibinfo{pages}{213106}
  (\bibinfo{year}{2017}).

\bibitem{Blake2020_SingleZnOOKGRelaxationOscillation}
\bibinfo{author}{Blake, J.~C.}, \bibinfo{author}{Nieto-Pescador, J.},
  \bibinfo{author}{Li, Z.} \& \bibinfo{author}{Gundlach, L.}
\newblock \bibinfo{title}{{Femtosecond Luminescence Imaging for Single
  Nanoparticle Characterization}}.
\newblock \emph{\bibinfo{journal}{J. Phys. Chem. A}}
  \textbf{\bibinfo{volume}{124}}, \bibinfo{pages}{4583--4593}
  (\bibinfo{year}{2020}).

\bibitem{Mayer2017_PumpProbeGaAsNWLaser}
\bibinfo{author}{Mayer, B.} \emph{et~al.}
\newblock \bibinfo{title}{{Long-term mutual phase locking of picosecond pulse
  pairs generated by a semiconductor nanowire laser}}.
\newblock \emph{\bibinfo{journal}{Nat. Commun.}} \textbf{\bibinfo{volume}{8}},
  \bibinfo{pages}{15521} (\bibinfo{year}{2017}).

\bibitem{Schneider1997_UltrafastDynamicsOfVCSELs}
\bibinfo{author}{Schneider, H.~C.}, \bibinfo{author}{Jahnke, F.} \&
  \bibinfo{author}{Koch, S.~W.}
\newblock \bibinfo{title}{{Microscopic theory of non-equilibrium microcavity
  laser dynamics}}.
\newblock \emph{\bibinfo{journal}{Quantum Semiclass. Opt.}}
  \textbf{\bibinfo{volume}{9}}, \bibinfo{pages}{693--711}
  (\bibinfo{year}{1997}).

\bibitem{Jahnke1995_QuantumStatisticalTheoryPaper}
\bibinfo{author}{Jahnke, F.} \& \bibinfo{author}{Koch, S.~W.}
\newblock \bibinfo{title}{{Many-body theory for semiconductor microcavity
  lasers}}.
\newblock \emph{\bibinfo{journal}{Phys. Rev. A}} \textbf{\bibinfo{volume}{52}},
  \bibinfo{pages}{1712--1727} (\bibinfo{year}{1995}).

\bibitem{Chow1994_SemiconductorLaserPhysicsBook}
\bibinfo{author}{Chow, W.~W.}, \bibinfo{author}{Koch, S.~W.} \&
  \bibinfo{author}{Sargent, M.}
\newblock \emph{\bibinfo{title}{{Semiconductor-Laser Physics}}}
  (\bibinfo{publisher}{Springer}, \bibinfo{address}{Berlin},
  \bibinfo{year}{1994}).

\bibitem{Roder2018b_ReviewUltrafastNWDynamics}
\bibinfo{author}{R{\"{o}}der, R.} \& \bibinfo{author}{Ronning, C.}
\newblock \bibinfo{title}{{Review on the dynamics of semiconductor nanowire
  lasers}}.
\newblock \emph{\bibinfo{journal}{Semicond. Sci. Technol.}}
  \textbf{\bibinfo{volume}{33}}, \bibinfo{pages}{033001}
  (\bibinfo{year}{2018}).

\bibitem{Loudon2000_TheQuantumTheoryOfLight}
\bibinfo{author}{Loudon, R.}
\newblock \emph{\bibinfo{title}{{The Quantum Theory of Light}}}
  (\bibinfo{publisher}{Oxford University Press}, \bibinfo{address}{Oxford},
  \bibinfo{year}{2000}).

\bibitem{Saxena2015ModeProfiling}
\bibinfo{author}{Saxena, D.} \emph{et~al.}
\newblock \bibinfo{title}{{Mode Profiling of Semiconductor Nanowire Lasers}}.
\newblock \emph{\bibinfo{journal}{Nano Lett.}} \textbf{\bibinfo{volume}{15}},
  \bibinfo{pages}{5342--5348} (\bibinfo{year}{2015}).

\bibitem{Leheny1979c_CoolingCurves}
\bibinfo{author}{Leheny, R.}, \bibinfo{author}{Shah, J.},
  \bibinfo{author}{Fork, R.}, \bibinfo{author}{Shank, C.} \&
  \bibinfo{author}{Migus, A.}
\newblock \bibinfo{title}{{Dynamics of hot carrier cooling in photo-excited
  GaAs}}.
\newblock \emph{\bibinfo{journal}{Solid State Commun.}}
  \textbf{\bibinfo{volume}{31}}, \bibinfo{pages}{809--813}
  (\bibinfo{year}{1979}).

\bibitem{Leo1987a_CoolingCurvesBulkGaAs}
\bibinfo{author}{Leo, K.} \& \bibinfo{author}{R{\"{u}}hle, W.}
\newblock \bibinfo{title}{{Influence of carrier lifetime on the cooling of a
  hot electron-hole plasma in GaAs}}.
\newblock \emph{\bibinfo{journal}{Solid State Commun.}}
  \textbf{\bibinfo{volume}{62}}, \bibinfo{pages}{659--662}
  (\bibinfo{year}{1987}).

\bibitem{Shah1969a_LOPhononScatteringDominant}
\bibinfo{author}{Shah, J.} \& \bibinfo{author}{Leite, R. C.~C.}
\newblock \bibinfo{title}{{Radiative Recombination from Photoexcited Hot
  Carriers in GaAs}}.
\newblock \emph{\bibinfo{journal}{Phys. Rev. Lett.}}
  \textbf{\bibinfo{volume}{22}}, \bibinfo{pages}{1304--1307}
  (\bibinfo{year}{1969}).

\bibitem{Shah1978a_CarrierRelaxationReview}
\bibinfo{author}{Shah, J.}
\newblock \bibinfo{title}{{Hot electrons and phonons under high intensity
  photoexcitation of semiconductors}}.
\newblock \emph{\bibinfo{journal}{Solid-State Electron.}}
  \textbf{\bibinfo{volume}{21}}, \bibinfo{pages}{43--50}
  (\bibinfo{year}{1978}).

\bibitem{Shah1999_UltrafastSpectroscopyBook}
\bibinfo{author}{Shah, J.}
\newblock \emph{\bibinfo{title}{{Ultrafast Spectroscopy of Semiconductors and
  Semiconductor Nanostructures}}}, vol. \bibinfo{volume}{115} of
  \emph{\bibinfo{series}{Springer Series in Solid-State Sciences}}
  (\bibinfo{publisher}{Springer}, \bibinfo{address}{Berlin},
  \bibinfo{year}{1999}).

\bibitem{Kesler1987_UltrafastTemperatureDynamicsLaser}
\bibinfo{author}{Kesler, M.~P.} \& \bibinfo{author}{Ippen, E.~P.}
\newblock \bibinfo{title}{{Subpicosecond gain dynamics in GaAlAs laser
  diodes}}.
\newblock \emph{\bibinfo{journal}{Appl. Phys. Lett.}}
  \textbf{\bibinfo{volume}{51}}, \bibinfo{pages}{1765--1767}
  (\bibinfo{year}{1987}).

\bibitem{Stix1986_GainStrongFunctionOfTemperature}
\bibinfo{author}{Stix, M.~S.}, \bibinfo{author}{Kesler, M.~P.} \&
  \bibinfo{author}{Ippen, E.~P.}
\newblock \bibinfo{title}{{Observations of subpicosecond dynamics in GaAlAs
  laser diodes}}.
\newblock \emph{\bibinfo{journal}{Appl. Phys. Lett.}}
  \textbf{\bibinfo{volume}{48}}, \bibinfo{pages}{1722--1724}
  (\bibinfo{year}{1986}).

\bibitem{Gomatam1990_ExtensionOfBimbergPaperToStimEm}
\bibinfo{author}{Gomatam, B.} \& \bibinfo{author}{DeFonzo, A.}
\newblock \bibinfo{title}{{Theory of Hot Carrier Effects on Nonlinear Gain in
  GaAs-GaAlAs Lasers and Amplifiers}}.
\newblock \emph{\bibinfo{journal}{IEEE J. Quantum Electron.}}
  \textbf{\bibinfo{volume}{26}}, \bibinfo{pages}{1689--1704}
  (\bibinfo{year}{1990}).

\bibitem{Jahnke1993a_CWSwitchOnVCSEL_Theoryof}
\bibinfo{author}{Jahnke, F.} \& \bibinfo{author}{Koch, S.~W.}
\newblock \bibinfo{title}{{Theory of carrier heating through injection pumping
  and lasing in semiconductor microcavity lasers}}.
\newblock \emph{\bibinfo{journal}{Opt. Lett.}} \textbf{\bibinfo{volume}{18}},
  \bibinfo{pages}{1438} (\bibinfo{year}{1993}).

\bibitem{Pompe1995_TransientResponseOfVCSELs}
\bibinfo{author}{Pompe, G.}, \bibinfo{author}{Rappen, T.} \&
  \bibinfo{author}{Wegener, M.}
\newblock \bibinfo{title}{{Transient response of an optically pumped
  short-cavity semiconductor laser}}.
\newblock \emph{\bibinfo{journal}{Phys. Rev. B}} \textbf{\bibinfo{volume}{51}},
  \bibinfo{pages}{7005--7009} (\bibinfo{year}{1995}).

\bibitem{Jahnke1996_UltrafastDynamicsOfQWVCSELs}
\bibinfo{author}{Jahnke, F.}, \bibinfo{author}{Schneider, H.~C.} \&
  \bibinfo{author}{Koch, S.~W.}
\newblock \bibinfo{title}{{Combined influence of design and carrier scattering
  on the ultrafast emission dynamics of quantum well microcavity lasers}}.
\newblock \emph{\bibinfo{journal}{Appl. Phys. Lett.}}
  \textbf{\bibinfo{volume}{69}}, \bibinfo{pages}{1185--1187}
  (\bibinfo{year}{1996}).

\bibitem{Michler1995_TransientPulseResponse}
\bibinfo{author}{Michler, P.}, \bibinfo{author}{Lohner, A.},
  \bibinfo{author}{R{\"{u}}hle, W.~W.} \& \bibinfo{author}{Reiner, G.}
\newblock \bibinfo{title}{{Transient pulse response of
  $\mathrm{In}_{0.2}\mathrm{Ga}_{0.8}\mathrm{As}/\mathrm{GaAs}$ microcavity
  lasers}}.
\newblock \emph{\bibinfo{journal}{Appl. Phys. Lett.}}
  \textbf{\bibinfo{volume}{66}}, \bibinfo{pages}{1599--1601}
  (\bibinfo{year}{1995}).

\bibitem{Michler1996_EmissionDynamicsOfVCSELs}
\bibinfo{author}{Michler, P.} \emph{et~al.}
\newblock \bibinfo{title}{{Emission dynamics of
  $\mathrm{In}_{0.2}\mathrm{Ga}_{0.8}\mathrm{As}/\mathrm{GaAs}$ $\lambda$ and
  2$\lambda$ microcavity lasers}}.
\newblock \emph{\bibinfo{journal}{Appl. Phys. Lett.}}
  \textbf{\bibinfo{volume}{68}}, \bibinfo{pages}{156--158}
  (\bibinfo{year}{1996}).

\bibitem{Coldren2012}
\bibinfo{author}{Coldren, L.~A.}, \bibinfo{author}{Corzine, S.~W.} \&
  \bibinfo{author}{Ma{\v{s}}anovi{\'{c}}, M.~L.}
\newblock \emph{\bibinfo{title}{{Diode Lasers and Photonic Integrated
  Circuits}}} (\bibinfo{publisher}{Wiley}, \bibinfo{address}{Hoboken},
  \bibinfo{year}{2012}).

\bibitem{maslov_modal_2004}
\bibinfo{author}{Maslov, A.} \& \bibinfo{author}{Ning, C.}
\newblock \bibinfo{title}{Modal gain in a semiconductor nanowire laser with
  anisotropic bandstructure}.
\newblock \emph{\bibinfo{journal}{IEEE J. Quantum Electron.}}
  \textbf{\bibinfo{volume}{40}}, \bibinfo{pages}{1389--1397}
  (\bibinfo{year}{2004}).

\bibitem{Ning2010a_ConfinementFactor}
\bibinfo{author}{Ning, C.~Z.}
\newblock \bibinfo{title}{{Semiconductor nanolasers}}.
\newblock \emph{\bibinfo{journal}{Phys. Status Solidi B}}
  \textbf{\bibinfo{volume}{247}}, \bibinfo{pages}{774--788}
  (\bibinfo{year}{2010}).

\bibitem{Li2000_UltrafastThzModulationLaser}
\bibinfo{author}{Li, J.} \& \bibinfo{author}{Ning, C.~Z.}
\newblock \bibinfo{title}{{Plasma heating and ultrafast semiconductor laser
  modulation through a terahertz heating field}}.
\newblock \emph{\bibinfo{journal}{J. Appl. Phys.}}
  \textbf{\bibinfo{volume}{88}}, \bibinfo{pages}{4933--4940}
  (\bibinfo{year}{2000}).

\bibitem{Yamaoka2021DirectModulation108GHz}
\bibinfo{author}{Yamaoka, S.} \emph{et~al.}
\newblock \bibinfo{title}{{Directly modulated membrane lasers with 108 GHz
  bandwidth on a high-thermal-conductivity silicon carbide substrate}}.
\newblock \emph{\bibinfo{journal}{Nat. Photonics}}
  \textbf{\bibinfo{volume}{15}}, \bibinfo{pages}{28--35}
  (\bibinfo{year}{2021}).

\bibitem{ning_self-consistent_1995}
\bibinfo{author}{Ning, C.~Z.}, \bibinfo{author}{Indik, R.~A.} \&
  \bibinfo{author}{Moloney, J.~V.}
\newblock \bibinfo{title}{Self-consistent approach to thermal effects in
  vertical-cavity surface-emitting lasers}.
\newblock \emph{\bibinfo{journal}{J. Opt. Soc. Am. B}}
  \textbf{\bibinfo{volume}{12}}, \bibinfo{pages}{1993--2004}
  (\bibinfo{year}{1995}).

\bibitem{Jahnke1993c_CWSwitchOnVCSEL_Transient}
\bibinfo{author}{Jahnke, F.}, \bibinfo{author}{Henneberger, K.},
  \bibinfo{author}{Sch{\"{a}}fer, W.} \& \bibinfo{author}{Koch, S.~W.}
\newblock \bibinfo{title}{{Transient nonequilibrium and many-body effects in
  semiconductor microcavity lasers}}.
\newblock \emph{\bibinfo{journal}{J. Opt. Soc. Am. B}}
  \textbf{\bibinfo{volume}{10}}, \bibinfo{pages}{2394} (\bibinfo{year}{1993}).

\bibitem{Grabmaier1991_DiffGainIncreasesWithIncreasingEnergy}
\bibinfo{author}{Grabmaier, A.} \emph{et~al.}
\newblock \bibinfo{title}{{Linewidth enhancement factor and carrier‐induced
  differential index in InGaAs separate confinement multi‐quantum‐well
  lasers}}.
\newblock \emph{\bibinfo{journal}{J. Appl. Phys.}}
  \textbf{\bibinfo{volume}{70}}, \bibinfo{pages}{2467--2469}
  (\bibinfo{year}{1991}).

\bibitem{Altug2006_UltrafastPhCLaser}
\bibinfo{author}{Altug, H.}, \bibinfo{author}{Englund, D.} \&
  \bibinfo{author}{Vu{\v{c}}kovi{\'{c}}, J.}
\newblock \bibinfo{title}{{Ultrafast photonic crystal nanocavity laser}}.
\newblock \emph{\bibinfo{journal}{Nat. Phys.}} \textbf{\bibinfo{volume}{2}},
  \bibinfo{pages}{484--488} (\bibinfo{year}{2006}).

\bibitem{romeira_purcell_2018}
\bibinfo{author}{Romeira, B.} \& \bibinfo{author}{Fiore, A.}
\newblock \bibinfo{title}{Purcell {Effect} in the {Stimulated} and
  {Spontaneous} {Emission} {Rates} of {Nanoscale} {Semiconductor} {Lasers}}.
\newblock \emph{\bibinfo{journal}{IEEE J. Quantum Electron.}}
  \textbf{\bibinfo{volume}{54}}, \bibinfo{pages}{1--12} (\bibinfo{year}{2018}).

\bibitem{khurgin_comparative_2014}
\bibinfo{author}{Khurgin, J.~B.} \& \bibinfo{author}{Sun, G.}
\newblock \bibinfo{title}{Comparative analysis of spasers, vertical-cavity
  surface-emitting lasers and surface-plasmon-emitting diodes}.
\newblock \emph{\bibinfo{journal}{Nat. Photonics}}
  \textbf{\bibinfo{volume}{8}}, \bibinfo{pages}{468--473}
  (\bibinfo{year}{2014}).

\bibitem{pellegrino_mode-field_2020}
\bibinfo{author}{Pellegrino, D.} \emph{et~al.}
\newblock \bibinfo{title}{Mode-field switching of nanolasers}.
\newblock \emph{\bibinfo{journal}{APL Photonics}} \textbf{\bibinfo{volume}{5}},
  \bibinfo{pages}{066109} (\bibinfo{year}{2020}).

\bibitem{suhr_modulation_2010}
\bibinfo{author}{Suhr, T.}, \bibinfo{author}{Gregersen, N.},
  \bibinfo{author}{Yvind, K.} \& \bibinfo{author}{Mørk, J.}
\newblock \bibinfo{title}{Modulation response of {nanoLEDs} and nanolasers
  exploiting {Purcell} enhanced spontaneous emission}.
\newblock \emph{\bibinfo{journal}{Opt. Express}} \textbf{\bibinfo{volume}{18}},
  \bibinfo{pages}{11230--11241} (\bibinfo{year}{2010}).

\bibitem{gregersen_quantum-dot_2012}
\bibinfo{author}{Gregersen, N.}, \bibinfo{author}{Suhr, T.},
  \bibinfo{author}{Lorke, M.} \& \bibinfo{author}{Mørk, J.}
\newblock \bibinfo{title}{Quantum-dot nano-cavity lasers with
  {Purcell}-enhanced stimulated emission}.
\newblock \emph{\bibinfo{journal}{Appl. Phys. Lett.}}
  \textbf{\bibinfo{volume}{100}}, \bibinfo{pages}{131107}
  (\bibinfo{year}{2012}).

\bibitem{Henneberger1992_SpectralHoleBurning}
\bibinfo{author}{Henneberger, K.} \emph{et~al.}
\newblock \bibinfo{title}{{Spectral hole burning and gain saturation in
  short-cavity semiconductor lasers}}.
\newblock \emph{\bibinfo{journal}{Phys. Rev. A}} \textbf{\bibinfo{volume}{45}},
  \bibinfo{pages}{1853--1859} (\bibinfo{year}{1992}).

\bibitem{Henneberger1992_ManyBodyEffects}
\bibinfo{author}{Henneberger, K.}, \bibinfo{author}{Jahnke, F.} \&
  \bibinfo{author}{Herzel, F.}
\newblock \bibinfo{title}{{Many-Body Effects and Multi-Mode Behaviour in
  Semiconductor Lasers}}.
\newblock \emph{\bibinfo{journal}{Phys. Status Solidi B}}
  \textbf{\bibinfo{volume}{173}}, \bibinfo{pages}{423--439}
  (\bibinfo{year}{1992}).

\bibitem{Jahnke1993b_CWSwitchOnVCSEL_DynamicResponse}
\bibinfo{author}{Jahnke, F.}, \bibinfo{author}{Koch, S.~W.} \&
  \bibinfo{author}{Henneberger, K.}
\newblock \bibinfo{title}{{Dynamic response of short‐cavity semiconductor
  lasers}}.
\newblock \emph{\bibinfo{journal}{Appl. Phys. Lett.}}
  \textbf{\bibinfo{volume}{62}}, \bibinfo{pages}{2313--2315}
  (\bibinfo{year}{1993}).

\end{thebibliography}


\begin{thebibliography}{10}
\expandafter\ifx\csname url\endcsname\relax
  \def\url#1{\texttt{#1}}\fi
\expandafter\ifx\csname urlprefix\endcsname\relax\def\urlprefix{URL }\fi
\providecommand{\bibinfo}[2]{#2}
\providecommand{\eprint}[2][]{\url{#2}}

\bibitem{Sidiropoulos2014_PumpProbePlasmonicLaser}
\bibinfo{author}{Sidiropoulos, T. P.~H.} \emph{et~al.}
\newblock \bibinfo{title}{{Ultrafast plasmonic nanowire lasers near the surface
  plasmon frequency}}.
\newblock \emph{\bibinfo{journal}{Nat. Phys.}} \textbf{\bibinfo{volume}{10}},
  \bibinfo{pages}{870--876} (\bibinfo{year}{2014}).

\bibitem{Coldren2012}
\bibinfo{author}{Coldren, L.~A.}, \bibinfo{author}{Corzine, S.~W.} \&
  \bibinfo{author}{Ma{\v{s}}anovi{\'{c}}, M.~L.}
\newblock \emph{\bibinfo{title}{{Diode Lasers and Photonic Integrated
  Circuits}}} (\bibinfo{publisher}{Wiley}, \bibinfo{address}{Hoboken},
  \bibinfo{year}{2012}).

\bibitem{Landsberg1966}
\bibinfo{author}{Landsberg, P.~T.}
\newblock \bibinfo{title}{{Electron Interaction Effects on Recombination
  Spectra}}.
\newblock \emph{\bibinfo{journal}{Phys. Status Solidi B}}
  \textbf{\bibinfo{volume}{15}}, \bibinfo{pages}{623--626}
  (\bibinfo{year}{1966}).

\bibitem{Martin1977}
\bibinfo{author}{Martin, R.} \& \bibinfo{author}{Stormer, H.}
\newblock \bibinfo{title}{{On the low energy tail of the electron-hole drop
  recombination spectrum}}.
\newblock \emph{\bibinfo{journal}{Solid State Commun.}}
  \textbf{\bibinfo{volume}{22}}, \bibinfo{pages}{523--526}
  (\bibinfo{year}{1977}).

\bibitem{Kalt1992}
\bibinfo{author}{Kalt, H.} \& \bibinfo{author}{Rinker, M.}
\newblock \bibinfo{title}{{Band-gap renormalization in semiconductors with
  multiple inequivalent valleys}}.
\newblock \emph{\bibinfo{journal}{Phys. Rev. B}} \textbf{\bibinfo{volume}{45}},
  \bibinfo{pages}{1139--1154} (\bibinfo{year}{1992}).

\bibitem{roder_continuous_2013}
\bibinfo{author}{Röder, R.} \emph{et~al.}
\newblock \bibinfo{title}{Continuous {Wave} {Nanowire} {Lasing}}.
\newblock \emph{\bibinfo{journal}{Nano Lett.}} \textbf{\bibinfo{volume}{13}},
  \bibinfo{pages}{3602--3606} (\bibinfo{year}{2013}).

\bibitem{Saxena2013_RTLaserGaAs}
\bibinfo{author}{Saxena, D.} \emph{et~al.}
\newblock \bibinfo{title}{{Optically pumped room-temperature GaAs nanowire
  lasers}}.
\newblock \emph{\bibinfo{journal}{Nat. Photonics}}
  \textbf{\bibinfo{volume}{7}}, \bibinfo{pages}{963--968}
  (\bibinfo{year}{2013}).

\bibitem{Saxena2015ModeProfiling}
\bibinfo{author}{Saxena, D.} \emph{et~al.}
\newblock \bibinfo{title}{{Mode Profiling of Semiconductor Nanowire Lasers}}.
\newblock \emph{\bibinfo{journal}{Nano Lett.}} \textbf{\bibinfo{volume}{15}},
  \bibinfo{pages}{5342--5348} (\bibinfo{year}{2015}).

\bibitem{Mayer2016_GaAsLaserOnSilicon}
\bibinfo{author}{Mayer, B.} \emph{et~al.}
\newblock \bibinfo{title}{{Monolithically Integrated High-$\beta$ Nanowire
  Lasers on Silicon}}.
\newblock \emph{\bibinfo{journal}{Nano Lett.}} \textbf{\bibinfo{volume}{16}},
  \bibinfo{pages}{152--156} (\bibinfo{year}{2016}).

\bibitem{Mayer2013_RTLaserGaAs}
\bibinfo{author}{Mayer, B.} \emph{et~al.}
\newblock \bibinfo{title}{{Lasing from individual GaAs-AlGaAs core-shell
  nanowires up to room temperature}}.
\newblock \emph{\bibinfo{journal}{Nat. Commun.}} \textbf{\bibinfo{volume}{4}}
  (\bibinfo{year}{2013}).

\bibitem{Passler1997}
\bibinfo{author}{P{\"{a}}ssler, R.} \& \bibinfo{author}{Oelgart, G.}
\newblock \bibinfo{title}{{Appropriate analytical description of the
  temperature dependence of exciton peak positions in
  $\mathrm{GaAs}/\mathrm{Al}_{x}\mathrm{Ga}_{1-x}\mathrm{As}$ multiple quantum
  wells and the $\Gamma_{8v}-\Gamma_{6c}$ gap of GaAs}}.
\newblock \emph{\bibinfo{journal}{J. Appl. Phys.}}
  \textbf{\bibinfo{volume}{82}}, \bibinfo{pages}{2611--2616}
  (\bibinfo{year}{1997}).

\bibitem{Reinhart2005a}
\bibinfo{author}{Reinhart, F.~K.}
\newblock \bibinfo{title}{{A heuristic approach to precisely represent optical
  absorption and refractive index data for photon energies below, at, and above
  the band gap of semiconductors: The case of high-purity GaAs. Part I}}.
\newblock \emph{\bibinfo{journal}{J. Appl. Phys.}}
  \textbf{\bibinfo{volume}{97}}, \bibinfo{pages}{123534}
  (\bibinfo{year}{2005}).

\bibitem{Reinhart2005b}
\bibinfo{author}{Reinhart, F.~K.}
\newblock \bibinfo{title}{{A heuristic approach to determine the modifications
  of electronic and optical properties of “intrinsic” GaAs under
  free-carrier injection. Part II}}.
\newblock \emph{\bibinfo{journal}{J. Appl. Phys.}}
  \textbf{\bibinfo{volume}{97}}, \bibinfo{pages}{123535}
  (\bibinfo{year}{2005}).

\bibitem{Chow1994_SemiconductorLaserPhysicsBook}
\bibinfo{author}{Chow, W.~W.}, \bibinfo{author}{Koch, S.~W.} \&
  \bibinfo{author}{Sargent, M.}
\newblock \emph{\bibinfo{title}{{Semiconductor-Laser Physics}}}
  (\bibinfo{publisher}{Springer}, \bibinfo{address}{Berlin},
  \bibinfo{year}{1994}).

\bibitem{Vurgaftman2001_PhysicalConstants}
\bibinfo{author}{Vurgaftman, I.}, \bibinfo{author}{Meyer, J.~R.} \&
  \bibinfo{author}{Ram-Mohan, L.~R.}
\newblock \bibinfo{title}{{Band parameters for III–V compound semiconductors
  and their alloys}}.
\newblock \emph{\bibinfo{journal}{J. Appl. Phys.}}
  \textbf{\bibinfo{volume}{89}}, \bibinfo{pages}{5815--5875}
  (\bibinfo{year}{2001}).

\bibitem{Samara1983DielectricConstants}
\bibinfo{author}{Samara, G.~A.}
\newblock \bibinfo{title}{{Temperature and pressure dependences of the
  dielectric constants of semiconductors}}.
\newblock \emph{\bibinfo{journal}{Phys. Rev. B}} \textbf{\bibinfo{volume}{27}},
  \bibinfo{pages}{3494--3505} (\bibinfo{year}{1983}).

\bibitem{Strauch1990LOPhononEnergy}
\bibinfo{author}{Strauch, D.} \& \bibinfo{author}{Dorner, B.}
\newblock \bibinfo{title}{{Phonon dispersion in GaAs}}.
\newblock \emph{\bibinfo{journal}{J. Phys. Condens. Matter}}
  \textbf{\bibinfo{volume}{2}}, \bibinfo{pages}{1457--1474}
  (\bibinfo{year}{1990}).

\bibitem{Jahnke1995a_UltrafastSwitching}
\bibinfo{author}{Jahnke, F.} \& \bibinfo{author}{Koch, S.~W.}
\newblock \bibinfo{title}{{Ultrafast intensity switching and nonthermal carrier
  effects in semiconductor microcavity lasers}}.
\newblock \emph{\bibinfo{journal}{Appl. Phys. Lett.}}
  \textbf{\bibinfo{volume}{67}}, \bibinfo{pages}{2278--2280}
  (\bibinfo{year}{1995}).

\bibitem{Oudar1985_ThermalizationAt15KGaAs}
\bibinfo{author}{Oudar, J.~L.}, \bibinfo{author}{Hulin, D.},
  \bibinfo{author}{Migus, A.}, \bibinfo{author}{Antonetti, A.} \&
  \bibinfo{author}{Alexandre, F.}
\newblock \bibinfo{title}{{Subpicosecond Spectral Hole Burning Due to
  Nonthermalized Photoexcited Carriers in GaAs}}.
\newblock \emph{\bibinfo{journal}{Phys. Rev. Lett.}}
  \textbf{\bibinfo{volume}{55}}, \bibinfo{pages}{2074--2077}
  (\bibinfo{year}{1985}).

\bibitem{Elsaesser1991_ThermalizationAt300KGaAs}
\bibinfo{author}{Elsaesser, T.}, \bibinfo{author}{Shah, J.},
  \bibinfo{author}{Rota, L.} \& \bibinfo{author}{Lugli, P.}
\newblock \bibinfo{title}{{Initial Thermalization of Photoexcited Carriers in
  GaAs Studied by Femtosecond Luminescence Spectroscopy}}.
\newblock \emph{\bibinfo{journal}{Phys. Rev. Lett.}}
  \textbf{\bibinfo{volume}{66}}, \bibinfo{pages}{1757--1760}
  (\bibinfo{year}{1991}).

\bibitem{Kane1994a_ThermalizationWithin150fs}
\bibinfo{author}{Kane, M.~G.}, \bibinfo{author}{Sun, K.~W.} \&
  \bibinfo{author}{Lyon, S.~A.}
\newblock \bibinfo{title}{{Ultrafast carrier-carrier scattering among
  photoexcited nonequilibrium carriers in GaAs}}.
\newblock \emph{\bibinfo{journal}{Phys. Rev. B}} \textbf{\bibinfo{volume}{50}},
  \bibinfo{pages}{7428--7438} (\bibinfo{year}{1994}).

\bibitem{Shah1999_UltrafastSpectroscopyBook}
\bibinfo{author}{Shah, J.}
\newblock \emph{\bibinfo{title}{{Ultrafast Spectroscopy of Semiconductors and
  Semiconductor Nanostructures}}}, vol. \bibinfo{volume}{115} of
  \emph{\bibinfo{series}{Springer Series in Solid-State Sciences}}
  (\bibinfo{publisher}{Springer}, \bibinfo{address}{Berlin},
  \bibinfo{year}{1999}).

\bibitem{Loudon2000_TheQuantumTheoryOfLight}
\bibinfo{author}{Loudon, R.}
\newblock \emph{\bibinfo{title}{{The Quantum Theory of Light}}}
  (\bibinfo{publisher}{Oxford University Press}, \bibinfo{year}{2000}).

\bibitem{Grabmaier1991_DiffGainIncreasesWithIncreasingEnergy}
\bibinfo{author}{Grabmaier, A.} \emph{et~al.}
\newblock \bibinfo{title}{{Linewidth enhancement factor and carrier‐induced
  differential index in InGaAs separate confinement multi‐quantum‐well
  lasers}}.
\newblock \emph{\bibinfo{journal}{J. Appl. Phys.}}
  \textbf{\bibinfo{volume}{70}}, \bibinfo{pages}{2467--2469}
  (\bibinfo{year}{1991}).

\bibitem{Michler1995_TransientPulseResponse}
\bibinfo{author}{Michler, P.}, \bibinfo{author}{Lohner, A.},
  \bibinfo{author}{R{\"{u}}hle, W.~W.} \& \bibinfo{author}{Reiner, G.}
\newblock \bibinfo{title}{{Transient pulse response of
  $\mathrm{In}_{0.2}\mathrm{Ga}_{0.8}\mathrm{As}/\mathrm{GaAs}$ microcavity
  lasers}}.
\newblock \emph{\bibinfo{journal}{Appl. Phys. Lett.}}
  \textbf{\bibinfo{volume}{66}}, \bibinfo{pages}{1599--1601}
  (\bibinfo{year}{1995}).

\bibitem{Michler1996_EmissionDynamicsOfVCSELs}
\bibinfo{author}{Michler, P.} \emph{et~al.}
\newblock \bibinfo{title}{{Emission dynamics of
  $\mathrm{In}_{0.2}\mathrm{Ga}_{0.8}\mathrm{As}/\mathrm{GaAs}$ $\lambda$ and
  2$\lambda$ microcavity lasers}}.
\newblock \emph{\bibinfo{journal}{Appl. Phys. Lett.}}
  \textbf{\bibinfo{volume}{68}}, \bibinfo{pages}{156--158}
  (\bibinfo{year}{1996}).

\bibitem{Jahnke1993c_CWSwitchOnVCSEL_Transient}
\bibinfo{author}{Jahnke, F.}, \bibinfo{author}{Henneberger, K.},
  \bibinfo{author}{Sch{\"{a}}fer, W.} \& \bibinfo{author}{Koch, S.~W.}
\newblock \bibinfo{title}{{Transient nonequilibrium and many-body effects in
  semiconductor microcavity lasers}}.
\newblock \emph{\bibinfo{journal}{J. Opt. Soc. Am. B}}
  \textbf{\bibinfo{volume}{10}}, \bibinfo{pages}{2394} (\bibinfo{year}{1993}).

\bibitem{korenman_nonequilibrium_1966}
\bibinfo{author}{Korenman, V.}
\newblock \bibinfo{title}{Nonequilibrium quantum statistics; application to the
  laser}.
\newblock \emph{\bibinfo{journal}{Ann. Phys.}} \textbf{\bibinfo{volume}{39}},
  \bibinfo{pages}{72--126} (\bibinfo{year}{1966}).

\bibitem{schaefer_approach_1986}
\bibinfo{author}{Schäfer, W.} \& \bibinfo{author}{Treusch, J.}
\newblock \bibinfo{title}{An approach to the nonequilibrium theory of highly
  excited semiconductors}.
\newblock \emph{\bibinfo{journal}{Z. Phys., B Condens. matter}}
  \textbf{\bibinfo{volume}{63}}, \bibinfo{pages}{407--426}
  (\bibinfo{year}{1986}).

\bibitem{henneberger_nonlinear_1988}
\bibinfo{author}{Henneberger, K.} \& \bibinfo{author}{Haug, H.}
\newblock \bibinfo{title}{Nonlinear optics and transport in laser-excited
  semiconductors}.
\newblock \emph{\bibinfo{journal}{Phys. Rev. B}} \textbf{\bibinfo{volume}{38}},
  \bibinfo{pages}{9759--9770} (\bibinfo{year}{1988}).

\bibitem{Henneberger1992_SpectralHoleBurning}
\bibinfo{author}{Henneberger, K.} \emph{et~al.}
\newblock \bibinfo{title}{{Spectral hole burning and gain saturation in
  short-cavity semiconductor lasers}}.
\newblock \emph{\bibinfo{journal}{Phys. Rev. A}} \textbf{\bibinfo{volume}{45}},
  \bibinfo{pages}{1853--1859} (\bibinfo{year}{1992}).

\bibitem{Henneberger1992_ManyBodyEffects}
\bibinfo{author}{Henneberger, K.}, \bibinfo{author}{Jahnke, F.} \&
  \bibinfo{author}{Herzel, F.}
\newblock \bibinfo{title}{{Many-Body Effects and Multi-Mode Behaviour in
  Semiconductor Lasers}}.
\newblock \emph{\bibinfo{journal}{Phys. Status Solidi B}}
  \textbf{\bibinfo{volume}{173}}, \bibinfo{pages}{423--439}
  (\bibinfo{year}{1992}).

\bibitem{herzel_semiconductor_1993}
\bibinfo{author}{Herzel, F.}, \bibinfo{author}{Henneberger, K.} \&
  \bibinfo{author}{Vogel, W.}
\newblock \bibinfo{title}{The semiconductor laser linewidth: a green's function
  approach}.
\newblock \emph{\bibinfo{journal}{{IEEE} J. Quantum Electron.}}
  \textbf{\bibinfo{volume}{29}}, \bibinfo{pages}{2891--2897}
  (\bibinfo{year}{1993}).

\bibitem{Jahnke1993a_CWSwitchOnVCSEL_Theoryof}
\bibinfo{author}{Jahnke, F.} \& \bibinfo{author}{Koch, S.~W.}
\newblock \bibinfo{title}{{Theory of carrier heating through injection pumping
  and lasing in semiconductor microcavity lasers}}.
\newblock \emph{\bibinfo{journal}{Opt. Lett.}} \textbf{\bibinfo{volume}{18}},
  \bibinfo{pages}{1438} (\bibinfo{year}{1993}).

\bibitem{Jahnke1993b_CWSwitchOnVCSEL_DynamicResponse}
\bibinfo{author}{Jahnke, F.}, \bibinfo{author}{Koch, S.~W.} \&
  \bibinfo{author}{Henneberger, K.}
\newblock \bibinfo{title}{{Dynamic response of short‐cavity semiconductor
  lasers}}.
\newblock \emph{\bibinfo{journal}{Appl. Phys. Lett.}}
  \textbf{\bibinfo{volume}{62}}, \bibinfo{pages}{2313--2315}
  (\bibinfo{year}{1993}).

\bibitem{mohideen_semiconductor_1994}
\bibinfo{author}{Mohideen, U.}, \bibinfo{author}{Slusher, R.~E.},
  \bibinfo{author}{Jahnke, F.} \& \bibinfo{author}{Koch, S.~W.}
\newblock \bibinfo{title}{Semiconductor microlaser linewidths}.
\newblock \emph{\bibinfo{journal}{Phys. Rev. Lett.}}
  \textbf{\bibinfo{volume}{73}}, \bibinfo{pages}{1785--1788}
  (\bibinfo{year}{1994}).

\bibitem{Jahnke1995_QuantumStatisticalTheoryPaper}
\bibinfo{author}{Jahnke, F.} \& \bibinfo{author}{Koch, S.~W.}
\newblock \bibinfo{title}{{Many-body theory for semiconductor microcavity
  lasers}}.
\newblock \emph{\bibinfo{journal}{Phys. Rev. A}} \textbf{\bibinfo{volume}{52}},
  \bibinfo{pages}{1712--1727} (\bibinfo{year}{1995}).

\bibitem{Schneider1997_UltrafastDynamicsOfVCSELs}
\bibinfo{author}{Schneider, H.~C.}, \bibinfo{author}{Jahnke, F.} \&
  \bibinfo{author}{Koch, S.~W.}
\newblock \bibinfo{title}{{Microscopic theory of non-equilibrium microcavity
  laser dynamics}}.
\newblock \emph{\bibinfo{journal}{Quantum Semiclass. Opt.}}
  \textbf{\bibinfo{volume}{9}}, \bibinfo{pages}{693--711}
  (\bibinfo{year}{1997}).

\bibitem{visser_confinement_1997}
\bibinfo{author}{Visser, T.}, \bibinfo{author}{Blok, H.},
  \bibinfo{author}{Demeulenaere, B.} \& \bibinfo{author}{Lenstra, D.}
\newblock \bibinfo{title}{Confinement factors and gain in optical amplifiers}.
\newblock \emph{\bibinfo{journal}{IEEE J. Quantum Electron.}}
  \textbf{\bibinfo{volume}{33}}, \bibinfo{pages}{1763--1766}
  (\bibinfo{year}{1997}).

\bibitem{maslov_modal_2004}
\bibinfo{author}{Maslov, A.} \& \bibinfo{author}{Ning, C.}
\newblock \bibinfo{title}{Modal gain in a semiconductor nanowire laser with
  anisotropic bandstructure}.
\newblock \emph{\bibinfo{journal}{IEEE J. Quantum Electron.}}
  \textbf{\bibinfo{volume}{40}}, \bibinfo{pages}{1389--1397}
  (\bibinfo{year}{2004}).

\bibitem{Ning2010a_ConfinementFactor}
\bibinfo{author}{Ning, C.~Z.}
\newblock \bibinfo{title}{{Semiconductor nanolasers}}.
\newblock \emph{\bibinfo{journal}{Phys. Status Solidi B}}
  \textbf{\bibinfo{volume}{247}}, \bibinfo{pages}{774--788}
  (\bibinfo{year}{2010}).

\bibitem{haug_quantum_2004}
\bibinfo{author}{Haug, H.} \& \bibinfo{author}{Koch, S.~W.}
\newblock \emph{\bibinfo{title}{Quantum Theory of the Optical and Electronic
  Properties of Semiconductors}} (\bibinfo{publisher}{World Scientific},
  \bibinfo{address}{Singapore}, \bibinfo{year}{2009}), \bibinfo{edition}{5th}
  edn.

\bibitem{collet_model_1986}
\bibinfo{author}{Collet, J.} \& \bibinfo{author}{Amand, T.}
\newblock \bibinfo{title}{Model calculation of the laser-semiconductor
  interaction in subpicosecond regime}.
\newblock \emph{\bibinfo{journal}{J. Phys. Chem. Solids}}
  \textbf{\bibinfo{volume}{47}}, \bibinfo{pages}{153--163}
  (\bibinfo{year}{1986}).

\bibitem{binder_carrier-carrier_1992}
\bibinfo{author}{Binder, R.} \emph{et~al.}
\newblock \bibinfo{title}{Carrier-carrier scattering and optical dephasing in
  highly excited semiconductors}.
\newblock \emph{\bibinfo{journal}{Phys. Rev. B}} \textbf{\bibinfo{volume}{45}},
  \bibinfo{pages}{1107--1115} (\bibinfo{year}{1992}).

\bibitem{collet_numerical_1983}
\bibinfo{author}{Collet, J.}, \bibinfo{author}{Amand, T.} \&
  \bibinfo{author}{Pugnet, M.}
\newblock \bibinfo{title}{Numerical approach to non-equilibrium carrier
  relaxation in picosecond and subpicosecond physics}.
\newblock \emph{\bibinfo{journal}{Phys. Lett. A}}
  \textbf{\bibinfo{volume}{96}}, \bibinfo{pages}{368--374}
  (\bibinfo{year}{1983}).

\bibitem{collins_generation_1984}
\bibinfo{author}{Collins, C.~L.} \& \bibinfo{author}{Yu, P.~Y.}
\newblock \bibinfo{title}{Generation of nonequilibrium optical phonons in
  {GaAs} and their application in studying intervalley electron-phonon
  scattering}.
\newblock \emph{\bibinfo{journal}{Phys. Rev. B}} \textbf{\bibinfo{volume}{30}},
  \bibinfo{pages}{4501--4515} (\bibinfo{year}{1984}).

\end{thebibliography}

\clearpage

\end{document}


\author{Andreas Thurn}\email{andreas.thurn@wsi.tum.de}
\author{Jochen Bissinger}%
\affiliation{Walter Schottky Institut, Technische Universität München, Am Coulombwall 4, 85748 Garching, Germany.}
\author{Stefan Meinecke}
\affiliation{Institut für Theoretische Physik, Technische Universität Berlin, Hardenbergstraße 36, 10623 Berlin, Germany.}
\author{Paul Schmiedeke}
\affiliation{Walter Schottky Institut, Technische Universität München, Am Coulombwall 4, 85748 Garching, Germany.}
\author{Sang Soon Oh}
\affiliation{School of Physics and Astronomy, Cardiff University, Cardiff CF24 3AA, UK.}
\author{Weng W. Chow}
\affiliation{Sandia National Laboratories, Albuquerque, New Mexico 87185-1086, USA.}
\author{Kathy Lüdge}
\affiliation{Institut für Physik, Technische Universität Ilmenau, Weimarer Straße 25, 98693 Ilmenau, Germany.}
\author{Gregor Koblmüller}
\author{Jonathan J. Finley}\email{finley@wsi.tum.de}
\affiliation{Walter Schottky Institut, Technische Universität München, Am Coulombwall 4, 85748 Garching, Germany.}
\title{Supplementary Information for: Self-induced ultrafast electron-hole plasma temperature oscillations in nanowire lasers}
\maketitle

\makeatletter
\renewcommand{\thesection}{\Roman{section}}
\renewcommand{\theequation}{S\arabic{equation}}
\renewcommand{\thefigure}{S\arabic{figure}}
\renewcommand{\thetable}{S\arabic{table}}
\renewcommand{\bibnumfmt}[1]{S#1.}
\renewcommand{\citenumfont}[1]{S#1}
\onecolumngrid
\newpage

\section{Single-pulse excitation and laser threshold}
\label{Sec:SinglePulseThreshold}
\begin{figure*}[ht]
    \includegraphics[keepaspectratio]{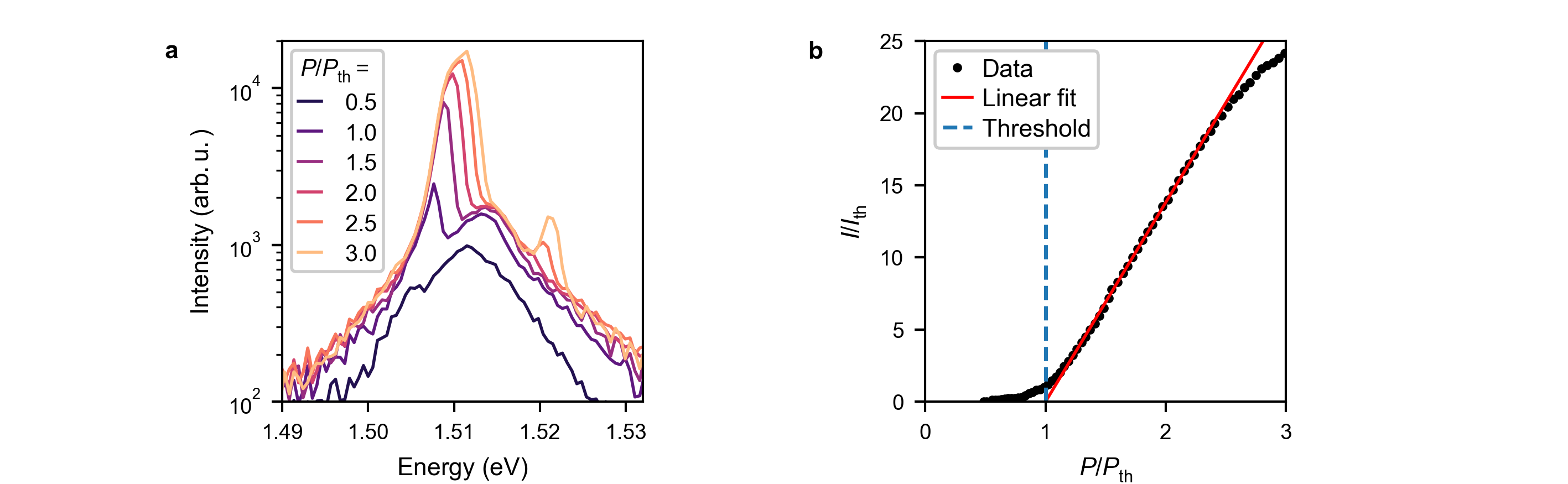}
    \caption{\textbf{Lasing spectra and light-in light-out curve for single-pulse excitation.} \textbf{a}, Spectra of the NW emission at selected excitation powers ($P$) relative to threshold ($P_{\mathrm{th}}$). The second mode at $\sim\SI{1.52}{eV}$ is more than one order of magnitude weaker than the main mode at $\sim\SI{1.51}{eV}$ and can therefore be neglected. \textbf{b}, Characteristic light-in light-out curve showing a clear transition into lasing. The intersection of the linear fit with the horizontal axis defines $P_{\mathrm{th}}$. The output intensity ($I$) is normalized to its threshold value ($I_{\mathrm{th}}$).}
    \label{Fig:S1_SinglePulsePowerseries}
\end{figure*}

The exemplary power series shown in Fig.~\ref{Fig:S1_SinglePulsePowerseries} was performed with single excitation pulses and otherwise the same experimental conditions as the measurement in Fig.~1 of the main manuscript. Figure~\ref{Fig:S1_SinglePulsePowerseries}a presents NW emission spectra at selected excitation powers ($P$) relative to threshold ($P_{\mathrm{th}}$). It shows that the NW laser remains single modal at an energy of $\sim\SI{1.51}{eV}$ for excitation powers $P/P_{\mathrm{th}}<2.5$. It further demonstrates that, for the range of excitation powers investigated, the mode at $\sim\SI{1.52}{eV}$ is always more than an order of magnitude weaker than the dominant mode. We can therefore neglect this weak second mode at higher energy and treat the laser as being single modal in all simulations. To obtain the characteristic light-in light-out curve we determined the output intensity ($I$) by spectrally integrating the dominant lasing peak at $\sim\SI{1.51}{eV}$. The resulting $I$ is normalized to its threshold value ($I_{\mathrm{th}}$) and presented in Figure~\ref{Fig:S1_SinglePulsePowerseries}b as a function of $P/P_{\mathrm{th}}$, demonstrating a clear transition into lasing. We fitted the output intensity above threshold with a linear function (solid red line) and determined a continuous wave (CW) equivalent threshold excitation power of $P_{\mathrm{th}}=\SI[separate-uncertainty=true]{0.84(4)}{mW}$ (dashed blue line) by the intersection of the fit with the horizontal axis. This corresponds to a threshold fluence of $F_{\mathrm{th}}=\SI[separate-uncertainty=true]{4.5(2)}{\micro J\ cm^{-2}}$. The error on $P_{\mathrm{th}}$ and $F_{\mathrm{th}}$ represents the $\SI{95}{\%}$ confidence interval (CI), obtained from the linear fit.

\clearpage
\section{Experimental determination of the turn-on time}
\begin{figure*}[ht]
    \includegraphics[keepaspectratio]{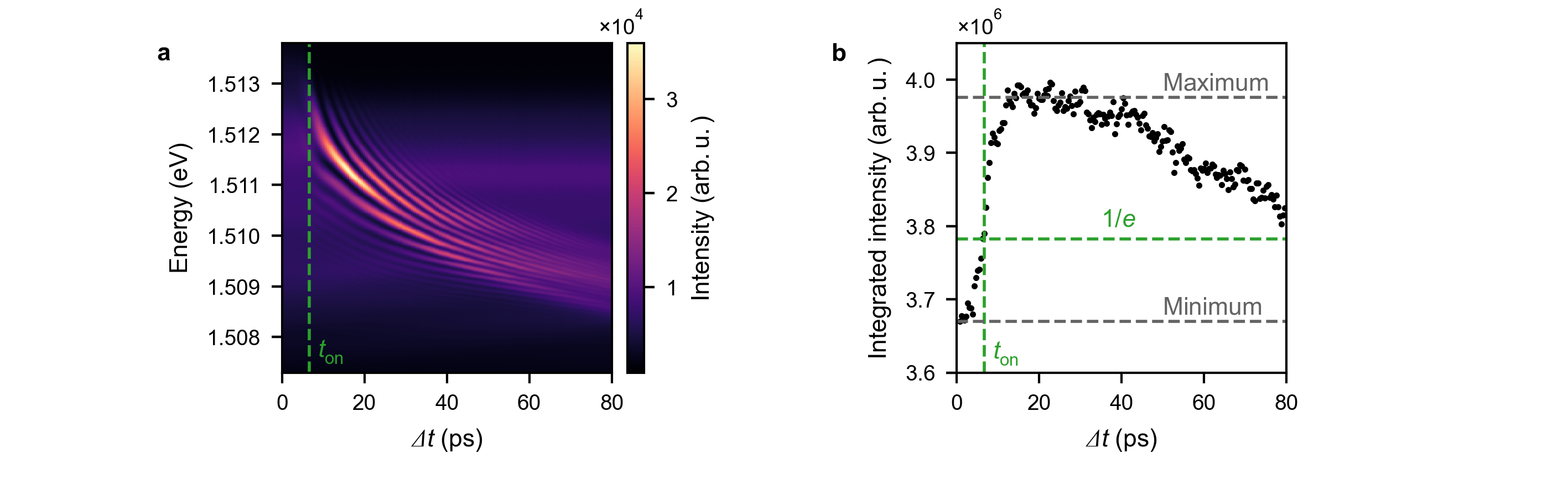}
    \caption{\textbf{Experimental determination of the turn-on time.} \textbf{a}, Time-integrated pump-probe spectra as a function of $\mathit{\Delta} t$. Integration over the energy axis yielded the integrated intensity shown in \textbf{b}. \textbf{b}, We defined the turn-on time ($t_{\mathrm{on}}$) as the time that it takes the integrated intensity to rise from the minimum to $1/e$ of the difference between the maximum and the minimum.}
    \label{Fig:S_DeterminationOfTurnOnTime_210111}
\end{figure*}

Here we discuss the method used to experimentally determine $t_{\mathrm{on}}$. Figure~\ref{Fig:S_DeterminationOfTurnOnTime_210111}a shows the time-integrated spectra of a selected pump-probe measurement belonging to Fig.~3a of the main manuscript. The measurement was performed with an excitation power of  $P_{\mathrm{pump}}/P_{\mathrm{th}}\sim\SI{2.9}{}$ and $P_{\mathrm{probe}}/P_{\mathrm{th}}\sim\SI{0.5}{}$, and otherwise the same experimental conditions as for Fig.~3a. We integrated all spectra along the energy axis and plotted the integrated intensity as a function of $\Delta t$ in Fig.~\ref{Fig:S_DeterminationOfTurnOnTime_210111}b. From these data a minimum and maximum value (dashed grey lines) was determined by averaging over a suitable number of data points around $\Delta t=\SI{0}{ps}$ and around the respective maximum. Using these values, we defined $t_{\mathrm{on}}$ as the time that it takes the integrated intensity to rise from the minimum to $1/e$ of the difference between the maximum and the minimum, as indicated by the horizontal dashed green line in Fig.~\ref{Fig:S_DeterminationOfTurnOnTime_210111}b. This method relies on the transient reduction of the probe pulse absorption due to a depletion of available states and its subsequent recovery~\cite{Sidiropoulos2014_PumpProbePlasmonicLaser}. We attribute this to a non-negligible overlap of the excitation laser (Energy $\sim\SI{1.588}{eV}$, $\mathrm{FWHM}\sim\SI{200}{fs}$) with the thermalized electron and hole distributions. In this particular example, this method yields $t_{\mathrm{on}}=\SI[separate-uncertainty=true]{6.3(4)}{ps}$, which is indicated by the vertical dashed green lines in Fig.~\ref{Fig:S_DeterminationOfTurnOnTime_210111}a,b. The error is an overestimate of the $95\%$ CI of the mean, which follows from the averaging used to obtain the minimum and maximum values, and the discreteness of the data (step size $\sim\SI{0.4}{ps}$). 

\clearpage
\section{Continuous wave (CW) lasing characteristics}
\label{Sec:CWcharacteristics}
For an accurate description of the ultrafast dynamical behaviour of NW lasers, finding reliable simulation parameters is crucial. We took advantage of the fact that important parameters such as the end-facet reflectivity ($R$) and the spontaneous emission factor ($\beta$) should be the same whether the NW is excited continuously or in a pulsed manner. We therefore performed power series with CW excitation on the same NW laser that was also investigated using pump-probe excitation and simulated the results with the quantum statistical model. Same as the experiments and simulations in Fig.~1, Fig.~2 and Fig.~3 of the main manuscript, the CW measurements were performed at $T_{\mathrm{L}}=\SI{10}{K}$ and with an electron excess energy of $\sim\SI{60}{meV}$. Under CW excitation the laser is in a steady state. This allows us to extract the excitation power dependence of properties such as the carrier density ($N$), carrier temperature ($T_{\mathrm{c}}$), renormalized band gap energy ($E_{\mathrm{g}}$) and the broadening of the energy states ($\gamma_{0}$) directly from time-integrated measurements of the spontaneous emission spectra. Together with the power dependence of the output intensity ($I$) and the peak energy of the lasing mode ($E_{\mathrm{p}}$), this provides an experimental data set that can be used to adjust and test the simulation parameters of the quantum statistical model. In the following, we first describe the approach used to extract all aforementioned parameters and then present the experimental and simulated results in detail.

\subsection{Theoretical model for spontaneous emission spectra}
For the purpose of extracting $N$, $T_{\mathrm{c}}$, $E_{\mathrm{g}}$ and $\gamma_{\mathrm{0}}$ from a measured spontaneous emission spectrum, we fitted it with the theoretical spontaneous emission spectrum given by~\cite{Coldren2012}
\begin{equation}
r_{\mathrm{sp}} (\omega) = \frac{n_{\mathrm{active}} e_{0}^{2} \omega}{\hbar \pi c_{0}^{3} \varepsilon_{0} m_{0}^{2}} |M_{\mathrm{T}}|^{2} 
\int_{E=0}^{\infty} \rho_{\mathrm{r}}(E)f_{\mathrm{c}}(E)[1-f_{\mathrm{v}}(E)]\mathcal{L}(E_{\mathrm{g}} + E -\hbar\omega) \mathrm{d}E \, .
\label{eq:spontaneous_emission_spectrum}
\end{equation}
We note that the model described by equation~\eqref{eq:spontaneous_emission_spectrum} assumes equal temperatures for electrons ($T_{\mathrm{e}}$) and holes ($T_{\mathrm{h}}$). Here, $\omega$ is the angular frequency, $n_{\mathrm{active}}$ is the refractive index of the NW material and $M_{\mathrm{T}}$ the transition matrix element. Furthermore, $e_{0}$ is the elementary charge, $\hbar$ the reduced Planck constant, $c_{0}$ the speed of light in vacuum, $\varepsilon_{0}$ the vacuum permittivity and $m_{0}$ the free electron mass. The reduced density of states ($\rho_{\mathrm{r}}$) and the Fermi functions in the valence ($f_{\mathrm{v}}$) and conduction band ($f_{\mathrm{c}}$) depend on the energy ($E$) and are determined by~\cite{Coldren2012}
\begin{flalign}
&&\rho_{\mathrm{r}}(E) &= \frac{1}{ 2 \pi^{2}} \left[ \frac{2 m_{\mathrm{r}}}{\hbar^{2}} \right]^{3/2} E^{1/2}\, ,&& \\
&&f_{\mathrm{c}}(E) &= \frac{1}{1 + \exp\{[E_{\mathrm{\mathrm{g}}} + (m_{\mathrm{r}}/m_{\mathrm{c}} E - \mu_{\mathrm{c}})]/\left[k_{\mathrm{B}}T_{\mathrm{c}}\right]\}}&&\\
\text{and} && \nonumber \\
&&f_{\mathrm{v}}(E) &= \frac{1}{1 + \exp\{[ - (m_{\mathrm{r}}/m_{\mathrm{v}} E - \mu_{\mathrm{v}})]/\left[k_{\mathrm{B}}T_{\mathrm{c}}\right]\}} \, ,&&
\end{flalign}
where $k_{\mathrm{B}}$ is the Boltzmann constant. The quasi-Fermi levels in the valence $(\mu_{\mathrm{v}})$ and conduction band $(\mu_{\mathrm{c}})$ are related to $N$ by the Fermi-Dirac integral. The reduced effective mass $m_{\mathrm{r}} = m_{\mathrm{c}}m_{\mathrm{v}}/(m_{\mathrm{c}} + m_{\mathrm{v}})$ is determined by the effective mass of the valence ($m_{\mathrm{v}}$) and conduction band ($m_{\mathrm{c}}$), respectively~\cite{Coldren2012}. The broadening of the energy states results mainly from scattering and recombination, and is included in equation \eqref{eq:spontaneous_emission_spectrum} by the lineshape function $\mathcal{L}(E)$, which is commonly expressed by a Lorentzian~\cite{Coldren2012}:
\begin{equation}
\mathcal{L}(E) = \frac{\gamma_{L}(E)}{2\pi} \frac{1}{E^{2} + (\gamma_{L}(E)/2)^{2}} \label{eq:Lorentz_broadening} \, .
\end{equation}

\begin{figure*}[ht]
    \includegraphics[keepaspectratio]{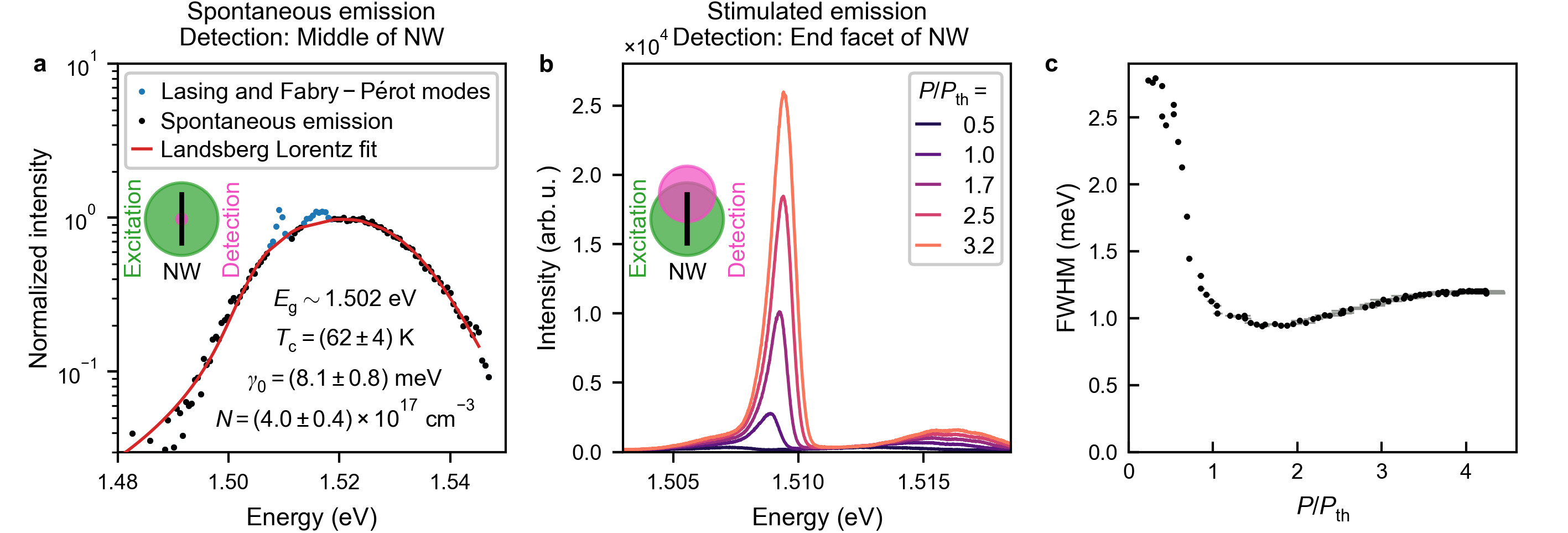}
    \caption{\textbf{Examples of CW spectra and linewidth narrowing}. \textbf{a}, Selected spontaneous emission spectrum of the NW laser under CW excitation on a semi-logarithmic scale for an excitation power ($P$) relative to threshold ($P_{\mathrm{th}}$) of $P/P_{\mathrm{th}}\sim1.7$. The spontaneous emission spectrum was obtained by centering the detection spot (pink) in the middle of the NW (black), as shown by the inset. This maximizes the amount of detected spontaneous emission and minimizes the amount of detected stimulated emission. This is because stimulated emission leaves the NW laser predominantly through its end facets and therefore most of it is not detected when the detection spot is centered in the middle of the NW. Since spontaneous emission is spectrally broad, the spectrum was measured using a broadband single spectrometer. The fit was obtained with the model described in equation~\eqref{eq:spontaneous_emission_spectrum}, after first removing the weak residual lasing peak and Fabry-P\'erot modes (blue) from the spontaneous emission spectrum. From the fit we obtained the renormalized band gap ($E_{\mathrm{g}}$), the carrier temperature ($T_{\mathrm{c}}$), the broadening ($\gamma_{0}$) and the carrier density ($N$). The best fit parameters together with their $\SI{95}{\%}$ CIs are given in the plot. \textbf{b}, Selected stimulated emission spectra of the NW laser under CW excitation on a linear scale. The stimulated emission spectra were obtained by positioning the detection spot (pink) on one of the NW (black) end facets while the excitation spot (green) is still covering the entire NW, as shown by the inset. This maximizes the amount of detected stimulated emission and minimizes the amount of detected spontaneous emission. Since stimulated emission is spectrally narrow, the spectra were measured using a high-resolution double spectrometer. The high resolution spectra of the lasing mode are shown for various excitation powers. \textbf{c}, Full width at half maximum (FWHM) of the lasing peak as a function of excitation power. The error bars represent $\SI{95}{\%}$ CIs of the mean and follow from the determination of $P_{\mathrm{th}}$.}
    \label{Fig:S_CW_SponEmFits_and_Spectra}
\end{figure*}

Figure~\ref{Fig:S_CW_SponEmFits_and_Spectra}a presents an exemplary spectrum of a NW on a semi-logarithmic scale, measured with a CW excitation power of $P/P_{\mathrm{th}}\sim1.7$. Here, the lasing peak ($\sim\SI{1.51}{eV}$) is relatively weak, since for this measurement the detection spot (diameter $\sim\SI{2}{\micro m}$) was centered in the middle of the NW resonator to maximize the amount of collected spontaneous emission and to minimize the amount of collected stimulated emission. The relative position of NW (black), excitation spot (green) and detection spot (pink) is illustrated by the inset. For the fitting procedure, the weak residual lasing peak and Fabry-P\'{e}rot modes (blue) were manually removed from the data, leaving only the spontaneous emission spectrum (black). Additionally, the spectrum was normalized to the highest spontaneous emission value. To properly fit both the low and high energy sides of the spectrum, both a Lorentzian line shape function and an energy dependent broadening factor $\gamma_{\mathrm{L}}(E)$ are required, as described by equation~\eqref{eq:Lorentz_broadening}. Phenomenologically, the broadening of the energy states is strong on the low- and weak on the high-energy side of the spectrum and can be described by Landsberg broadening \cite{Landsberg1966, Martin1977, Kalt1992}
\begin{equation}
\frac{\gamma_{\mathrm{L}}}{\gamma_{0}} = 1 - 2.229 \left(\frac{E}{\mu_{\mathrm{a}}}\right)^{1} 
+ 1.458 \left(\frac{E}{\mu_{\mathrm{a}}} \right)^{2}
- 0.229 \left(\frac{E}{\mu_{\mathrm{a}}} \right)^{3} \, ,
\label{eq:Landsberg_broadening}
\end{equation} 
where $a=\mathrm{c},\mathrm{v}$. The Landsberg broadening is valid for $0\leq E \leq \mu_{\mathrm{a}}$, reaches a maximum value of $\gamma_{0}$ at the band edge and approaches zero at the quasi-Fermi level. Figure~\ref{Fig:S_CW_SponEmFits_and_Spectra}a shows that the fitted Landsberg broadened spectrum exhibits excellent agreement with the measured data. By fitting each spectrum of an excitation power series in this way, we obtained the values and the power dependence of $N$, $T_{\mathrm{c}}$, $E_{\mathrm{g}}$ and $\gamma_{\mathrm{0}}$. The best fit parameters for the selected example are listed in Fig.~\ref{Fig:S_CW_SponEmFits_and_Spectra}a.

Additional information about the CW lasing characteristics of NW lasers is obtained from detailed measured spectra of the lasing peak. Of particular interest for comparison with the simulation is how the laser output power and $E_{\mathrm{p}}$ change with excitation power. For a better resolution of the lasing mode, we repeated the CW excitation measurement from before on the same nanowire but this time detected the NW emission with a double spectrometer, allowing a spectral resolution of $\Delta E< \SI{40}{\micro\eV}$. For this measurement, the detection spot was positioned on one of the end-facets of the NW laser to maximize the amount of collected stimulated emission while minimizing the amount of collected spontaneous emission. The amount of light coupled into the microscope objective can depend strongly on the exact relative position of the NW end facet and the detection spot if it is small ($\sim\SI{2}{\micro m}$). To avoid this problem and to make the measurements less sensitive to mechanical noise, here we used a large detection spot with a diameter of $\sim\SI{11}{\micro m}$. A small spot size is not a problem for the measurements with the detection spot centered in the middle of the NW, since the radiation characteristic of spontaneous emission is isotropic. Figure~\ref{Fig:S_CW_SponEmFits_and_Spectra}b shows how the NW emission changes when the excitation power is increased from $P/P_{\mathrm{th}}\sim0.5$ to $P/P_{\mathrm{th}}\sim3.2$. The spectra in Fig.~\ref{Fig:S_CW_SponEmFits_and_Spectra}b show a clear blue shift of the lasing peak and an increase in intensity with increasing excitation power. To determine $I$, the spontaneous emission was removed from the spectrum and the area under the peak was integrated. The threshold of the NW under CW excitation was determined in the same way as in Fig.~\ref{Fig:S1_SinglePulsePowerseries}c. For completeness, Fig.~\ref{Fig:S_CW_SponEmFits_and_Spectra}c shows the characteristic drop of the NW laser linewidth as the excitation power is increased from below to above threshold.

\begin{figure*}[htp]
    \includegraphics[keepaspectratio]{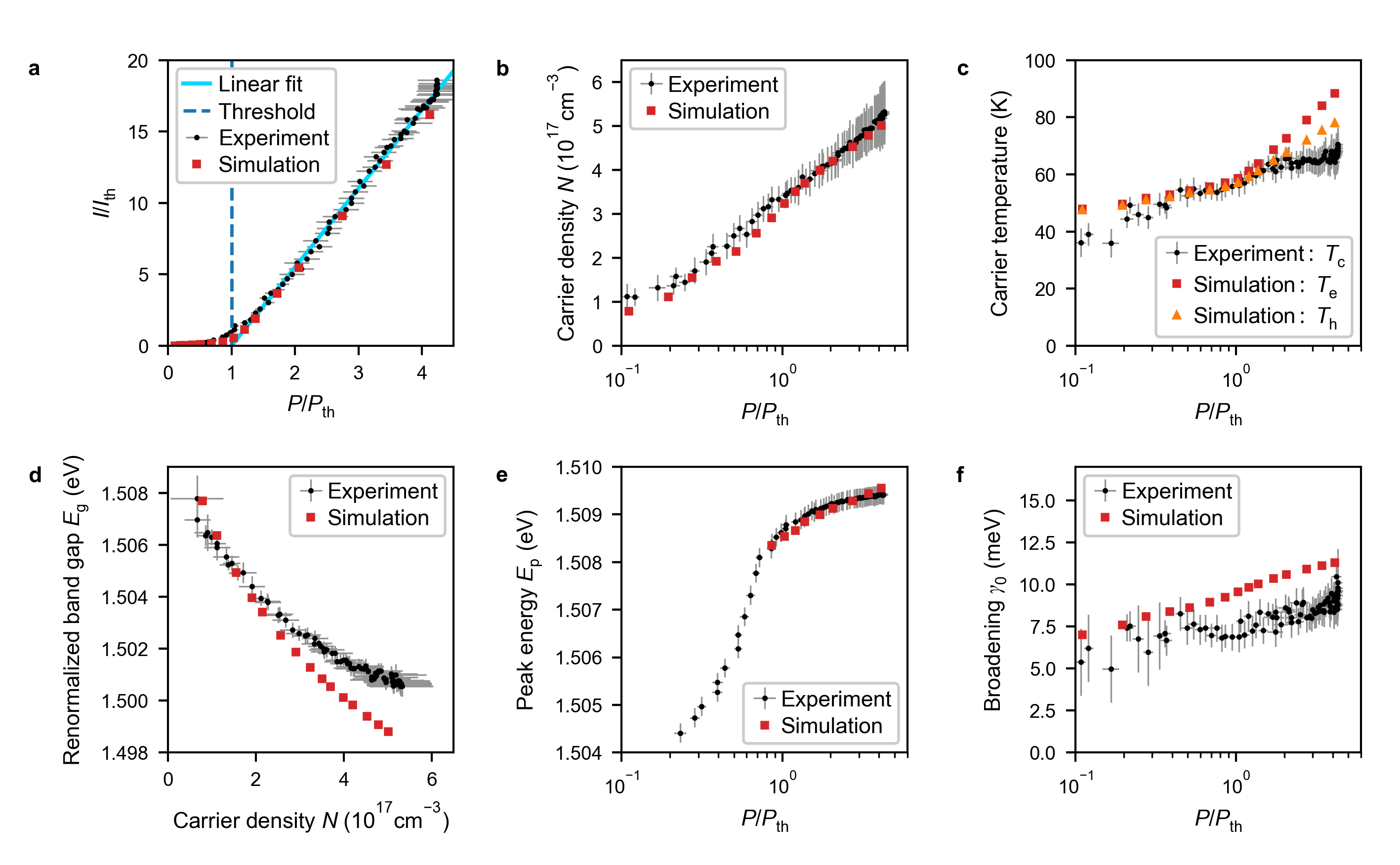}
    \caption{\textbf{CW lasing characteristics.} \textbf{a}, Light-in light-out curve on a linear scale. The integrated intensity ($I$) and excitation power ($P$) are normalized to their respective threshold values $I_{\mathrm{th}}$ and $P_{\mathrm{th}}$. The threshold was determined by the intersection of the linear fit with the horizontal axis. \textbf{b},\textbf{c}, The carrier density ($N$) and the carrier temperatures ($T_{\mathrm{c}}$, $T_{\mathrm{e}}$ and $T_{\mathrm{h}}$) are only partially clamped for $P/P_{\mathrm{th}}>1$ due to carrier heating via stimulated emission. The apparent saturation of $T_{\mathrm{c}}$ in \textbf{c} for $P/P_{\mathrm{th}}>2$ is a measurement artefact. \textbf{d}, The band gap $E_{\mathrm{g}}$ redshifts with increasing $N$. \textbf{e}, Changes in the refractive index lead to a strong blueshift of the peak energy $E_{\mathrm{p}}$, which weakens for $P/P_{\mathrm{th}}>1$ due to partial gain clamping. \textbf{f}, The simulated broadening $\gamma_{0}$ is in good agreement with the values obtained from experiment using the phenomenological Landsberg model, as described by equation~\eqref{eq:Landsberg_broadening}. All error bars represent $\SI{95}{\%}$ CIs of the mean. In \textbf{e}, the error bars reflect the calibration of the spectrometer, whereas in \textbf{b}-\textbf{d} and \textbf{f} the uncertainties result from the fit of equation~\eqref{eq:spontaneous_emission_spectrum} to the measured spontaneous emission spectra. Furthermore, for all panels, the uncertainty of the excitation power follows from the determination of $P_{\mathrm{th}}$, see Supplementary Section~\ref{Sec:SinglePulseThreshold}.}
    \label{Fig:S_CW_Data_and_Simulation}
\end{figure*}

In the following sections below we describe the experimental results, obtained with the procedure outlined above, and discuss the corresponding simulated data, both of which are presented in Fig.~\ref{Fig:S_CW_Data_and_Simulation}.

\subsection{Light-in light-out curve}
Figure~\ref{Fig:S_CW_Data_and_Simulation}a shows the measured (black) and simulated (red) $I$ as a function of normalized excitation power ($P/P_{\mathrm{th}}$). To allow comparison between experiment and simulation, $I$ was also normalized to its threshold value ($I_{\mathrm{th}}$). We fitted the output intensity above threshold with a linear function (solid cyan line) and determined a CW threshold excitation power of $P_{\mathrm{th}}=\SI[separate-uncertainty=true]{4.36(23)}{mW}$ by the intersection of the fit with the horizontal axis. With a spot diameter of $\sim\SI{14}{\micro m}$ for the CW excitation laser, this corresponds to a power density of $p_{\mathrm{th}}=\SI[separate-uncertainty=true]{2.83(15)}{kW\ cm^{-2}}$, which is comparable to the CW threshold power density of tin-doped cadmium
sulfide NW lasers~\cite{roder_continuous_2013}. The measured light-in light-out curve shows a pronounced nonlinearity around threshold and a linear dependence above, which is a clear indication of lasing and is typical for this type of NW laser. For reasons described in the following, we adjusted the simulated $I/I_{\mathrm{th}}$ to the experimental data for one data point above $P_{\mathrm{th}}$, where spontaneous emission is negligible. The resulting agreement of the simulation with the experiment is excellent for all data points above threshold. At and below threshold, the measured intensities are slightly larger ($\sim\SI{10}{\%}$) than the values predicted by the simulation. There are two main reasons for this behaviour. The first is the measurement geometry. We excited the NW, that was lying horizontally on a substrate, from the top, using the same objective for excitation and detection. Since spontaneous emission has an isotropic radiation characteristic compared to stimulated emission, which is directed along the NW axis, this measurement geometry leads to an overestimation of the amount of spontaneous emission into the lasing mode. The second reason is that the determination of the emission into the laser mode below threshold is challenging. The shape of the simulated data can be adjusted by $\beta$. For this NW laser we achieved best agreement with the experimental data above $P_{\mathrm{th}}$ for $\beta=0.1$, which is in full accord with values found in literature~\cite{Saxena2013_RTLaserGaAs,Saxena2015ModeProfiling,Mayer2016_GaAsLaserOnSilicon}.

\subsection{Carrier density and carrier temperature}
\label{Subsec:CarrierDensityAndTemperature}
In Fig.~\ref{Fig:S_CW_Data_and_Simulation}b we observe that $N$ increases with $P/P_{\mathrm{th}}$. It increases even above $P_{\mathrm{th}}$, although with a slightly reduced slope. Thereby, $N$ changes from $N = \SI[separate-uncertainty = true]{3.4(3)e17}{\per\cubic\cm}$ at $P_{\mathrm{th}}$ to $N = \SI[separate-uncertainty = true]{5.3(6)e17}{\per\cubic\cm}$ at $P/P_{\mathrm{th}}\sim4.3$. For the whole range of excitation powers the simulated carrier densities are in excellent quantitative agreement with the experimental data.

Figure~\ref{Fig:S_CW_Data_and_Simulation}c shows that, similar to $N$, also the measured $T_{\mathrm{c}}$ increases with $P/P_{\mathrm{th}}$, even above threshold. It ranges from $T_{\mathrm{c}} = \SI[separate-uncertainty = true]{36(5)}{K}$ at $P/P_{\mathrm{th}}\sim\SI{0.11}{}$ to $T_{\mathrm{c}} = \SI[separate-uncertainty = true]{70(6)}{K}$ at $P/P_{\mathrm{th}}\sim\SI{4.3}{}$. The apparent saturation of $T_{\mathrm{c}}$ for $P/P_{\mathrm{th}}>2$ is a measurement artefact due to a longpass filter that was used to remove the excitation laser. This filter sharply cuts off the NW emission at $\sim\SI{1.55}{eV}$, leading to a systematic underestimation of $T_{\mathrm{c}}$ in this range by a few $\si{K}$, while leaving the other fit parameters unaffected. This argument is substantiated by the fact that for $P/P_{\mathrm{th}}>2$ the carrier density $N$ in Fig.~\ref{Fig:S_CW_Data_and_Simulation}b keeps increasing, which is only possible if the $T_{\mathrm{c}}$ increases simultaneously. This increase in $T_{\mathrm{c}}$ is caused by the increased stimulated emission rate with increased $P/P_{\mathrm{th}}$, which heats up the carrier distribution, as discussed in the main manuscript. Due to this photoinduced heating of the carrier distribution, an increase in $N$ is necessary to maintain the gain required for lasing. Thus, the gain is only partially clamped above threshold.

For excitation powers from $P/P_{\mathrm{th}}\sim0.2$ to $P/P_{\mathrm{th}}\sim2$, the simulated electron ($T_{\mathrm{e}}$) and hole ($T_{\mathrm{h}}$) temperatures are in excellent quantitative agreement with the measurement, whereas for $P/P_{\mathrm{th}}<0.2$ they slightly overestimate the experimental results. For high excitation powers ($P/P_{\mathrm{th}}>2$) the simulated carrier temperatures are observed to further increase, as expected from the behaviour of the simulated $N$ in Fig.~\ref{Fig:S_CW_Data_and_Simulation}b. Over the investigated range of excitation powers, $T_{\mathrm{e}}$ and $T_{\mathrm{h}}$ reach maximum values of $T_{\mathrm{e}}=\SI{88}{K}$ and $T_{\mathrm{h}}=\SI{78}{K}$, respectively. The model additionally predicts that $T_{\mathrm{e}}$ is slightly larger than $T_{\mathrm{h}}$ with an increasing gap between the two as $P/P_{\mathrm{th}}$ increases. However, this can not be tested with our experimental approach.

The main simulation parameter which influences the behaviour of $T_{\mathrm{e}}$ and $T_{\mathrm{h}}$ and also the power dependence of the $N$ is the end-facet reflectivity $R$ of the NW. A higher reflectivity leads to an increased stimulated emission rate, which results in a stronger increase of $T_{\mathrm{e}}$, $T_{\mathrm{h}}$ and $N$ as a function of $P/P_{\mathrm{th}}$. Best agreement with the experiment was achieved for an end-facet reflectivity of $R=0.5$, which is in full accord with values in literature~\cite{Mayer2013_RTLaserGaAs,Saxena2013_RTLaserGaAs}. 

\subsection{Band gap narrowing}
Figure~\ref{Fig:S_CW_Data_and_Simulation}d shows the measured $E_{\mathrm{g}}$ (black) as a function of $N$ taken from Fig.~\ref{Fig:S_CW_Data_and_Simulation}b. As $N$ is increased from $\SI[separate-uncertainty=true]{0.7(6)e17}{\per\cubic\cm}$ to $\SI[separate-uncertainty = true]{5.3(7)e17}{\per\cubic\cm}$, the measured $E_{\mathrm{g}}$ decreases from $\sim\SI{1.508}{eV}$ to $\sim\SI{1.501}{eV}$. A similar decrease of $E_{\mathrm{g}}$ is calculated with the simulation (red). We observe very good quantitative agreement between experiment and theory, although for $N>\SI[separate-uncertainty=true]{2.5e17}{\per\cubic\cm}$ the simulated band gap narrowing is slightly overestimated, with a maximum deviation of $\sim\SI{2}{meV}$. The band gap energy of unexcited bulk GaAs at $T_{\mathrm{L}} = \SI{10}{K}$ was taken from literature~\cite{Passler1997}, see Table~\ref{Table:SimulationParameters}. The simulated band gap narrowing is determined purely from many-body Coulomb interactions and is not directly adjusted by any simulation parameter. Thus, the comparison with the experiment in Fig.~\ref{Fig:S_CW_Data_and_Simulation}d provides a good validation of our simulation model.

\subsection{Peak energy}
From spectra such as in Fig.~\ref{Fig:S_CW_SponEmFits_and_Spectra}b, $E_{\mathrm{p}}$ was determined as a function of $P/P_{\mathrm{th}}$. In the experimental data this was reliably possible for excitation powers $P/P_{\mathrm{th}}>0.2$. In Fig.~\ref{Fig:S_CW_Data_and_Simulation}e, we observe a strong increase of $E_{\mathrm{p}}$ from $\SI[separate-uncertainty=true]{1.5044(2)}{eV}$ to $\SI[separate-uncertainty=true]{1.5086(2)}{eV}$ as the excitation power is increased from $P/P_{\mathrm{th}}\sim\SI{0.2}{}$ up to threshold. The reason for this increase is explained by a decrease in refractive index, which is caused by the combined interplay of band gap renormalization and the increased gain with larger $N$. Above threshold, the gain is partially clamped, resulting in a reduced blueshift, with $E_{\mathrm{p}}$ further increasing up to $\SI[separate-uncertainty=true]{1.5094(2)}{eV}$ at $P/P_{\mathrm{th}}\sim\SI{4.2}{}$. For the simulations, we used a heuristic refractive index model of bulk GaAs from literature~\cite{Reinhart2005a, Reinhart2005b}. Its input parameters do not include any external fit parameters and were exclusively calculated using the quantum statistical model, i.e. a shift in $E_{\mathrm{p}}$ is caused entirely by a change in the microscopic dynamic variables. Only the absolute value of the simulated $E_{\mathrm{p}}$ at threshold was adjusted to the experiment. We achieved this by first setting the laser length to $L\sim\SI{10}{\micro m}$ using the microscope image in Fig.~1d of the main manuscript and then slightly fine-tuning it to $L\sim\SI{10.15}{\micro m}$. This last step had no effect on the lasing dynamics. As can be seen in Fig.~\ref{Fig:S_CW_Data_and_Simulation}e, our approach resulted in excellent quantitative agreement with the experiment.

\subsection{Broadening}
Figure~\ref{Fig:S_CW_Data_and_Simulation}f shows the experimentally determined $\gamma_{0}$ as a function of $P/P_{\mathrm{th}}$. The values range from $\gamma_{0}=\SI[separate-uncertainty=true]{5.0(20)}{meV}$ at $P/P_{\mathrm{th}}\sim\SI{0.17}{}$ to $\gamma_{0}=\SI[separate-uncertainty=true]{10.5(10)}{meV}$ at $P/P_{\mathrm{th}}\sim\SI{4.2}{}$.  It is observed that $\gamma_{0}$ does not increase continuously with increasing $P/P_{\mathrm{th}}$, but exhibits a small decrease to $\gamma_{0}=\SI[separate-uncertainty=true]{6.9(9)}{meV}$ near threshold.

In the quantum statistical model, the energy broadening
\begin{equation}
    \gamma_{\mathrm{eh}}(\mathbf{k}) = \frac{1}{2}\left[\gamma_{\mathrm{e}}(\mathbf{k}) + \gamma_{\mathrm{h}}(\mathbf{k})\right]
    \label{eq:Broadening}
\end{equation}
results mainly from carrier-carrier and carrier-phonon scattering. Thus, the energies $\gamma_{\mathrm{e}}$ and $\gamma_{\mathrm{h}}$ are related to the corresponding scattering rates into ($\Sigma_{\mathrm{a}}^{\mathrm{in}}(\mathbf{k})$) and out of ($\Sigma_{\mathrm{a}}^{\mathrm{out}}(\mathbf{k})$) a given state with wavevector $\mathbf{k}$, written as~\cite{Chow1994_SemiconductorLaserPhysicsBook}
\begin{equation}
\gamma_{\mathrm{a}}(\mathbf{k}) = \hbar[\Sigma_{\mathrm{a}}^{\mathrm{in}}(\mathbf{k}) + \Sigma_{\mathrm{a}}^{\mathrm{out}}(\mathbf{k})]\, ,
\end{equation}
with $\mathrm{a}=\mathrm{e},\mathrm{h}$ for electrons and holes, respectively. Similar to the Landsberg model, the broadening calculated within the quantum statistical model is in general energy dependent. In Fig.~\ref{Fig:S_CW_Data_and_Simulation}f we therefore only compare the broadening values near the band edge. Over the investigated range of $P/P_{\mathrm{th}}$, the simulated $\gamma_{\mathrm{eh}}(0)=\gamma_{0}$ exhibits a continuous increase from $\sim\SI{7}{meV}$ to $\sim\SI{11.3}{meV}$. Remarkably, the simulated results are very close to the experimental values that were determined using the phenomenological Landsberg model and also exhibit a very similar behaviour above threshold. We emphasize that the simulated $\gamma_{0}$ contained no fit parameters, as it was calculated solely from the scattering rates of the quantum statistical model.

\subsection{Simulation parameters}
Table~\ref{Table:SimulationParameters} summarizes the parameters used for all simulations in the main manuscript and the Supplementary Information. These were used for both the quantum statistical and semiconductor Bloch model. We note that $\beta$ and $R$ were the only parameters that could be freely chosen.

\begin{table}
\centering
\begin{tabular}{|c|c|c|c|} 
\hline
Name                                        & Symbol                                & Value                                 & Source   \\ 
\hline
Electron effective mass                     & $m_{\mathrm{e}}$                      & $0.067\times m_{0}$                   & Literature\cite{Chow1994_SemiconductorLaserPhysicsBook,Vurgaftman2001_PhysicalConstants}\\
Hole effective mass                         & $m_{\mathrm{h}}$                      & $0.377\times m_{0}$                   & Literature\cite{Chow1994_SemiconductorLaserPhysicsBook,Vurgaftman2001_PhysicalConstants}\\
Transition matrix element                   & $6|M_{\mathrm{T}}|^{2}/m_{0}$         & $\SI{28.8}{eV}$                       & Literature\cite{Coldren2012}\\
Bandgap at $\SI{10}{K}$                     & $E_{\mathrm{g}0,10\,\mathrm{K}}$                     & $\SI{1.5192}{eV}$                     & Literature\cite{Passler1997}\\
Relative static dielectric constant                  & $\varepsilon_{r}$                        & $12.9$            & Literature\cite{Samara1983DielectricConstants}\\
Relative optical dielectric constant                 & $\varepsilon_{\infty}$                       & $10.89$          & Literature\cite{Samara1983DielectricConstants}\\
Bohr radius                                 & $a_{0}$                               & $\SI{12.43}{nm}$                      & Literature\cite{Chow1994_SemiconductorLaserPhysicsBook}\\
LO phonon energy                            & $\hbar\omega_{\mathrm{LO}}$           & $\SI{36.5}{meV}$                      & Literature\cite{Strauch1990LOPhononEnergy}\\
Confinement factor                          & $\mathit{\Gamma}$                              & $1.2$                                 & COMSOL Multiphysics\\
Excitation pulse FWHM                       & $\Delta\tau_{\mathrm{FWHM}}$          & $\SI{200}{fs}$                        & Measurement\\
Nanowire length                             & $L$                                   & $\SI{10}{\micro m}$                   & Microscope images\\
End facet reflectivity                      & $R$                                   & $0.5$                                 & Free parameter\\
Spontaneous emission factor                 & $\beta$                               & $0.1$                                 & Free parameter\\
\hline
\end{tabular}
\caption{\textbf{Simulation parameters.} The listed parameters were used for both the quantum statistical and the semiconductor Bloch model. The constants $m_{0}$ and $\varepsilon_{0}$ are the free electron mass and vacuum permittivity, respectively.}
\label{Table:SimulationParameters}
\end{table}

\clearpage
\section{Semiconductor Bloch simulation of the time-resolved pump-probe response}
\label{Sec:BlochModel}
\begin{figure*}[ht]
    \includegraphics[keepaspectratio]{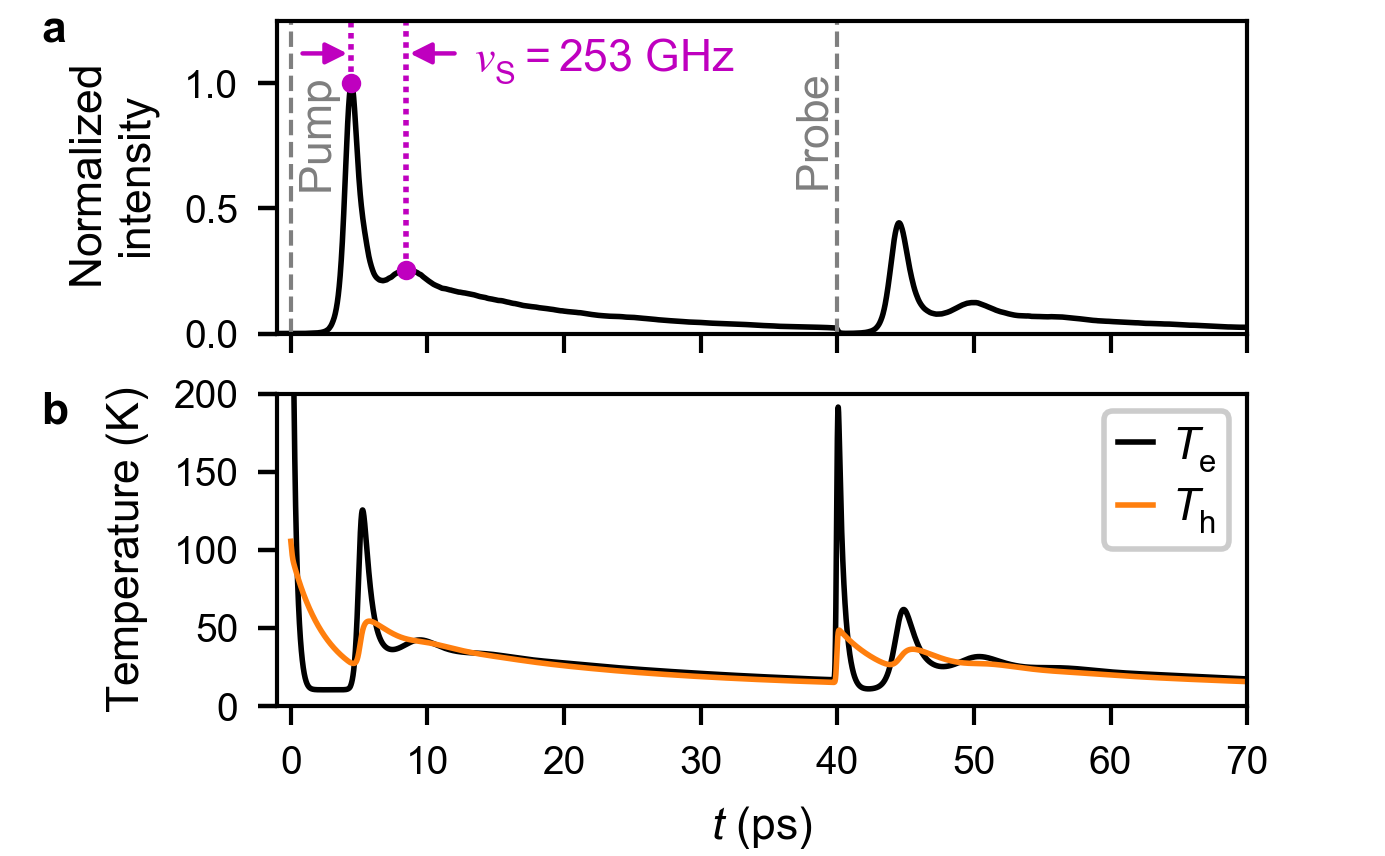}
    \caption{\textbf{Semiconductor Bloch simulation of the time-resolved pump-probe response.} \textbf{a}, Output intensity as a function of time for pump-probe excitation with $\mathit{\Delta} t=\SI{40}{ps}$. The oscillation frequency ($\nu_{\mathrm{S}}$) of the first output pulse is indicated in magenta. \textbf{b}, Corresponding time dependence of the electron ($T_{\mathrm{e}}$) and hole ($T_{\mathrm{h}}$) temperatures, reflecting the intensity dynamics in \textbf{a}.}
    \label{Fig:S_TimeSeriesBlochModel}
\end{figure*}

Here we discuss the time-resolved pump-probe response of a NW laser simulated using the semiconductor Bloch model. Figure~\ref{Fig:S_TimeSeriesBlochModel}a presents the output intensity as a function of time for a pump-probe delay of $\Delta t=\SI{40}{ps}$ (grey dashed lines), where we used the same excitation conditions as for Fig.~1e,f and Fig.~2 in the main manuscript. The NW output pulses are strongly asymmetric in time with a pronounced initial peak, following oscillations ($\nu_{\mathrm{S}}=\SI{253}{GHz}$, magenta) and a slow decay. Figure~\ref{Fig:S_TimeSeriesBlochModel}b reveals that these intensity oscillations result from oscillations of the carrier temperatures, which mirror the time dependence of the intensity in Fig.~\ref{Fig:S_TimeSeriesBlochModel}a. The sudden increase of $T_{\mathrm{e}}$ and $T_{\mathrm{h}}$ at $t=\SI{40}{ps}$ is due to transient heating caused by the injection of additional carriers~\cite{Jahnke1995a_UltrafastSwitching}. The above observations are in qualitative agreement with the quantum statistical simulation in Fig.~2a,b, although here the oscillations are more strongly damped. A possible reason for this could be the weaker manifestation of oscillations in $T_{\mathrm{e}}$ and $T_{\mathrm{h}}$ compared to the quantum statistical model. Especially $T_{\mathrm{h}}$ shows only very weak and strongly damped oscillations. This can have several reasons. Firstly, the quantum statistical model self-consistently calculates time-dependent scattering rates, while the scattering rates of the semiconductor Bloch model are kept constant. Secondly, the electron-hole scattering included in the Boltzmann equations used to calculate the scattering rates allows for an energetic exchange of the two subsystems. Thereby, the periodic oscillation of $T_{\mathrm{e}}$ is partially transferred to the holes in the valence band. This demonstrates the limitations of the relaxation rate approximation and the importance of treating carrier-carrier and carrier-LO-phonon scattering microscopically. Nevertheless, $\nu_{\mathrm{S}}$ is not strongly affected by these limitations and the resulting stronger damping, as it mainly reflects the time dependence of $T_{\mathrm{e}}$ in the present case.

In the relaxation rate approximation of the semiconductor Bloch model we used piecewise constant scattering rates, which we chose with guidance from the rates calculated within the quantum statistical simulation (see Fig.~2c). We used two sets of scattering rates, one for the initial relaxation of the photoexcited carriers and another for the lasing process. The former gets an additional factor of 1.5 to approximately account for the faster scattering during the initial carrier relaxation (see Fig.~2c). Both sets of scattering rates are summarized in Table~\ref{Table:ScatteringRatesBloch}. Since the k-resolved and time-dependent scattering rates of the quantum statistical model do not directly translate to the relaxation rate approximation of the semiconductor Bloch model, we added a factor $\xi$ for adjustment. We chose $\xi$ such that the simulated $t_{\mathrm{on}}$ and $f_{\mathrm{S}}$ approximately match the experiment. For simulations, $\xi=10$ was used. All other simulation parameters are listed in Table~\ref{Table:SimulationParameters}.

\begin{table}[ht]
\begin{tabular}{|c|c|c|c|}
\hline
Symbol & Value during relaxation & Value during lasing & Unit \\ \hline
$\gamma_{\mathrm{cc},\mathrm{e}}$ & $1.5\cdot\xi\cdot8.5$  & $\xi\cdot8.5$  & $\si{ps^{-1}}$\\
$\gamma_{\mathrm{cc},\mathrm{h}}$ & $1.5\cdot\xi\cdot16.4$ & $\xi\cdot16.4$ & $\si{ps^{-1}}$\\
$\gamma_{\mathrm{ep}}$ & $1.5\cdot\xi\cdot0.4$  & $\xi\cdot0.4$  & $\si{ps^{-1}}$\\
$\gamma_{\mathrm{hp}}$ & $1.5\cdot\xi\cdot0.03$ & $\xi\cdot0.03$ & $\si{ps^{-1}}$\\ \hline
\end{tabular}
\caption{\textbf{Scattering rates used in the semiconductor Bloch model.} Summary of the rates for carrier-carrier scattering of electrons ($\gamma_{\mathrm{cc},\mathrm{e}}$), carrier-carrier scattering of holes ($\gamma_{\mathrm{cc},\mathrm{h}}$), electron-LO-phonon scattering ($\gamma_{\mathrm{ep}}$) and hole-LO-phonon scattering ($\gamma_{\mathrm{hp}}$). Two sets of constant scattering rates were used. One for the initial relaxation of the photoexcited carriers and another for the lasing process. For our simulations we used $\xi=10$.}
\label{Table:ScatteringRatesBloch}
\end{table}

\clearpage
\section{Determination of the carrier temperature}
\begin{figure*}[ht]
    \includegraphics[keepaspectratio]{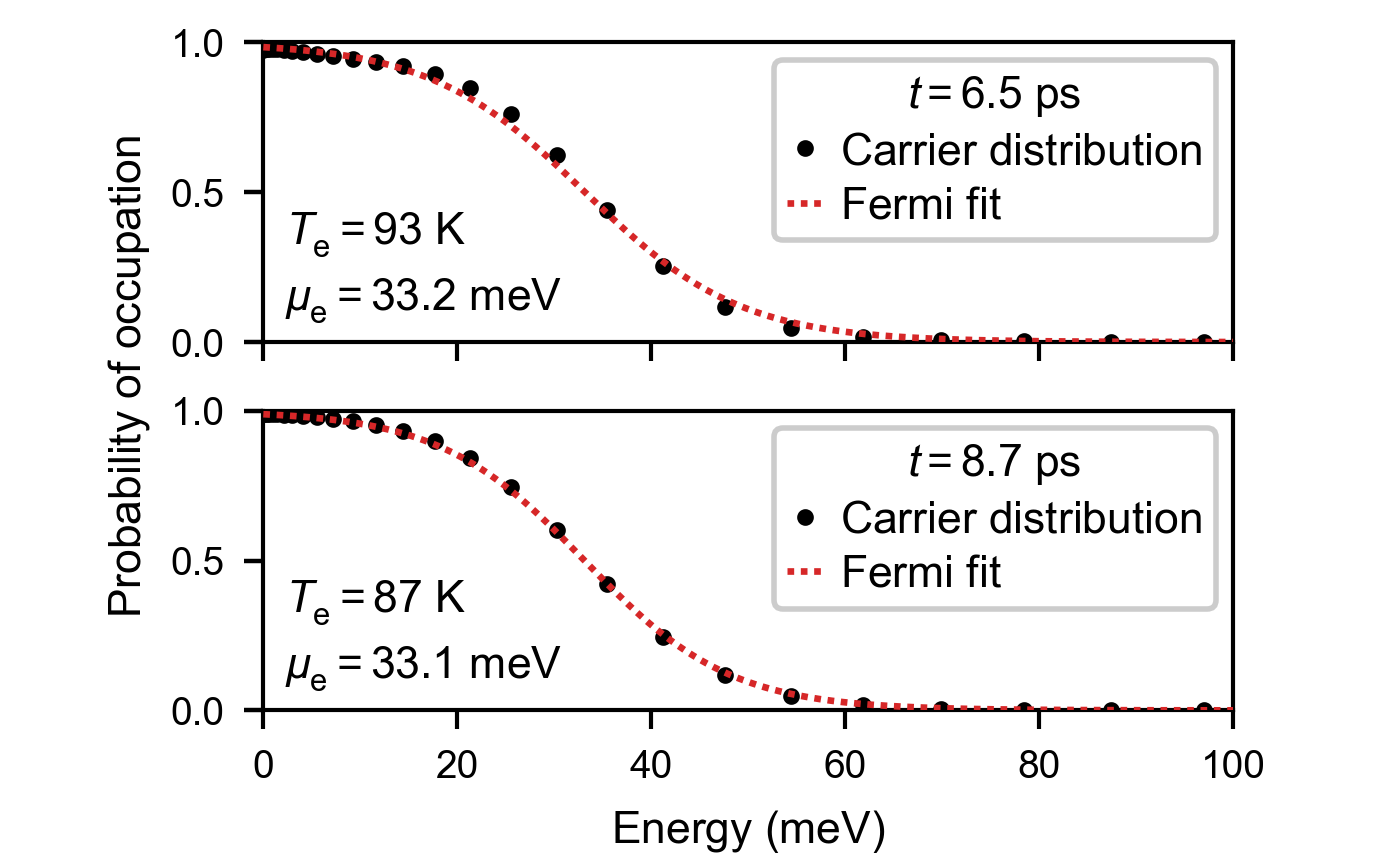}
    \caption{\textbf{Determination of the carrier temperature.} Two exemplary electron distributions for $t=\SI{6.5}{ps}$ (upper panel) and $t=\SI{8.7}{ps}$ (lower panel). In each case, the fitted Fermi-Dirac functions are a good approximation. The respective fit parameters (electron temperature $T_{\mathrm{e}}$ and Fermi energy $\mu_{\mathrm{e}}$) are given in each plot. The carrier density in the upper panel is $N = \SI{5.12e17}{\per\cubic\cm}$, whereas in the lower panel it is $N = \SI[separate-uncertainty = true]{5.06e17}{\per\cubic\cm}$.}
    \label{Fig:S_FermiFunctionFit}
\end{figure*}

Here we discuss how the carrier temperatures are determined and show that they are well-defined parameters. The data presented here belong to the quantum statistical simulation in Fig.~2 of the main manuscript. Figure~\ref{Fig:S_FermiFunctionFit} shows two selected electron distributions for $t=\SI{6.5}{ps}$ (upper panel) and $t=\SI{8.7}{ps}$ (lower panel). At $t=\SI{6.5}{ps}$ the laser intensity reaches its first maximum (see Fig.~2a) and therefore spectral hole burning, if it plays a role, is expected to be strong for this distribution. In contrast, at $t=\SI{8.7}{ps}$ the intensity is at the following local minimum, which means that spectral hole burning should be weak in this case. We fitted each carrier distribution with a Fermi-Dirac function, whereby the carrier density $N$ and the total energy of the carriers served as a constraint. The resulting fits are shown in Fig.~\ref{Fig:S_FermiFunctionFit} with red dotted lines, together with their respective fit parameters. In both cases, the fit provides a good approximation to the carrier distribution, demonstrating that spectral hole burning does not play a significant role. Thus, for our NW laser $T_{\mathrm{e}}$ is a well defined parameter for all time steps. The only exception to this are very short timeframes after excitation, since the photoexcited carrier distributions need a few hundred femtoseconds to thermalize~\cite{Oudar1985_ThermalizationAt15KGaAs,Elsaesser1991_ThermalizationAt300KGaAs,Kane1994a_ThermalizationWithin150fs}. A similar discussion to the above applies to holes.

\clearpage
\section{Carrier relaxation}
\begin{figure*}[ht]
    \includegraphics[keepaspectratio]{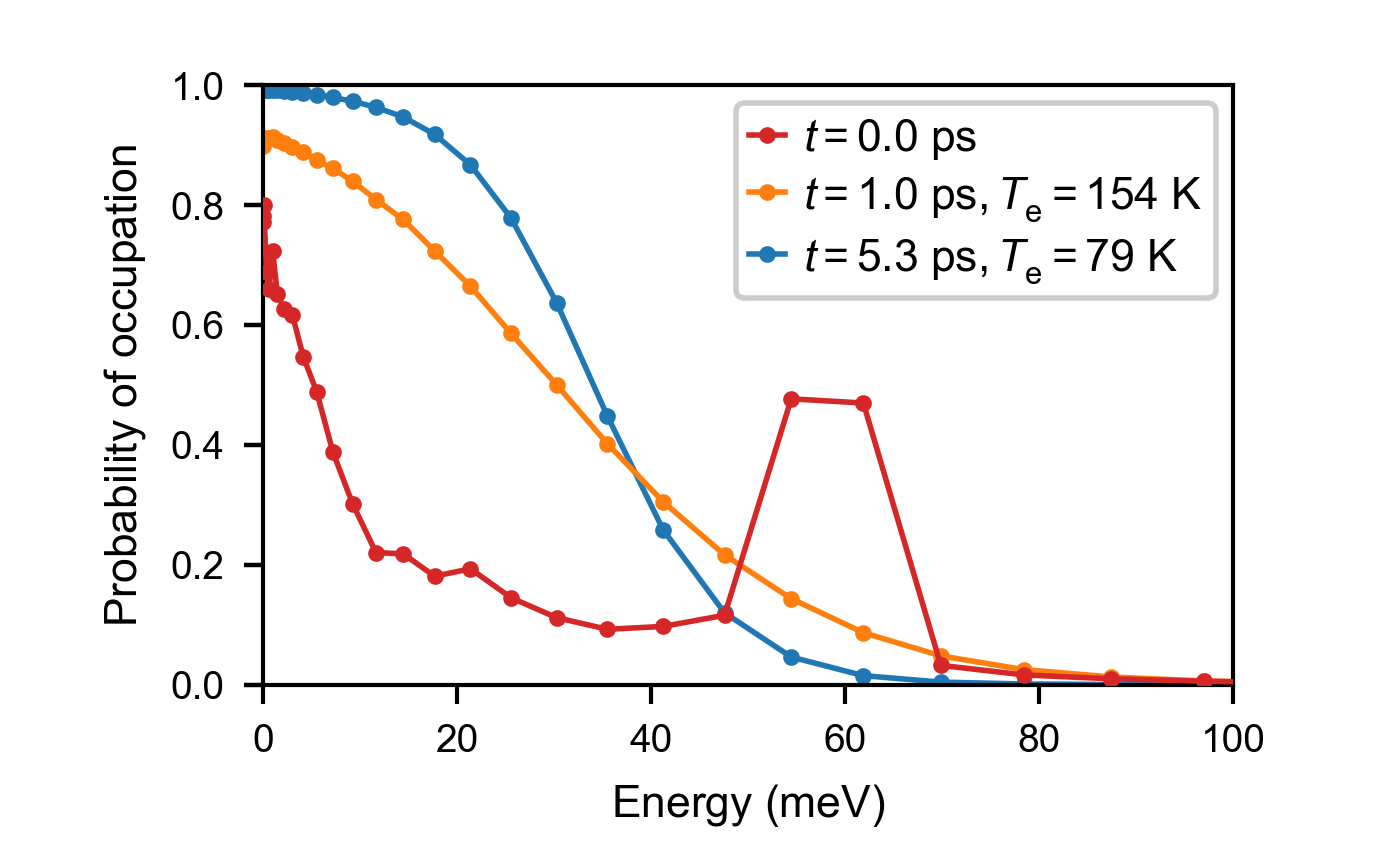}
    \caption{\textbf{Electron relaxation.} Presented are three selected electron distributions, illustrating the initial relaxation process. We extracted the distributions from the quantum statistical simulation at $t=\SI{0.0}{ps}$, $t=\SI{1.0}{ps}$ and $t=\SI{5.3}{ps}$. The filled circles indicate the numerical discretization of the conduction band in the model. The temporal origin $t=\SI{0.0}{ps}$ is defined as the moment in time where the maximum of the excitation pulse arrives at the NW.} 
    \label{Fig:S_CarrierRelaxation}
\end{figure*}

The data presented here belong to the quantum statistical simulation in Fig.~2 of the main manuscript. Figure~\ref{Fig:S_CarrierRelaxation} shows the relaxation of photoexcited electrons in the conduction band at three different time steps with $t=\SI{0.0}{ps}$ (red), $t=\SI{1.0}{ps}$ (orange) and $t=\SI{5.3}{ps}$ (blue). The electrons are excited with a $\mathrm{sech}^2$ pulse ($\mathrm{FWHM}=\SI{200}{fs}$) and an initial excess energy of $\sim\SI{60}{meV}$ with respect to the non-renormalized band gap ($E_{\mathrm{g}0,10\,\mathrm{K}}=\SI{1.5192}{eV}$~\cite{Passler1997}). Due to the finite discretization in k-space, this excitation injects the carriers into two k-points. During the excitation process, the carrier distribution (red) is clearly non-thermal, exhibiting a peak where the carriers are injected, a very weak LO-phonon replica at $\sim\SI{20}{meV}$ and a certain fraction of electrons that already relaxed down to the conduction band edge. At $t=\SI{1.0}{ps}$ the electron distribution (orange) is already thermalized and takes on the form of a hot Fermi-Dirac distribution with a temperature of $T_{\mathrm{e}}=\SI{154}{K}$. This is consistent with literature, where thermalization would be expected within a few hundred femtoseconds~\cite{Oudar1985_ThermalizationAt15KGaAs,Elsaesser1991_ThermalizationAt300KGaAs,Kane1994a_ThermalizationWithin150fs}. The subsequent cooling of this hot, thermalized electron distribution is dominated by carrier-LO-phonon scattering~\cite{Shah1999_UltrafastSpectroscopyBook}, as already mentioned in the main manuscript. This cooling leads to an accumulation of electrons at the bottom of the conduction band and thus gradually changes the distribution from the orange to the blue curve ($t=\SI{5.3}{ps}$, $T_{\mathrm{e}}=\SI{79}{K}$). This increases the gain near the band edge and enables the laser to turn-on shortly after at $t_{\mathrm{on}}=\SI{5.7}{ps}$. A similar discussion to the above also applies to the holes in the valence band.

\clearpage
\section{Influence of the confinement factor on the lasing dynamics}
\begin{figure*}[ht]
    \includegraphics[keepaspectratio]{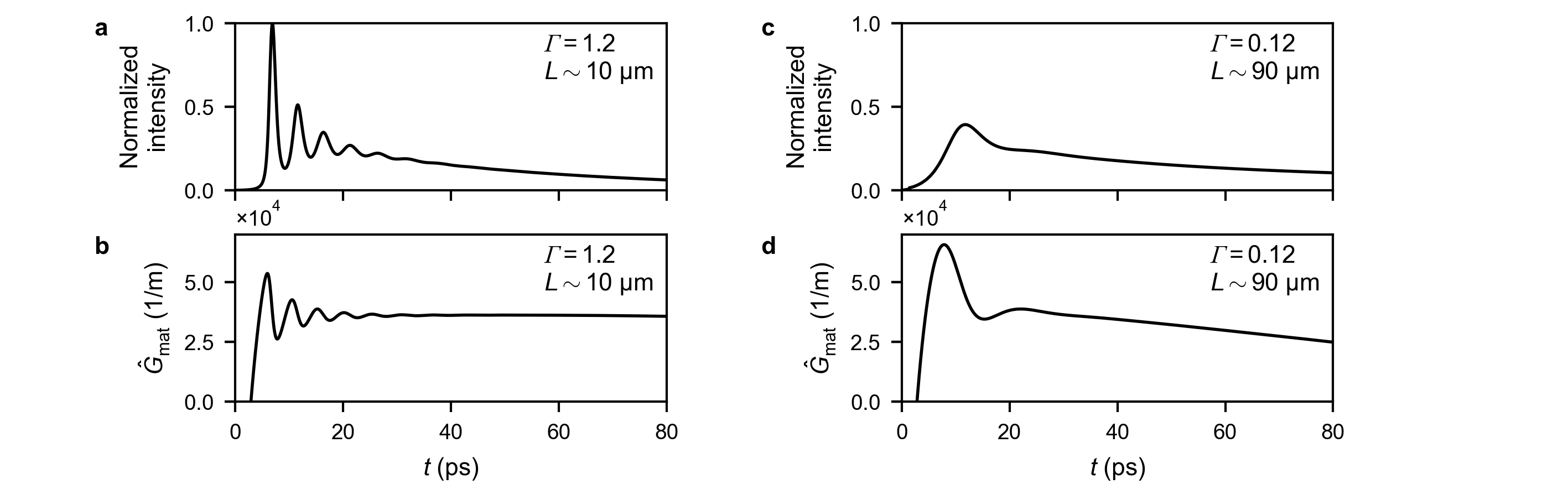}
    \caption{\textbf{Confinement factor and lasing dynamics.} \textbf{a},\textbf{b}, Intensity and peak material gain ($\hat{G}_{\mathrm{mat}}$) of a NW laser with a confinement factor of $\mathit{\Gamma}=1.2$ and a length of $L\sim\SI{10}{\micro m}$, exhibiting clear oscillations in the laser output intensity. \textbf{c},\textbf{d}, Dynamics of a similar laser with $\mathit{\Gamma}=0.12$ and $L\sim\SI{90}{\micro m}$, showing no oscillations in the laser output. The intensities in \textbf{a},\textbf{c} were normalized to the maximum value in \textbf{a}.}
    \label{Fig:S_ConfinementFactorAndLaserDynamics}
\end{figure*}

Using the quantum statistical model, we illustrate the strong influence of $\mathit{\Gamma}$ on the lasing dynamics by comparing two similar lasers with markedly different $\mathit{\Gamma}$. Both lasers were excited with single pulses and an excitation power of $P/P_{\mathrm{th}}=2.5$. Figure~\ref{Fig:S_ConfinementFactorAndLaserDynamics}a shows the dynamics of the NW investigated in this work, with a confinement factor of $\mathit{\Gamma}=1.2$ and a length of $L\sim\SI{10}{\micro m}$, see Table~\ref{Table:SimulationParameters}. The corresponding peak material gain ($\hat{G}_{\mathrm{mat}}$) is presented in Fig.~\ref{Fig:S_ConfinementFactorAndLaserDynamics}b. For comparison, Fig.~\ref{Fig:S_ConfinementFactorAndLaserDynamics}c and Fig.~\ref{Fig:S_ConfinementFactorAndLaserDynamics}d show the intensity and $\hat{G}_{\mathrm{mat}}$ of a laser with $\mathit{\Gamma}=0.12$, respectively. A similar threshold and $\hat{G}_{\mathrm{mat}}$ as for the case with high $\mathit{\Gamma}$ was achieved by adjusting the losses via $L$. No intensity oscillations are observed, demonstrating the important role $\mathit{\Gamma}$ plays for the ultrafast dynamics in Fig.~\ref{Fig:S_ConfinementFactorAndLaserDynamics}a. Except $\mathit{\Gamma}$ and $L$, all parameters were the same for both simulations and are listed in Table~\ref{Table:SimulationParameters}.

\clearpage
\section{Determination of the oscillation frequency}
\label{Sec:OscillationFrequency}
\begin{figure*}[ht]
    \includegraphics[keepaspectratio]{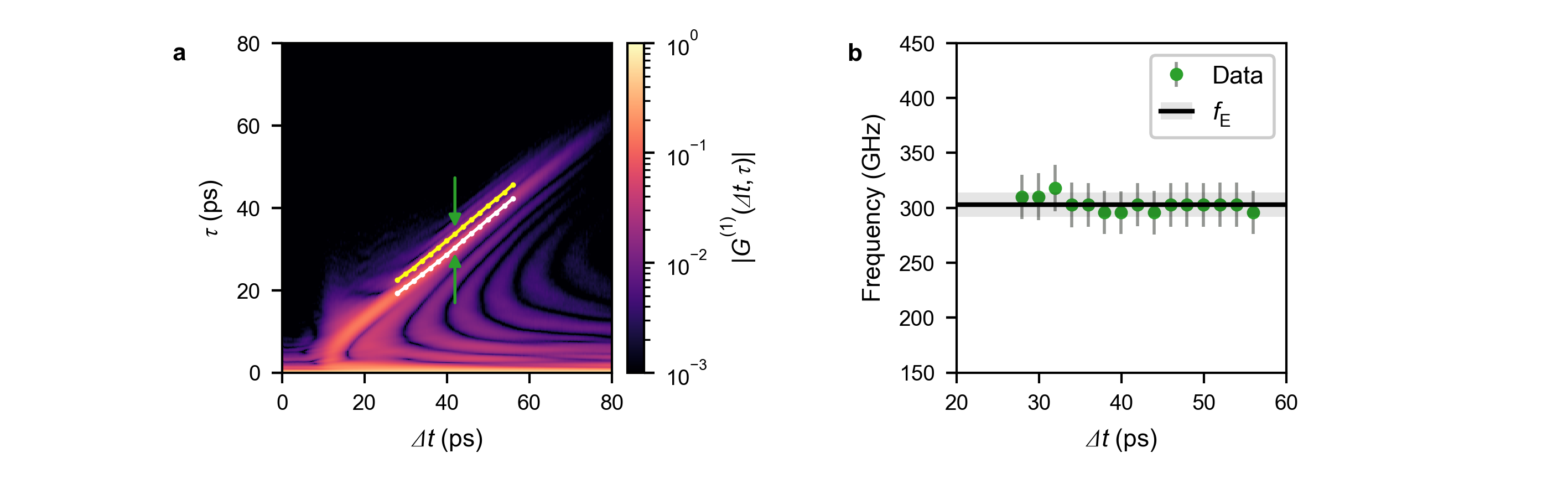}
    \caption{\textbf{Determination of the oscillation frequency.} \textbf{a}, Measured $|G^{(1)}(\mathit{\Delta} t, \tau)|$ as a function of pump-probe delay $\mathit{\Delta}t$ and time shift $\tau$. For a selected range of $\mathit{\Delta}t$, the white line indicates the maximum of the main sideband, whereas the yellow line indicates the maximum of the first oscillation above. For each selected $\mathit{\Delta}t$ value (indicated by points), we calculated the oscillation frequency by evaluating the distance between the white and yellow maxima, as indicated by the green arrows. \textbf{b}, Resulting oscillation frequencies as a function of $\mathit{\Delta}t$. For easier comparison, the frequency axis is scaled in the same way as in Fig.~3b of the main manuscript. Error bars represent the standard deviation resulting from the Fourier transform of the spectral data. We obtained the oscillation frequency $f_{\mathrm{E}}$ by averaging over all green data points and indicated the corresponding $\SI{95}{\%}$ CI with the grey shaded area.}
    \label{Fig:S_DeterminationOfOscillationFrequency_210219}
\end{figure*}

Here, we discuss the procedure that is used to determine the experimental ($f_{\mathrm{E}}$) and simulated ($f_{\mathrm{S}}$) oscillation frequency from the respective magnitude of the electric field autocorrelation $|G^{(1)}(\mathit{\Delta}t,\tau)|$. In the following, we use an example that was taken from the experimental excitation power series in Fig.~3b of the main manuscript. The selected $|G^{(1)}(\mathit{\Delta}t,\tau)|$ is presented in Figure~\ref{Fig:S_DeterminationOfOscillationFrequency_210219}a and was measured with an excitation power of $P_{\mathrm{pump}}/P_{\mathrm{th}}\sim\SI{2.50}{}$ and $P_{\mathrm{probe}}/P_{\mathrm{th}}\sim\SI{0.63}{}$. We calculated $f_{\mathrm{E}}$ from the time interval (indicated by green arrows) between the main sideband (white) and the first oscillation above (yellow). There are three reasons for this. Firstly, this oscillation is mainly influenced by the first output pulse (see Supplementary Section~\ref{Sec:ComparisonElectricVsIntensityAC}). Secondly, there the changes with $\mathit{\Delta} t$ are small, and thirdly, this feature is also pronounced in the simulated $|G^{(1)}(\mathit{\Delta}t,\tau)|$ (see Fig.~1f). To avoid outliers, we determined the oscillation frequency over a range of different $\mathit{\Delta}t$. In this example, the selected region ranged from $\mathit{\Delta} t\sim\SI{28}{ps}$ to $\mathit{\Delta}t\sim\SI{56}{ps}$ with a step size of $\SI{2}{ps}$, as indicated in Fig.~\ref{Fig:S_DeterminationOfOscillationFrequency_210219}a. The obtained oscillation frequencies are presented in Fig.~\ref{Fig:S_DeterminationOfOscillationFrequency_210219}b as a function of $\mathit{\Delta}t$ (green), whereby the error bars represent an estimate of the standard deviation resulting from the Fourier transform of the spectral data. By averaging, we obtained $f_{\mathrm{E}}=\SI[separate-uncertainty=true]{302(16)}{GHz}$ (black line). The error represents the $\SI{95}{\%}$ CI of the mean and is indicated in Fig.~\ref{Fig:S_DeterminationOfOscillationFrequency_210219}b by the grey shaded area. The same procedure, as discussed above, is used to determine $f_{\mathrm{S}}$ from the simulated $|G^{(1)}(\mathit{\Delta}t,\tau)|$.

\clearpage
\section{Comparison of electric field and intensity autocorrelation}
\label{Sec:ComparisonElectricVsIntensityAC}
Here, we compare the electric field autocorrelation $G^{(1)}(\mathit{\Delta}t,\tau)$ with the intensity autocorrelation $G^{(2)}(\mathit{\Delta}t,\tau)$. To facilitate this, we define the emitted electric field as $E(t)=A(t)\cdot\mathrm{exp}[i\mathit{\Phi}(t)]$, where $A(t)$ and $\mathit{\Phi}(t)$ denote the time-dependent amplitude and phase, respectively. With this definition one can write $G^{(1)}(\mathit{\Delta}t,\tau)$ as~\cite{Loudon2000_TheQuantumTheoryOfLight} 
\begin{equation}
    G^{(1)}(\mathit{\Delta}t,\tau)=\frac{1}{T}\int\limits_{T}E^{*}(t)E(t+\tau)dt=\frac{1}{T}\int\limits_{T}A(t)A(t+\tau)e^{i[\mathit{\Phi}(t+\tau)-\mathit{\Phi}(t)]}dt
    \label{eq:G1}
\end{equation}
and $G^{(2)}(\mathit{\Delta}t,\tau)$ as
\begin{equation}
    G^{(2)}(\mathit{\Delta}t,\tau)=\frac{1}{T}\int\limits_{T}I(t)I(t+\tau)dt=\frac{1}{T}\int\limits_{T}[A(t)A(t+\tau)]^2dt
    \label{eq:G2}
\end{equation}
where $T$ is the range of integration and $I(t)=|E(t)|^{2}$ the time-dependent laser intensity. From equation \eqref{eq:G1} and \eqref{eq:G2} it is apparent that $G^{(1)}(\mathit{\Delta}t,\tau)$ depends on both $A(t)$ and $\mathit{\Phi(t)}$, whereas $G^{(2)}(\mathit{\Delta}t,\tau)$ is only influenced by $A(t)$. Here, it is important to note that $A(t)$ and $\mathit{\Phi}(t)$ are not independent of each other, but coupled. This is because the carrier temperature oscillations (as discussed in the main manuscript and Supplementary Section~\ref{Sec:BlochModel}) modulate the complex susceptibility whose real and imaginary parts are related by Kramers-Kronig relations. Thus, the carrier temperature modulates both $A(t)$ and $\mathit{\Phi(t)}$ with the same periodicity, which then also leads to oscillations in $G^{(1)}(\mathit{\Delta}t,\tau)$ and $G^{(2)}(\mathit{\Delta}t,\tau)$. 

\begin{figure*}[ht]
    \includegraphics[keepaspectratio]{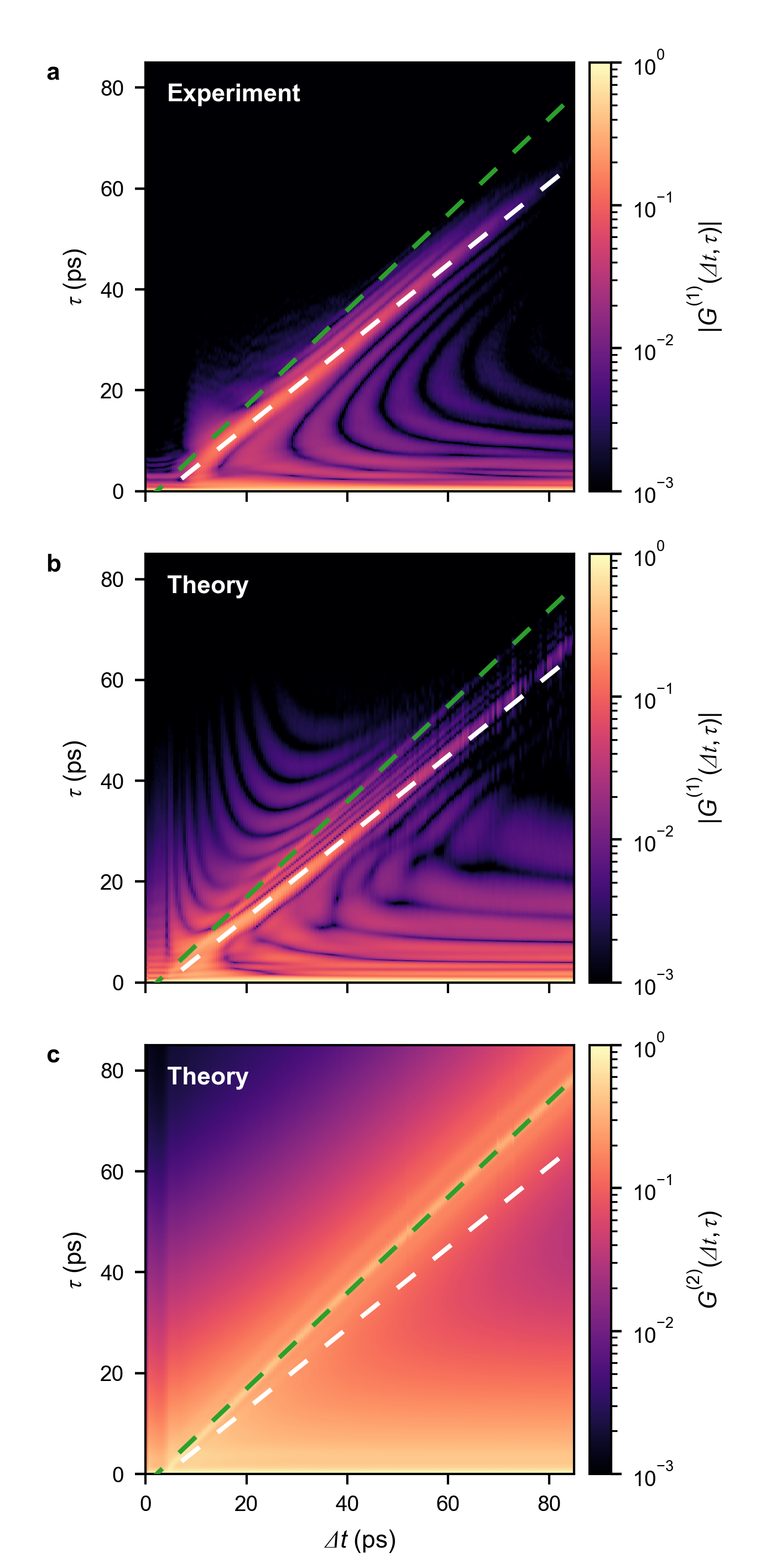}
    \caption{\textbf{Comparison of electric field and intensity autocorrelation.} \textbf{a},\textbf{b} Experimental and simulated $|G^{(1)}(\mathit{\Delta} t,\tau)|$. \textbf{c}, Simulated intensity autocorrelation $G^{(2)}(\mathit{\Delta} t,\tau)$ obtained from the same simulation as in \textbf{b}. In all panels, the white and green dashed lines indicate the scaling of the main sideband of $|G^{(1)}(\mathit{\Delta} t,\tau)|$ and $G^{(2)}(\mathit{\Delta} t,\tau)$ with $\mathit{\Delta} t$, respectively.}
    \label{Fig:S_Comparison_G1_vs_G2}
\end{figure*}

An example where this can be observed is shown in Fig.~\ref{Fig:S_Comparison_G1_vs_G2}a and Fig.~\ref{Fig:S_Comparison_G1_vs_G2}b, where we reproduced the normalized magnitude of the measured and simulated $G^{(1)}(\mathit{\Delta}t,\tau)$ from Fig.~1 of the main manuscript. The corresponding simulated $G^{(2)}(\mathit{\Delta}t,\tau)$ is presented in Fig.~\ref{Fig:S_Comparison_G1_vs_G2}c and, as expected, oscillations near the main sideband are also observed there. Comparing Fig.~\ref{Fig:S_Comparison_G1_vs_G2}b and Fig.~\ref{Fig:S_Comparison_G1_vs_G2}c, the influence of the phase-dependent part of equation~\eqref{eq:G1} is directly apparent, since it is naturally absent in equation~\eqref{eq:G2} and thus Fig.~\ref{Fig:S_Comparison_G1_vs_G2}c. To allow for a more detailed comparison, we marked the main sideband of $|G^{(1)}(\mathit{\Delta}t,\tau)|$ and $G^{(2)}(\mathit{\Delta}t,\tau)$ in all panels with white and green dashed lines, respectively. As expected, the sideband of $G^{(2)}(\mathit{\Delta}t,\tau)$, which is only influenced by $A(t)$, scales linearly with $\mathit{\Delta t}$ and has a slope close to $1$. A faint remnant of this sideband is visible in Fig.~\ref{Fig:S_Comparison_G1_vs_G2}b, whereas in Fig.~\ref{Fig:S_Comparison_G1_vs_G2}a it is completely gone. In contrast, the clearly visible sideband of $|G^{(1)}(\mathit{\Delta}t,\tau)|$ in Fig.~\ref{Fig:S_Comparison_G1_vs_G2}a,b only has a slope of $\sim0.8$ (white dashed line). This shows that, in the present case, $G^{(1)}(\mathit{\Delta}t,\tau)$ is strongly influenced by the phase-dependent factor in equation~\eqref{eq:G1} and less so by its amplitude-dependent part. In Fig.~\ref{Fig:S_Comparison_G1_vs_G2}a we furthermore observe that $|G^{(1)}(\mathit{\Delta}t,\tau)|$ almost completely vanishes above the green dashed line. Due to the long output pulse lengths of up to $\sim\SI{76}{ps}$ (see Fig.~1 and Fig.~2 of the main manuscript), one would expect a signal in this region, as can be seen when comparing with Fig.~\ref{Fig:S_Comparison_G1_vs_G2}c. This reveals that the second output pulse quickly looses its phase coherence with the first output pulse. The simulation in Fig.~\ref{Fig:S_Comparison_G1_vs_G2}b shows a similar behaviour, although not as pronounced as in the experiment in Fig.~\ref{Fig:S_Comparison_G1_vs_G2}a. 

As a consequence, the oscillations near the main sideband of $|G^{(1)}(\mathit{\Delta}t,\tau)|$ are mainly determined by the first output pulse. Using the method described in Supplementary Section~\ref{Sec:OscillationFrequency}, we obtained an oscillation frequency of $f_{\mathrm{S}}=\SI{350}{GHz}$ for the simulated $|G^{(1)}(\mathit{\Delta}t,\tau)|$ in Fig.~\ref{Fig:S_Comparison_G1_vs_G2}b. Among other effects, $f_{\mathrm{S}}$ does not directly correspond to the intensity oscillation frequency $\nu_{\mathrm{S}}$ of the laser output, due to the influence of the carrier density dependent refractive index on the phase evolution. For comparison, in Supplementary Section~\ref{Sec:BlochModel} we determined $\nu_{\mathrm{S}}=\SI{253}{GHz}$. The semiconductor Bloch model therefore predicts that, in the present case, the frequencies $f_{\mathrm{S}}$ and $\nu_{\mathrm{S}}$ are related by $f_{\mathrm{S}}=1.38\cdot\nu_{\mathrm{S}}$, which is confirmed by Fig.~3b and Fig.~4e in the main manuscript.

\clearpage

\clearpage
\section{Output pulse length as a function of lattice temperature}
\begin{figure*}[ht]
    \includegraphics[keepaspectratio]{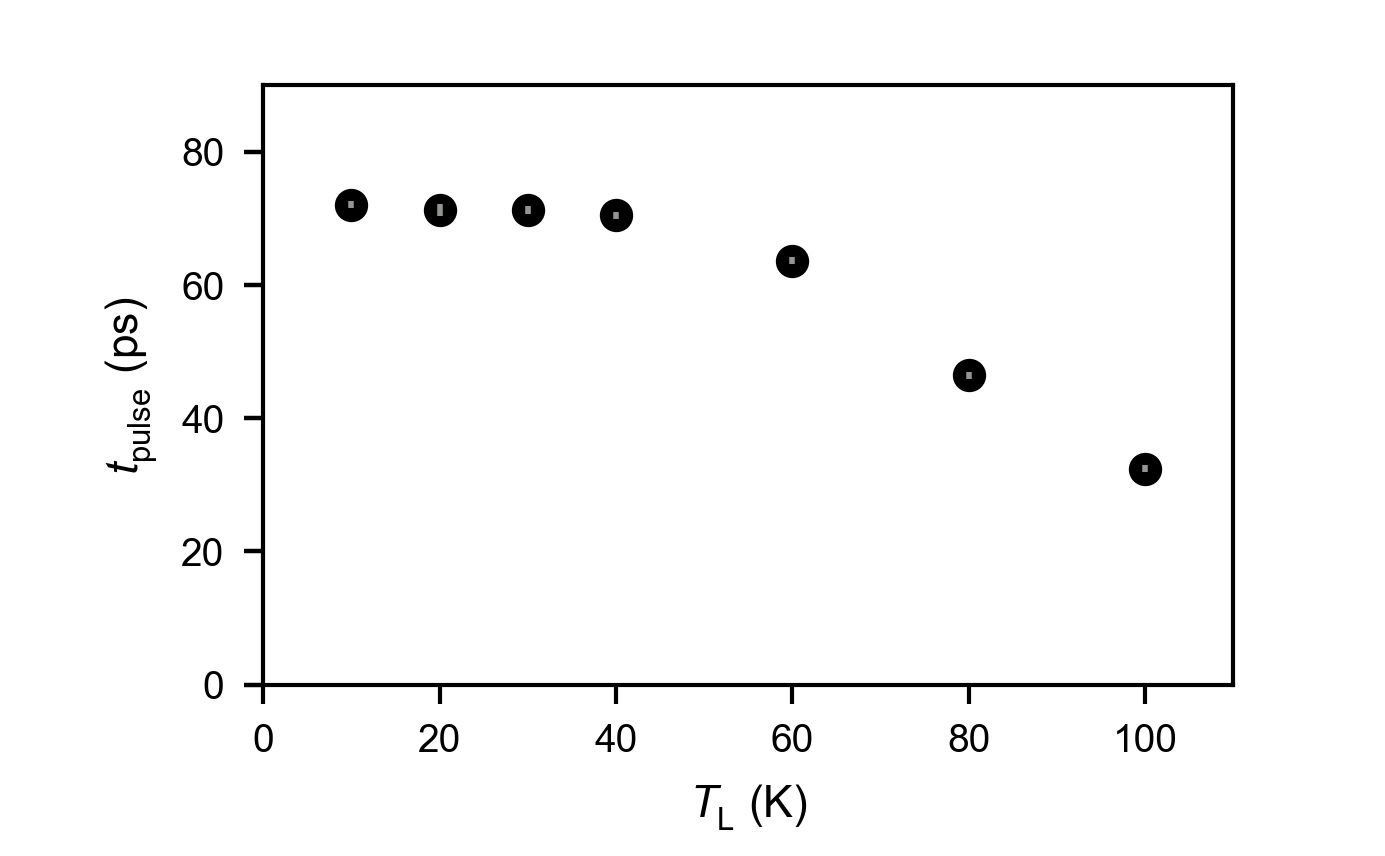}
    \caption{\textbf{Lattice temperature dependence of the output pulse length.} As the lattice temperature ($T_{\mathrm{L}}$) increases, the output pulse length ($t_{\mathrm{pulse}}$) is at first constant and then decreases for $T_{\mathrm{L}}>\SI{40}{K}$. The error bars (grey) represent $\SI{95}{\%}$ CIs of the mean. They result from the errors that are associated with $t_{\mathrm{off}}$ and $t_{\mathrm{on}}$, and are smaller than the black data points.}
    \label{Fig:S_TpulseVsLatticeTemperature}
\end{figure*}

Fig.~\ref{Fig:S_TpulseVsLatticeTemperature} presents the dependence of $t_{\mathrm{pulse}}\sim t_{\mathrm{off}}-t_{\mathrm{on}}$ on $T_{\mathrm{L}}$. The data correspond to the $T_{\mathrm{L}}$ series shown in Fig.~4 of the main manuscript and were measured with the excitation powers fixed at $P_{\mathrm{pump}}/P_{\mathrm{th}}\sim2.5$ and $P_{\mathrm{probe}}/P_{\mathrm{th}}\sim0.5$. For $T_{\mathrm{L}}\leq\SI{40}{K}$, $t_{\mathrm{pulse}}$ remains constant at $\sim\SI{71}{ps}$ and then decreases down to a value of $\SI[separate-uncertainty = true]{32.4(6)}{ps}$ at $T_{\mathrm{L}}=\SI{100}{K}$. This decrease is connected to the shift of the lasing mode towards the high energy side of the gain spectrum as $T_{\mathrm{L}}$ increases (see discussion of Fig.~4). There, the lasing mode experiences a larger differential material gain with respect to carrier density ($\partial G_{\mathrm{mat}} / \partial N$)~\cite{Grabmaier1991_DiffGainIncreasesWithIncreasingEnergy}, which leads to a decrease of $t_{\mathrm{pulse}}$. This explains the behaviour seen in Fig.~\ref{Fig:S_TpulseVsLatticeTemperature} and is in accord with previous work on microcavity lasers~\cite{Michler1995_TransientPulseResponse,Michler1996_EmissionDynamicsOfVCSELs}.

\clearpage
\section{Semiconductor Bloch model with relaxation rate approximation} 
The conduction and valence band are described in the parabolic band approximation. Furthermore, the equations are formulated in k-space, with a non-equidistant discretization that has an increased density close to the band edge. In the following, the carrier type is denoted by $a=e,h$ for electrons and holes, respectively, and their wavevector by $\mathbf{k}$. Both carrier-carrier and carrier-LO-phonon scattering are treated in the relaxation time approximation with fixed scattering rates $\gamma_{\mathrm{cc},\mathrm{a}}$ and $\gamma_{\mathrm{ap}}$, respectively, see Table~\ref{Table:ScatteringRatesBloch}. The laser dynamics are then described by the occupation probabilities $n_{\mathrm{a}\mathbf{k}}$, microscopic polarization amplitudes $p_{\mathbf{k}}$ and the slowly varying electric field envelope $E$. The equations for the occupation probabilities $n_{\mathrm{a}\mathbf{k}}$ read~\cite{Chow1994_SemiconductorLaserPhysicsBook}
\begin{equation}
    \frac{d}{dt}n_{\mathrm{a}\mathbf{k}}=2\operatorname{Im}(p_{\mathbf{k}}\Omega_{\mathbf{k}}^{*})-\gamma_{1}n_{\mathrm{a}\mathbf{k}}+\gamma_{\mathrm{cc},\mathrm{a}}[f(E_{\mathrm{a}\mathbf{k}},T_{\mathrm{a}},\mu_{\mathrm{ac}})-n_{\mathrm{a}\mathbf{k}}]+\gamma_{\mathrm{ap}}[f(E_{\mathrm{a}\mathbf{k}},T_{\mathrm{L}},\mu_{\mathrm{aL}})-n_{\mathrm{a}\mathbf{k}}].
\end{equation}
The first term represents stimulated recombination treated in the semiconductor Maxwell-Bloch equation framework, where $\Omega_{\mathbf{k}}$ is the renormalized Rabi frequency. The second term represents losses due to spontaneous emission and non-radiative processes with the rate $\gamma_{1}$, which is determined by the inverse of the carrier lifetime. The third term describes carrier-carrier scattering which conserves energy and drives the carriers towards a quasi-Fermi distribution $f$ with the carrier temperature $T_{\mathrm{a}}$ and Fermi energy $\mu_{\mathrm{ac}}$. The last term represents carrier-LO-phonon scattering which does not conserve energy and drives the carriers towards a quasi-Fermi distribution with lattice temperature $T_{\mathrm{L}}$ and Fermi energy $\mu_{\mathrm{aL}}$. The equation for the microscopic polarization amplitude is given by~\cite{Chow1994_SemiconductorLaserPhysicsBook}
\begin{equation}
    \frac{d}{dt}p_{\mathbf{k}}=-(\gamma_{2}+i\omega_{\mathbf{k}})p_{\mathbf{k}}+i(n_{\mathrm{e}\mathbf{k}}+n_{\mathrm{h}\mathbf{k}}-1)\Omega_{\mathbf{k}},
\end{equation}
where $\gamma_{2}$ is the dephasing rate and $\omega_{\mathbf{k}}$ the transition frequency. The transition energy $\hbar\omega_{\mathbf{k}}$ is given by equation~\eqref{eq:QM_TransitionFrequency}. The dynamics of the complex electric field envelope $E$ are determined by~\cite{Chow1994_SemiconductorLaserPhysicsBook} 
\begin{equation}
    \frac{d}{dt}E=\frac{i\omega_{0}\mathit{\Gamma}}{\varepsilon_{0}\varepsilon_{\mathrm{b}}}P-(\kappa+i\Delta\omega)E,
\end{equation}
with the optical frequency at the band edge $\omega_{0}$, the optical confinement factor $\mathit{\Gamma}$, the passive background permittivity of the host medium $\varepsilon_{\mathrm{b}}$, the macroscopic polarization $P$, the detuning of the lasing mode from the band edge $\Delta\omega$ and the field loss rate $\kappa=\frac{1}{2}\tau_{\mathrm{p}}$, where $\tau_{\mathrm{p}}$ is the photon lifetime. Finally, the macroscopic polarization is computed by~\cite{Chow1994_SemiconductorLaserPhysicsBook}
\begin{equation}
    P=2\sum_{\mathbf{k}}\mu_{\mathbf{k}}^{*}p_{\mathbf{k}},
\end{equation}
with the dipole moment $\mu_{\mathbf{k}}$. The above set of differential equations is solved with a fourth order Runge-Kutta method. The convergence is limited by the large transition frequencies of the higher k-states, which results in a required time discretization of $\Delta t_{\mathrm{disc}} \approx\SI{1e-16}{s}$.
\clearpage

\section{Quantum statistical model}
In this chapter we briefly summarize the equations that were used in the quantum statistical model. Thereby, we closely follow previous work~\cite{Jahnke1993c_CWSwitchOnVCSEL_Transient} and slightly adapted the equations and notation for our purposes. Within this model, both the electromagnetic field and the semiconductor gain medium are treated quantum mechanically. The theoretical framework is based on a non-equilibrium Green's function technique. For an extensive discussion of this technique and its application to semiconductor lasers we refer the reader to the literature \cite{korenman_nonequilibrium_1966,schaefer_approach_1986,henneberger_nonlinear_1988,Henneberger1992_SpectralHoleBurning,Henneberger1992_ManyBodyEffects,herzel_semiconductor_1993,Jahnke1993c_CWSwitchOnVCSEL_Transient,Jahnke1993a_CWSwitchOnVCSEL_Theoryof,Jahnke1993b_CWSwitchOnVCSEL_DynamicResponse,mohideen_semiconductor_1994,Jahnke1995_QuantumStatisticalTheoryPaper,Jahnke1995a_UltrafastSwitching,Schneider1997_UltrafastDynamicsOfVCSELs,Chow1994_SemiconductorLaserPhysicsBook}.
\subsection{Photon kinetics}
The model describes the temporal and spectral evolution of the laser intensity
\begin{equation}
    I(\omega,t)=\frac{\mu_{0}e_{0}^{2}}{2m_{0}^{2}}|M_{\mathrm{T}}|^{2}\frac{1}{V_{\mathrm{NW}}}\sum_{\mathbf{q}_{\mathrm{l}}}I_{\mathbf{q}_{\mathrm{l}}}(\omega,t),
\end{equation}
where $\mu_{0}$ is the vacuum permeability and $V_{\mathrm{NW}}$ the volume of the nanowire. The transition matrix element $e_{0}|M_{\mathrm{T}}|$ is related to the dipole moment $|\mu_{\mathbf{k}}|$ via
\begin{equation}
    e_{0}|M_{\mathrm{T}}|=m_{0}\omega|\mu_{\mathbf{k}}|.
\end{equation}
The laser intensity $I_{\mathbf{q}_{\mathrm{l}}}(\omega,t)$ for a given laser mode with wavevector $\mathbf{q}_{\mathrm{l}}$ is given by 
\begin{equation}
    I_{\mathbf{q}_{\mathrm{l}}}(\omega,t)=\int e^{i\omega\tau}\bar{I}_{\mathbf{q}_{\mathrm{l}}}(\tau,t)d\tau,
\end{equation}
where
\begin{equation}
    \bar{I}_{\mathbf{q}_{\mathrm{l}}}(\tau,t)=\langle A_{\mathbf{q}_{\mathrm{l}}}(t+\tau/2)A_{\mathbf{q}_{\mathrm{l}}}(t-\tau/2)\rangle
\end{equation}
is the field correlation function with respect to $\tau$. Here, $\mathbf{A}_{\mathbf{q}_{\mathrm{l}}}(t)$ is the vector potential of the quantized laser field. The kinetic equation for the laser intensity is
\begin{equation}
    \frac{1}{\tilde{F}(\omega)}\frac{1}{c}\frac{\partial I(\omega)}{\partial t}=\left[G_{\mathrm{mod}}(\omega)-\frac{2\kappa}{\tilde{F}(\omega)}\right]I(\omega)+\beta W(\omega)S(\omega),
\end{equation}
where $\tilde{F}(\omega)$ is an approximation for the cavity response function $F(\mathbf{q}_{\mathrm{l}})$\cite{Jahnke1993c_CWSwitchOnVCSEL_Transient}, $c=c_{0}/n(\omega)$ the renormalized speed of light with the frequency-dependent refractive index $n(\omega)$~\cite{Reinhart2005a, Reinhart2005b}, $G_{\mathrm{mod}}$ the modal intensity gain, $\kappa$ the amplitude loss rate, $\beta$ the spontaneous emission factor, $W(\omega)$ the spontaneous emission rate and $S(\omega)$ the photon density of states. The photon density of states is given by
\begin{equation}
    S(\omega)=\frac{\mu_{0}e_{0}^{2}}{2m_{0}^{2}}|M_{\mathrm{T}}|^{2}\frac{1}{V_{\mathrm{NW}}}\sum_{\mathbf{q}_{\mathrm{l}}}S_{\mathbf{q}_{\mathrm{l}}}(\omega),
\end{equation}
with the photon lineshape function
\begin{equation}
    S_{\mathbf{q}_{\mathrm{l}}}(\omega)=F(\mathbf{q}_{\mathrm{l}})\frac{2\chi_{\mathbf{q}_{\mathrm{l}}}(\omega)}{\left(\mathbf{q}_{\mathrm{l}}^2-\left(\frac{\omega}{c}\right)^{2}\right)^{2}+\chi_{\mathbf{q}_{\mathrm{l}}}(\omega)^{2}},
\end{equation}
where $\chi_{\mathbf{q}_{\mathrm{l}}}$ is the photon damping. The cavity response function $F(\mathbf{q}_{\mathrm{l}})$ and its approximation $\tilde{F}(\omega)$ are given by
\begin{equation}
    F(\mathbf{q}_{\mathrm{l}})=\frac{C_{\mathrm{FP}}}{2}\left[\frac{1}{C_{\mathrm{FP}}^2\cos^2\left(\mathrm{q}_{\mathrm{l}}L/2\right)+\sin^2\left(\mathrm{q}_{\mathrm{l}}L/2\right)}\right].
\end{equation}
and
\begin{equation}
    \tilde{F}(\omega)=\frac{2\pi}{q_{\mathrm{app}}(\omega)}S(\omega),
\end{equation}
with $q_{\mathrm{app}}(\omega)=\omega/c$ and $C_{\mathrm{FP}}=(1+R)/(1-R)$. The photon damping is described by
\begin{equation}
    \chi_{\mathbf{q}_{\mathrm{l}}}(\omega)=\frac{\omega}{c}\left[F(\mathbf{q}_{\mathrm{l}})G_{\mathrm{mod}}-2\kappa\right],
\end{equation}
with
\begin{equation}
    \kappa=-\frac{1}{2L}\ln\left(R\right),
\end{equation}
where $L$ is the length of the nanowire and $R=r^{2}$, with $r$ being the amplitude reflection coefficient. The modal gain of the laser intensity is connected to the confinement factor $\Gamma$ and the material gain $G_{\mathrm{mat}}$ as following
\begin{equation}
    G_{\mathrm{mod}}=\Gamma\cdot G_{\mathrm{mat}}.
\end{equation}
Here, the confinement factor is calculated by\cite{visser_confinement_1997,maslov_modal_2004,Ning2010a_ConfinementFactor}
\begin{equation}
    \Gamma=\frac{2\varepsilon_{0}c_{0}n_{\mathrm{active}}\int\int_{\mathrm{active}}\mathrm{d}x\mathrm{d}y|\mathbf{E}|^2}{\int\int \mathrm{d}x\mathrm{d}y\left[\mathbf{E}\times\mathbf{H}^{*}+\mathbf{E}^{*}\times\mathbf{H}\right]\cdot\hat{z}},
\end{equation}
where $\mathbf{E}$ and $\mathbf{H}$ is the electric and magnetic field of the mode, $n_{\mathrm{active}}$ the refractive index of the active region, and $\hat{z}$ the unit vector along the nanowire axis. The material gain of the semiconductor medium is described by
\begin{equation}
    G_{\mathrm{mat}}(\omega)=\frac{\mu_{0}e_{0}^{2}}{2m_{0}^{2}}\frac{c}{\omega}\sum_{\mathbf{k}}|M_{\mathrm{T}}|^{2}\mathcal{L}_{\mathbf{k}}\left(\omega\right)\left[n_{\mathrm{e}\mathbf{k}}+n_{\mathrm{h}\mathbf{k}}-1\right]\left[1+2w_{\mathbf{k}}(\omega)\right],
\end{equation}
where $n_{a\mathbf{k}}$ ($a=e,h$) are the carrier distribution functions, $\mathcal{L}_{\mathbf{k}}\left(\omega\right)$ is a Lorentzian lineshape function and $w_{\mathbf{k}}(\omega)$ the Coulomb enhancement factor. The lineshape function is described by
\begin{equation}
    \mathcal{L}_{\mathbf{k}}=2\frac{\gamma_{\mathrm{eh}}(\mathbf{k})}{\left(\hbar\omega_{\mathbf{k}}-\hbar\omega\right)^{2}+\gamma^{2}_{\mathrm{eh}}\left(\mathbf{k}\right)},
\end{equation}
whereby $\gamma_{\mathrm{eh}}(\mathbf{k})$ is the broadening as defined in equation~\eqref{eq:Broadening}. The Coulomb enhancement is given by
\begin{equation}
    w_{\mathbf{k}}(\omega)=\sum_{\mathbf{k}^{'}\neq\mathbf{k}}\frac{\hbar\omega_{\mathbf{k}^{'}}-\hbar\omega}{\left(\hbar\omega_{\mathbf{k}^{'}}-\hbar\omega\right)^{2}+\gamma_{\mathrm{eh}}^{2}(k)}\left[1-n_{\mathrm{e}\mathbf{k}^{'}}-n_{\mathrm{h}\mathbf{k}^{'}}\right]V_{\mathrm{s}\mathbf{q}}(k-k^{'}),
\end{equation}
where
\begin{equation}
    V_{\mathrm{s}\mathbf{q}}=V_{\mathbf{q}}\left[1-\frac{\omega^{2}_{\mathrm{pl}}}{\omega_\mathbf{q}^{2}}\right]
\end{equation}
is an approximation for the screened Coulomb potential energy. Here,
\begin{equation}
    V_{\mathbf{q}}=\frac{1}{V_{\mathrm{NW}}}\int d^{3}r e^{-i\mathbf{q}\cdot\mathbf{r}}V(r)
\end{equation}
is the Fourier transform of the Coulomb potential energy $V(r)$. The plasma frequency $\omega_{\mathrm{pl}}$ is expressed as~\cite{Chow1994_SemiconductorLaserPhysicsBook,haug_quantum_2004} 
\begin{equation}
    \omega_{\mathrm{pl}}^{2}=\frac{N e_{0}^{2}}{\varepsilon_{r}\varepsilon_{0}m_r},
\end{equation}
with the total carrier density $N$, the reduced electron-hole mass $m_{\mathrm{r}}=m_{\mathrm{e}}m_{\mathrm{h}}/\left(m_{\mathrm{e}}+m_{\mathrm{h}}\right)$ and the relative static dielectric constant $\varepsilon_{\mathrm{r}}$. The effective plasmon frequency $\omega_{\mathbf{q}}$ is given by\cite{Chow1994_SemiconductorLaserPhysicsBook}
\begin{equation}
    \omega^{2}_{\mathbf{q}}=\omega^2_{\mathrm{pl}}\left(1+\frac{\mathbf{q}^2}{\kappa_{\mathrm{sc}}^{2}}\right)+C\left(\frac{\hbar\mathbf{q}^2}{4 m_{\mathrm{r}}}\right)^{2},
\end{equation}
where $C$ is a numerical constant that can have a value between $1$ and $4$~\cite{Chow1994_SemiconductorLaserPhysicsBook}, and $\kappa_{\mathrm{sc}}$ is the inverse static screening length defined by\cite{Chow1994_SemiconductorLaserPhysicsBook} 
\begin{equation}
    \kappa_{\mathrm{sc}}^{2}=-\frac{e_{0}^{2}}{\varepsilon_{\mathrm{r}}\varepsilon_{0}}\sum_{\mathbf{k},\mathrm{a}=\mathrm{e},\mathrm{h}}\frac{\partial n_{\mathrm{a}\mathbf{k}}}{\partial E_{\mathrm{a}\mathbf{k}}}.
\end{equation}
Here, $E_{\mathrm{a}\mathbf{k}}=\hbar^2\mathbf{k}^2/(2m_{\mathrm{e}})$ is the respective $\mathbf{k}$-dependent energy of electrons and holes. For high carrier densities, the transition energy $\hbar\omega_{\mathbf{k}}$ is influenced by band gap renormalization effects, which results in
\begin{equation}
    \hbar\omega_{\mathbf{k}}=\frac{\hbar^2k^2}{2m_{\mathrm{r}}}+E_{\mathrm{g}0}+\mathit{\Delta}E_{\mathrm{CH}}+\mathit{\Delta}E_{\mathrm{SX},\mathbf{k}},
    \label{eq:QM_TransitionFrequency}
\end{equation}
where $E_{\mathrm{g}0}$ is the unexcited band gap energy. For better readability, indices are separated by a comma when necessary. The Coulomb-hole self energy can be calculated by\cite{Chow1994_SemiconductorLaserPhysicsBook}
\begin{equation}
    \mathit{\Delta}E_{\mathrm{CH}}=\sum_{\mathbf{q}\neq 0}\left[V_{\mathrm{s}\mathbf{q}}-V_{\mathbf{q}}\right]=-2E_{\mathrm{R}}\frac{\alpha_{0}\kappa_{\mathrm{sc}}}{\sqrt{1+\sqrt{C}\alpha_{0}^{2}\kappa_{\mathrm{sc}}^{2}E_{\mathrm{R}}/(\hbar\omega_{\mathrm{pl}})}},
\end{equation}
where $E_{\mathrm{R}}$ is the exciton Rydberg energy and $\alpha_{0}$ the exciton Bohr radius, while the screened exchange energy is given by by\cite{Chow1994_SemiconductorLaserPhysicsBook}
\begin{equation}
    \mathit{\Delta}E_{\mathrm{SX},\mathbf{k}}=\sum_{\mathbf{k}^{'}\neq\mathbf{k}}V_{\mathrm{s},|\mathbf{k}^{'}-\mathbf{k}|}\left[n_{\mathrm{e}\mathbf{k}^{'}}+n_{\mathrm{h}\mathbf{k}^{'}}\right].
\end{equation}
Lastly, the frequency dependent spontaneous emission rate is expressed as
\begin{equation}
    W(\omega)=\frac{\mu_{0}e_{0}^{2}}{2m_{0}^{2}}\frac{c}{\omega}\sum_{\mathbf{k}}|M_{\mathrm{T}}|^{2}\mathcal{L}_{\mathbf{k}}(\omega)n_{\mathrm{e}\mathbf{k}}n_{\mathrm{h}\mathbf{k}}\left[1+2w_{\mathbf{k}}\left(\omega\right)\right].
\end{equation}
\subsection{Carrier kinetics}
The temporal evolution of the $\mathbf{k}$-dependent carrier distribution functions $n_{a\mathbf{k}}$ is described by the carrier kinetic equation:
\begin{equation}
    \frac{\partial}{\partial t}n_{\mathrm{a}\mathbf{k}}\left(t\right)=\left[\frac{\partial}{\partial t}n_{\mathrm{a}\mathbf{k}}\left(t\right)\right]_{\mathrm{relax}}+\left[\frac{\partial}{\partial t}n_{\mathrm{a}\mathbf{k}}\left(t\right)\right]_{\mathrm{pump}}+\left[\frac{\partial}{\partial t}n_{\mathrm{a}\mathbf{k}}\left(t\right)\right]_{\mathrm{spon}}+\left[\frac{\partial}{\partial t}n_{\mathrm{a}\mathbf{k}}\left(t\right)\right]_{\mathrm{stim}}.
\end{equation}
The spontaneous and stimulated emission rates are given by
\begin{equation}
    \left[\frac{\partial}{\partial t}n_{\mathrm{a}\mathbf{k}}\left(t\right)\right]_{\mathrm{spon}}=-\frac{n_{\mathrm{e}\mathbf{k}}n_{\mathrm{h}\mathbf{k}}}{\tau_{\mathrm{spon}}(\mathbf{k})}
\end{equation}
and
\begin{equation}
    \left[\frac{\partial}{\partial t}n_{\mathrm{a}\mathbf{k}}\left(t\right)\right]_{\mathrm{stim}}=-\frac{n_{\mathrm{e}\mathbf{k}}+n_{\mathrm{h}\mathbf{k}}-1}{\tau_{\mathrm{stim}}(\mathbf{k})},
\end{equation}
respectively, whereby the spontaneous and stimulated emission lifetimes are defined as
\begin{equation}
    \tau_{\mathrm{spon}}^{-1}(\mathbf{k})=\int\frac{\mathrm{d}\omega}{2\pi}\mathcal{L}_{\mathbf{k}}(\omega)S(\omega)\left[1+w_{\mathbf{k}(\omega)}\right]
\end{equation}
and
\begin{equation}
    \tau_{\mathrm{stim}}^{-1}(\mathbf{k})=\int\frac{\mathrm{d}\omega}{2\pi}\mathcal{L}_{\mathbf{k}}(\omega)I(\omega)\left[1+w_{\mathbf{k}(\omega)}\right].
\end{equation}
Carrier-carrier and carrier-LO-phonon scattering is described by a Boltzmann equation
\begin{equation}
    \left[\frac{\partial}{\partial t}n_{\mathrm{a}\mathbf{k}}\left(t\right)\right]_{\mathrm{relax}}=-n_{\mathrm{a}\mathbf{k}}\Sigma_{\mathrm{a}}^{\mathrm{out}}(\mathbf{k})+\left(1-n_{\mathrm{a}\mathbf{k}}\right)\Sigma_{\mathrm{a}}^{\mathrm{in}}(\mathbf{k}),
\end{equation}
where $\Sigma^{\mathrm{in}}_{\mathrm{a}}(\mathbf{k})$ and $\Sigma^{\mathrm{out}}_{\mathrm{a}}(\mathbf{k})$ are the scattering rates in and out of a certain state with wavevector $\mathbf{k}$. Carrier-carrier scattering is treated in the $2^{\mathrm{nd}}$ Born approximation and the rates in and out of a state $\mathbf{k}$ are determined by~\cite{collet_model_1986,binder_carrier-carrier_1992}
\begin{equation}
    \begin{split}
        \Sigma^{\mathrm{in}}_{\mathrm{a}}(\mathbf{k})&=\frac{2\pi}{\hbar}\sum_{\mathbf{k}^{'},\mathbf{q}_{\mathrm{t}},b}\left[1-n_{b\mathbf{k}^{'}}\right]n_{b,\mathbf{k}^{'}+\mathbf{q}_{\mathrm{t}}}n_{\mathrm{a},\mathbf{k}-\mathbf{q}_{\mathrm{t}}}\\&\times\left[V_{\mathrm{s}\mathbf{q}_{\mathrm{t}}}^{2}+\left(V_{\mathrm{s}\mathbf{q}_{\mathrm{t}}}-\delta_{\mathrm{ab}}V^{\mathrm{ex}}_{\mathbf{k}-\mathbf{k}^{'}-\mathbf{q}_{\mathrm{t}}}\right)^{2}\right]\\&\times\delta\left[E_{\mathrm{a}\mathbf{k}}+E_{\mathrm{b}\mathbf{k}^{'}}-E_{\mathrm{a},\mathbf{k}-\mathbf{q}_{\mathrm{t}}}-E_{\mathrm{b},\mathbf{k}^{'}+\mathbf{q}_{\mathrm{t}}}\right]
    \end{split}
    \label{eq:Carrier-carrier_in}
\end{equation}
and
\begin{equation}
    \begin{split}
        \Sigma^{\mathrm{out}}_{\mathrm{a}}(\mathbf{k})&
        =\frac{2\pi}{\hbar}\sum_{\mathbf{k}^{'},\mathbf{q}_{\mathrm{t}},b}n_{\mathrm{b}\mathbf{k}^{'}}[1-n_{\mathrm{b},\mathbf{k}^{'}+\mathbf{q}_{\mathrm{t}}}]\left[1-n_{\mathrm{a},\mathbf{k}-\mathbf{q}_{\mathrm{t}}}\right]\\&    \times\left[V_{\mathrm{s}\mathbf{q}_{\mathrm{t}}}^{2}+\left(V_{\mathrm{s}\mathbf{q}_{\mathrm{t}}}-\delta_{\mathrm{ab}}V^{\mathrm{ex}}_{\mathbf{k}-\mathbf{k}^{'}-\mathbf{q}_{\mathrm{t}}}\right)^{2}\right]\\&
        \times\delta\left[E_{\mathrm{a}\mathbf{k}}+E_{\mathrm{b}\mathbf{k}^{'}}-E_{\mathrm{a},\mathbf{k}-\mathbf{q}_{\mathrm{t}}}-E_{\mathrm{b},\mathbf{k}^{'}+\mathbf{q}_{\mathrm{t}}}\right],
    \end{split}
    \label{eq:Carrier-carrier_out}
\end{equation}
respectively, with a,b = e,h. Here, $\delta_{\mathrm{ab}}$ is the Kronecker delta, $\delta$ the Dirac delta distribution, and $\mathbf{q}_{\mathrm{t}}$ describes the change in wavevector for two charge carriers with initial wavevectors $\mathbf{k}$ and $\mathbf{k}^{'}$. The exchange term $V^{\mathrm{ex}}_{\mathbf{k}}$ can be approximated by~\cite{collet_model_1986}
\begin{equation}
    V^{\mathrm{ex}}_{\mathbf{k}}=\frac{e_{0}^{2}}{\varepsilon_{0}\varepsilon_{r}V_{\mathrm{NW}}}\frac{1}{|\mathbf{k}|^{2}+\kappa_{\mathrm{sc}}^{2}}.
\end{equation}
Equations \eqref{eq:Carrier-carrier_in} and \eqref{eq:Carrier-carrier_out} describe electron-electron, electron-hole and hole-hole scattering. The rates for carrier-LO-phonon scattering are calculated by~\cite{collet_numerical_1983,collins_generation_1984,collet_model_1986}
\begin{equation}
    \begin{split}
    \Sigma_{\mathrm{a}}^{\mathrm{in}}(\mathbf{k})
    &
    =\frac{2\pi}{\hbar}\sum_{\mathbf{q}_{\mathrm{PN}}}M^{2}(\mathbf{q}_{\mathrm{PN}})\big\{n_{\mathrm{a},\mathbf{k}+\mathbf{q}_{\mathrm{PN}}}\left[1+n_{\mathrm{PN}}(\mathbf{q}_{\mathrm{PN}})\right]
    \\&
    \times\delta\left[E_{\mathrm{a},\mathbf{k}+\mathbf{q}_{\mathrm{PN}}}-E_{\mathrm{a}\mathbf{k}}-\hbar\omega_{\mathrm{LO}}(\mathbf{q}_{\mathrm{PN}})\right]
    \\&
    +n_{\mathrm{a},\mathbf{k}-\mathbf{q}_{\mathrm{PN}}}n_{\mathrm{PN}}(\mathbf{q}_{\mathrm{PN}})
    \\&
    \times\delta\left[E_{\mathrm{a},\mathbf{k}-\mathbf{q}_{\mathrm{PN}}}-E_{\mathrm{a}\mathbf{k}}+\hbar\omega_{\mathrm{LO}}(\mathbf{q}_{\mathrm{PN}})\right]\big\}
    \end{split}
\end{equation}
and
\begin{equation}
    \begin{split}
    \Sigma_{\mathrm{a}}^{\mathrm{out}}(\mathbf{k})
    &
    =\frac{2\pi}{\hbar}\sum_{\mathbf{q}_{\mathrm{PN}}}M^{2}(\mathbf{q}_{\mathrm{PN}})\big\{\left[1-n_{\mathrm{a},\mathbf{k}+\mathbf{q}_{\mathrm{PN}}}\right]n_{\mathrm{PN}}(\mathbf{q}_{\mathrm{PN}})
    \\&
    \times\delta\left[E_{\mathrm{a},\mathbf{k}+\mathbf{q}_{\mathrm{PN}}}-E_{\mathrm{a}\mathbf{k}}-\hbar\omega_{\mathrm{LO}}(\mathbf{q}_{\mathrm{PN}})\right]
    \\&
    +\left[1-n_{\mathrm{a},\mathbf{k}-\mathbf{q}_{\mathrm{PN}}}\right]\left[1+n_{\mathrm{PN}}(\mathbf{q}_{\mathrm{PN}})\right]
    \\&
    \times\delta\left[E_{\mathrm{a},\mathbf{k}-\mathbf{q}_{\mathrm{PN}}}-E_{\mathrm{a}\mathbf{k}}+\hbar\omega_{\mathrm{LO}}(\mathbf{q}_{\mathrm{PN}})\right]\big\}.
    \end{split}
\end{equation}
Here,
\begin{equation}
    M^{2}(\mathbf{q}_{\mathrm{PN}})=V_{\mathrm{s}\mathbf{q}}\left(\frac{1}{\varepsilon_{\infty}}-\frac{1}{\varepsilon_{\mathrm{r}}}\right)\frac{\hbar\omega_{\mathrm{LO}}(\mathbf{q}_{\mathrm{PN}})}{2}
\end{equation}
is the Fröhlich coupling and $\hbar\omega_{\mathrm{LO}}(\mathbf{q}_{\mathrm{PN}})$ the longitudinal optical phonon energy with wavevector $\mathbf{q}_{\mathrm{PN}}$. Under the assumption that the LO-phonons are always thermalized with a lattice temperature $T_{\mathrm{L}}$, the phonon population can be expressed by
\begin{equation}
    n_{\mathrm{PN}}(\mathbf{q}_{\mathrm{PN}})=\frac{1}{e^{\eta\hbar\omega_{\mathrm{LO}}(\mathbf{q}_{\mathrm{PN}})}-1},
\end{equation}
with $\eta=1/\left(k_{\mathrm{B}}T_{\mathrm{L}}\right)$. Depending on the experiment that is to be simulated the pumping term is either continuous wave
\begin{equation}
    \left[\frac{\partial}{\partial t}n_{\mathrm{a\mathbf{k}}}(t)\right]_{\mathrm{pump}}=P_{\mathrm{pump}}\left[1-n_{\mathrm{e\mathbf{k}}}(t)\right]\left[1-n_{\mathrm{h}\mathbf{k}}(t)\right]
\end{equation}
or pulsed
\begin{equation}
    \left[\frac{\partial}{\partial t}n_{\mathrm{a\mathbf{k}}}(t)\right]_{\mathrm{pump}}=P_{\mathrm{pump}}\left[1-n_{\mathrm{e\mathbf{k}}}(t)\right]\left[1-n_{\mathrm{h}\mathbf{k}}(t)\right]\mathrm{sech}^{2}\left(1.76\cdot\frac{t-t_{0}}{\mathit{\Delta}t_{\mathrm{pump}}}\right),
\end{equation}
where $P_{\mathrm{pump}}$ is the pump rate, $t_{0}$ the start of the optical excitation, and $\Delta t_{\mathrm{pump}}$ the FWHM of the excitation pulse. All equations were solved with a fourth order Runge-Kutta method with step size predictions. The integrals were numerically evaluated using a gaussian quadrature mesh.
\clearpage

\bibliography{supplementary}